\documentclass[12pt,headinclude,footinclude,twoside]{article}

\usepackage[dvips]{graphicx}
\usepackage{amsmath}
\usepackage{amsfonts}
\usepackage{placeins}
\usepackage{afterpage}
\usepackage{flafter}     
\usepackage{bbm}   

\usepackage[nosort]{cite}  
 
\usepackage{scrpage}

\usepackage[a4paper]{typearea}  
  \areaset[1.25cm]
          {17.5cm}{27cm}        

\renewpagestyle{headings}{(\textwidth,0.0pt)%
                {\itshape\headmark\hfill}
                {\itshape\headmark\hfill}
                {\itshape\hfill\headmark\hfill}%
                (\textwidth,.4pt)}%
               {(\textwidth,.0pt)%
                {\hfill\pagemark\hfill}%
                {\hfill\pagemark\hfill}%
                {}%
                (\textwidth,0.0pt)}

\newcommand{\PSImagxy}[3]{\includegraphics[width=#2,height=#3]{psplotssmall/#1}} 
\newcommand{\PSImagx}[2]{\includegraphics[width=#2]{psplotssmall/#1}}

\newcommand{\BILD}[4]{\begin{figure}[#1]%

     #2

     \centerline{\parbox{15cm}{\caption[.]{#3} \label{#4}}}
     \end{figure} }

\newcommand{\BILDD}[4]{\begin{figure}[#1]%

     #2

     \centerline{\parbox{17.25cm}{\caption[.]{#3} \label{#4}}}
     \end{figure} }

\newcommand{\Int}{\int\limits}

\newcommand{\ud}{\text{d}}
\newcommand{\ui}{\text{i}}
\newcommand{\ue}{\text{e}}

\newcommand{\R}{\mathbb{R}}
\newcommand{\Z}{\mathbb{Z}}
\newcommand{\N}{\mathbb{N}}

\newcommand{\setsep}{ \;\; | \;\;}

\begin{document}

\thispagestyle{empty}


\vspace*{-2cm}

\newcommand{\titel}{\hspace*{-0.5cm}Numerical aspects of eigenvalue and 
   eigenfunction\\[1.5ex] 
   computations for chaotic quantum systems}

\vspace{1.5cm}

\begin{center}  \Large \bf

\titel
\end{center}

\vspace{0.5cm}

\begin{center}

          {\large Arnd B\"acker%
\footnote[1]{E-mail address: {\tt arnd.baecker@physik.uni-ulm.de},
homepage: {\tt www.uni-ulm.de/theo/qc/baec}.}

\vspace*{1cm}
\normalsize 

Abteilung Theoretische Physik\\[0.5ex]
   Universit\"at Ulm, 
Albert-Einstein-Allee 11\\[0.5ex]
D-89081 Ulm, Germany\\ 
          }

    \vspace{1.0cm}
\end{center}


\newcommand{\PBProceedingsfootnote}[1]{}

\newcommand{\noBFx}{{q}}

\newcommand{\Imag}{\text{Im}}
\newcommand{\Real}{\text{Re}}

\newcommand{\cH}{{\cal H}}

\newcommand{\cP}{{\cal P}}
\newcommand{\BFN}{{\boldsymbol N}}    
\newcommand{\BFT}{{\boldsymbol T}} 
   
\newcommand{\BFv}{{\boldsymbol v}}    
\newcommand{\BFr}{{\boldsymbol r}}    
\newcommand{\pO}{\partial\Omega}

\newcommand{\ADD}[1]{{\sf ADD: #1}}
\newcommand{\TrA}{\text{Tr}\,A}
\newcommand{\zbar}{{\overline{z}}}

\newcommand{\oInt}{\oint\limits}

\newcommand{\TS}{\textstyle}
\newcommand{\DS}{\displaystyle}
\newcommand{\halb}{{\TS \frac{1}{2}}}
\newcommand{\bfn}{{\boldsymbol n}}
\newcommand{\bfu}{{\boldsymbol u}}

\newcommand{\CA}{{\cal A}}
\newcommand{\CL}{{\cal L}}
\newcommand{\CC}{{\cal C}}
\newcommand{\cA}{{\cal A}}
\newcommand{\cL}{{\cal L}}
\newcommand{\cC}{{\cal C}}

\newcommand{\TODO}{\fbox{\sf \bf TODO} }
\newcommand{\BFq}{{\boldsymbol q}}
\newcommand{\BFx}{{\boldsymbol q}}
\newcommand{\BFp}{{\boldsymbol p}}
\newcommand{\noBFq}{q}
\newcommand{\bfq}{\BFq}
\newcommand{\limacon}{lima\c{c}on }

\newcommand{\BFc}{\boldsymbol{c}}
\newcommand{\erf}{\text{erf}}

\newcommand{\C}{\mathbb{C}}

\newcommand{\T}{\mathbb{T}}
\newcommand{\twovec}[2]{\left(\begin{array}{c} #1 \\ #2 \end{array}\right)}
\newcommand{\matII}[2]{\left(\begin{array}{cc} #1 \\ #2 \end{array}\right)}

\newcommand{\Op}{\text{Op}}

\vspace*{1cm}

\leftline{\bf Abstract:}

\vspace*{0.25cm}

\noindent
We give an introduction to some of the numerical aspects
in quantum chaos. The classical dynamics of two--dimensional
area--preserving maps on the torus
is illustrated using the standard map
and a perturbed cat map. The quantization of area--preserving
maps given by their generating function is discussed
and for the computation of the eigenvalues
a computer program in Python is presented.
We illustrate the eigenvalue distribution for two types of perturbed
cat maps, one leading to COE and the other to CUE statistics.
For the eigenfunctions of quantum maps we study the
distribution of the eigenvectors and compare them with
the corresponding random matrix distributions.
The Husimi representation allows for a direct comparison
of the localization of the eigenstates in phase space
with the corresponding classical structures.
Examples for a perturbed cat map and the standard map with
different parameters are shown.

Billiard systems and the corresponding quantum billiards
are another important class of systems 
(which are also relevant to applications, 
for example in mesoscopic physics).
We provide a detailed exposition of the boundary integral method,
which is one important method to determine the eigenvalues
and eigenfunctions of the Helmholtz equation.
We discuss several methods to determine the eigenvalues
from the Fredholm equation and illustrate them
for the stadium billiard. The occurrence of spurious
solutions is discussed in detail and
illustrated for the circular billiard, the stadium billiard,
and the annular sector billiard.

We emphasize the role of the normal derivative function to
compute the normalization of eigenfunctions, momentum
representations or autocorrelation functions in
a very efficient and direct way. Some examples
for these quantities are given and discussed.

\newpage

\section{Introduction}

In this text, which is an expanded version of lectures held at a
summer school in Bologna in 2001, we give
an introduction to some of the numerical aspects
in quantum chaos;
some of the sections on the boundary integral method 
contain more advanced material.
In quantum chaos one studies quantum systems
whose classical limit is (in some sense) chaotic.
In this subject 
computer experiments play an important role. 
For integrable systems the eigenvalues and eigenfunctions
can be determined either explicitly or 
as solutions of simple equations.
In contrast, for chaotic systems there
are no explicit formulae for eigenvalues and eigenfunctions
such that numerical methods have to be used.
In many cases numerical observations have lead to the formulation
of important conjectures.
Such numerical computations also allow us to test analytical results
which have been derived under certain assumptions or by using
approximations.

An important class of systems for the study of classical
chaos are area--preserving maps as several types of different
dynamical behaviour like integrable motion, mixed dynamics,
ergodicity, mixing or Anosov systems can be found.
We discuss the numerics for the corresponding quantum maps
and illustrate some of the methods and results
using the standard map and the perturbed cat map
as prominent examples.

Another important class of systems are classical
billiards and the corresponding quantum billiards.
In section 3 we discuss in detail the boundary integral method,
which is one of the main methods for the solution
of the Helmholtz equation, which is the time--independent
Schr\"odinger equation for these systems.

\section{Area preserving maps}

\subsection{Some examples}

We will restrict ourselves to area--preserving maps on the two--torus 
\begin{align} 
 P:\T^2  &\to \T^2 \\
   (q,p) &\mapsto (q',p') \;\;, 
\end{align}
where $\T^2 \simeq \R^2/\Z^2$, 
i.e.\ the map is defined on a square
with opposite sides identified.
The requirement that the map $P$ is area--preserving
is equivalent to the condition that $\det DP=1$, where $DP$ is
the linearization of the map $P$.
The natural invariant measure on $\T^2$ 
is the Lebesgue  measure $\ud \mu = \ud q  \ud p$.

\newcommand{\zweibild}[4]{
\begin{center}
\begin{minipage}{8.3cm}
\centerline{$\kappa=#2$}

\PSImagx{section_standard_map_#1.ps}{8.25cm}
\end{minipage}
\begin{minipage}{8.3cm}
\centerline{$\kappa=#4$}

\PSImagx{section_standard_map_#3.ps}{8.25cm}
\end{minipage}
\end{center}}

\BILD{t}
  {
  \zweibild{}{0.5} 
           {}{1.0} 

  \zweibild{}{1.5} 
           {}{3.0} 
  }
  {Examples of orbits in the standard map 
   for different parameters $\kappa$.}           
  {fig:orbits-standard-map}

As a first example let us consider the so-called {\it standard map}, 
defined by
\begin{equation} \label{eq:standard-map}
  \twovec{q'}
         {p'} = 
  \twovec{q+p-\frac{\kappa}{2\pi} \sin(2\pi q ) } 
         {p  -\frac{\kappa}{2\pi} \sin(2\pi q ) } \mod 1 \;\;.
\end{equation}
One easily checks that this map is area--preserving.
Fig.~\ref{fig:orbits-standard-map} shows some orbits
(i.e.\ for different initial points $(q,p)$ the points $(q_n,p_n)=P^n(q,p)$
are plotted for $n\le 1000$)
of the standard map for different parameters $\kappa$.
For $\kappa=0$ an initial point $(q,p)$ stays on the horizontal
line and in $q$ it rotates with frequency $p$. 
So for irrational $p$ the corresponding line is filled densely.
For $\kappa>0$,
the lines with rational $p$ break up into an island-chain structure
composed of (initially) stable orbits and their corresponding
unstable (hyperbolic) partner. For small enough perturbation
there are invariant (Kolmogorov--Arnold--Moser or short KAM) 
curves which are absolute barriers to the motion
(for a more detailed discussion of these aspects
the review \cite{Mei92} is a good starting point).
For stronger perturbations, e.g.\ $\kappa=1$
or $\kappa=1.5$, the stochastic bands become larger
and for even stronger perturbation (see the picture for $\kappa=3.0$)
there appears to be just one quite big stochastic region
together with the elliptic island.
The elliptic islands coexist with regions
of irregular motion,  therefore the standard map is an example
of a  so-called system with mixed phase space or, more briefly,
a {\it mixed system}.
Whether the motion in those stochastic regions
is ergodic is one of the big unsolved problems, see \cite{Str91}
for a review on the coexistence problem.
For some recent results on the classical dynamics of
the standard map, in particular at large parameters, 
see \cite{Dua94,GioLaz2000,Laz2000}.

An alternative way to specify a map $P:\T^2\to\T^2$ 
is to use a generating function $S(q',q)$, from which the map
is obtained  by
\begin{equation} \label{eq:gen-fct}
  p= - \frac{\partial S(q',q)}{\partial q}
 \qquad \qquad
  p'= \frac{\partial S(q',q)}{\partial q'} \;\;.
\end{equation}
One easily checks that 
\begin{equation}
  S(q',q)= \frac{1}{2} (q-q')^2 +\frac{\kappa}{4\pi^2} \cos(2\pi q) \;\;,
\end{equation}
is a generating function for the standard map \eqref{eq:standard-map}.

\renewcommand{\zweibild}[4]{
\begin{center}
\begin{minipage}{8.3cm}
a) \centerline{$\kappa=#2$}

\PSImagx{section_perturbed_cat_#1.ps}{8.25cm}
\end{minipage}
\begin{minipage}{8.3cm}
b) \centerline{$\kappa=#4$}

\PSImagx{section_perturbed_cat_#3.ps}{8.25cm}
\end{minipage}
\end{center}}

\BILD{bt}
  {
  \zweibild{}{0.3}   
           {}{6.5} 
  }
  {Examples of orbits in the perturbed cat map \eqref{eq:p-cat} 
   for $\kappa=0.3$
   and $\kappa=6.5$.}           
  {fig:orbits-perturbed-cat}

Another important class are perturbed cat maps 
\cite{BasOzo95,BoaKea95}, like
\begin{equation} \label{eq:p-cat-general}
  \twovec{q'}{p'} = A \twovec{q}{p} + \kappa G(q) \twovec{A_{12}}{A_{22}} 
                \qquad   \mod 1 \;\;,
\end{equation}
where 
\begin{equation}
  A = \matII{ A_{11} & A_{12} }{ A_{21} & A_{22} }
\end{equation}
is a matrix with integer entries
(ensuring the continuity of the map), $\det A=1$ (area preservation)
and $\TrA>2$ (hyperbolicity).
The perturbation $G(q)$ is a smooth
periodic function on $[0,1[$.
For $\kappa=0$ the mapping is Anosov (see e.g.\ \cite{ArnAve68}),
in particular it is ergodic and mixing. Moreover,
following from the the Anosov theorem 
the map \eqref{eq:p-cat-general} is structurally stable,
i.e.\ it stays Anosov as long
\begin{equation}
  \kappa \le \kappa_{\text{max}} := 
                        \frac{\sqrt{(\TrA)^2-4}-\TrA+2}
                             {2 \max_q |G'(q)| \sqrt{1+A_{22}^2}} \;\;;
\end{equation}
in particular 
the orbits are topologically conjugate to those of the unperturbed
cat map.
For larger parameters there are typically elliptic islands,
so it becomes a mixed system.

A common choice for $A$ and the perturbation is 
\begin{equation} \label{eq:p-cat}
  \twovec{q'}{p'} =   \matII{ 2 & 1 } { 3 & 2 }   
  \twovec{q}{p} + \frac{\kappa}{2\pi} \cos(2\pi q) \twovec{1}{2} 
                \qquad   \mod 1 \;\;.
\end{equation}
 For $\kappa\le \kappa_{\text{max}} = (\sqrt{3}-1)/\sqrt{5}= 0.33\dots$ 
 the map is Anosov.
The corresponding generating function is given by
\begin{equation} \label{eq:gen-fct-p-cat}
 S(q',q) = {q'}^2 - qq' +q^2 + \frac{\kappa}{4\pi^2} \sin(2\pi q) \;\;.
\end{equation}
In fig.~\ref{fig:orbits-perturbed-cat}a) one
orbit for 20\,000 iterations for the perturbed cat map \eqref{eq:p-cat}
with $\kappa=0.3$ is shown. The orbit appears to fill
the torus in a uniform way, as it has to be asymptotically
for almost all initial conditions because of the ergodicity of the map.
For $\kappa=6.5$ fig.~\ref{fig:orbits-perturbed-cat}b) shows
one orbit  (20\,000 iterates) in the irregular component and 
some orbits (1000 iterations) in the elliptic islands.

\FloatBarrier
\subsection{Quantization of area--preserving maps}

For the quantization of area--preserving
maps exist several approaches, see for example
\cite{BerBalTabVor79,HanBer80,BalVor89,Sar90,Esp93,BasOzo95,DegGraIso95,DeBDegGia96,Zel97c}; 
a detailed account can be found in \cite{Haa99:Dip},
and \cite{Bie2001} provides a pedagogical introduction to the subject.
First one has to find a suitable Hilbert space
which incorporates the topology of the torus $\T^2$, i.e.\
the eigenfunctions in position and momentum have to fulfil
\begin{align}
 \psi(q+j)           &= \ue^{\frac{\ui}{\hbar}j \theta_2} \psi(q) 
          \qquad ; \;\; j\in  \N \\
 \widehat{\psi}(p+k) &= \ue^{-\frac{\ui}{\hbar}k \theta_1} \widehat{\psi}(p)
          \!\!\qquad ; \;\; k\in  \N  \;\;.
\end{align}
These conditions imply that Planck's constant
$\hbar$ can only take the values $\hbar=\frac{1}{2\pi N}$
with $N\in\N$.
Thus the semiclassical limit $\hbar\to 0$ corresponds
to $N\to\infty$.
The phases $(\theta_1,\theta_2)\in[0,1[^2$ are at first arbitrary;
for $\theta_1=\theta_2=0$ one obtains periodic boundary conditions.
For each $N$ one has a Hilbert space $\cH_N$
of finite dimension $N$. Observables $f\in C^\infty(\T^2)$
can be quantized analogous to the Weyl quantization
to give an operator Op$(f)$ on $\cH_N$.
Finally, a quantum map is a sequence of unitary operators $U_N$, $N\in \N$
on a Hilbert space $\cH_N$. The quantum map is a quantization of
a classical map $P$ on $\T^2$, if the so--called Egorov property 
is fulfilled, i.e.\
\begin{equation} \label{eq:Egorov}
  \lim_{N\to\infty} || U_N^{-1} \Op(f) U_N - \Op(f \circ P) || =0
       \qquad \forall f \in C^\infty(\T^2) \;\;.
\end{equation}
This means that semiclassically quantum time evolution
and classical time evolution commute.

So the aim is to find for a given classical
map a corresponding sequence of unitary operators.
Unfortunately, this is not 
as straight forward as the quantization of Hamiltonian systems
and a lot of information on this can be found
in the above cited literature and references therein.
One of the simplest approaches to determine $U_N$
corresponding to a given
area--preserving map uses its generating function to define
\begin{equation} \label{eq:generating-function-quantization}
 (U_N)_{j',j} := \langle q_{j'} | U_N | q_j \rangle =
         \frac{1}{\sqrt{N}} 
     \left| \frac{\partial^2 S(\tilde{q}',\tilde{q})}
                 {\partial \tilde{q}' \partial\tilde{q}} 
          \right|^{1/2}_{\tilde{q}'=q_{j'},\tilde{q}=q_j}
        \exp\left( 2\pi\ui N S\left( q_{j'},q_{j} \right) \right)
\end{equation}
with $q_j=j/N$, $q_{j'}=j'/N$, where $j,j'=0,1,\ldots,N-1$.
In the same way one may (and for certain maps which cannot
represented in terms of $S(q',q)$ one has to) use 
other generating functions
such as $S(p',p)$ or $S(q,p)$; usually these will lead
to different eigenvalues and eigenfunctions.
The question is to determine conditions on the generating function $S(q',q)$
such that $U_N$ is unitary and fulfils the Egorov property \eqref{eq:Egorov}.
To my knowledge this question has not yet been fully explored,
even though the quantum maps studied in the literature
provide both examples and counterexamples.
We will leave this as an interesting open question.

For the examples introduced before the quantization via
\eqref{eq:generating-function-quantization} can be used.
For the standard map we get 
\begin{equation} \label{eq:quantized-standard}
  (U_N)_{j',j} = \frac{1}{\sqrt{N}} 
                \exp\left[ \frac{\ui \pi}{N} (j'-j)^2 
                         +\frac{\ui \kappa N }{2 \pi} 
                          \cos\left(\frac{2\pi}{N} j \right)
                     \right]
\end{equation}
with $j,j'=0,\ldots,N-1$.
A quantization of the standard map which takes the
symmetries into account can be found in \cite{ProRob94}.
For the perturbed cat map \eqref{eq:p-cat} one gets
using its generating function \eqref{eq:gen-fct-p-cat}
\begin{equation} \label{eq:quantized-p-cat}
   (U_N)_{j',j} = \frac{1}{\sqrt{N}} 
       \exp\left( \frac{2\pi\ui}{N} ({j'}^2-j'j +j^2 +\kappa/(4\pi^2) 
                   \sin(2\pi j/N) ) \right)  \;\;.
\end{equation}

\BILD{t}
     {
     \begin{center}
       \PSImagx{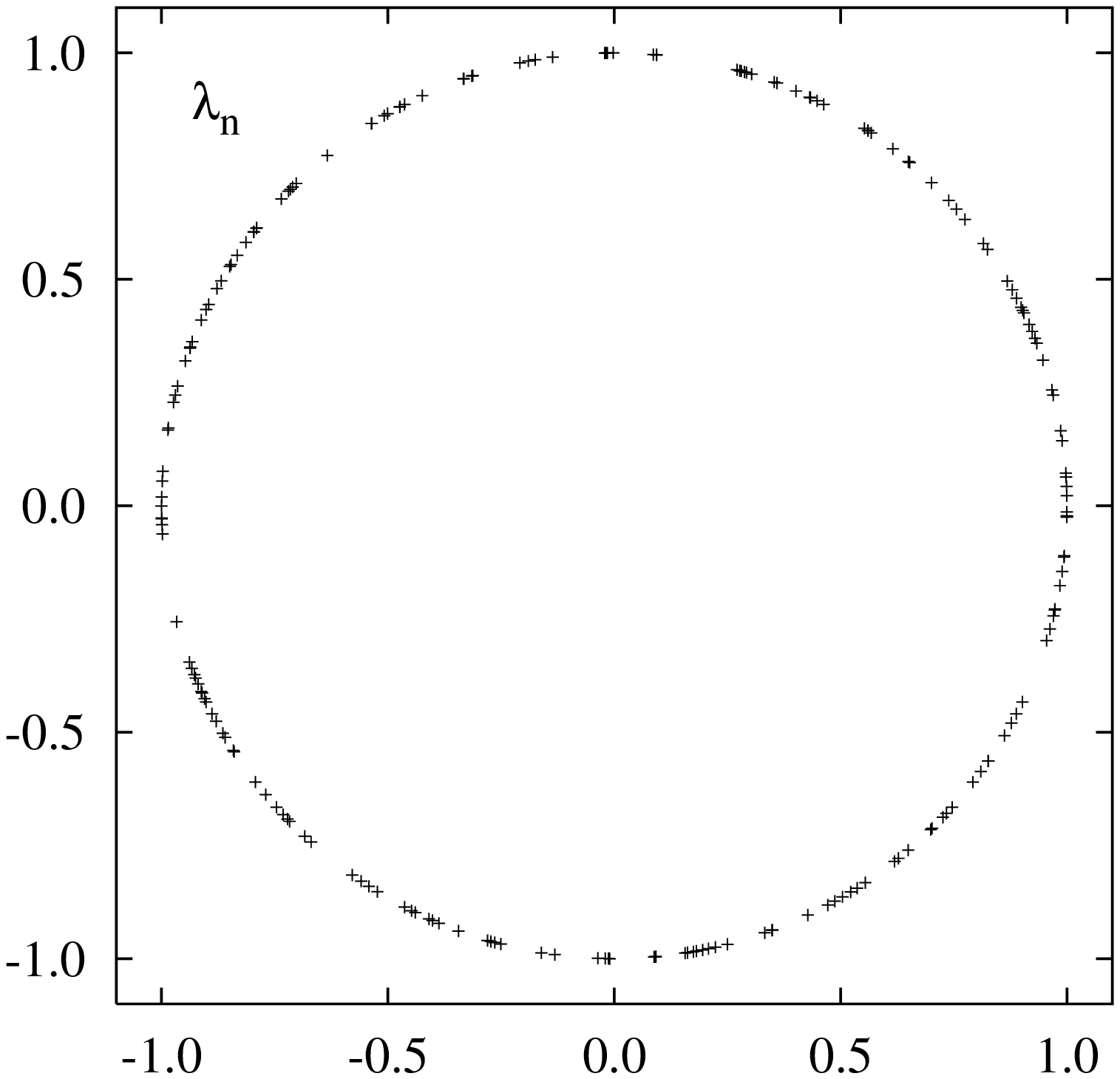}{12cm}
     \end{center}
     }
     {Plotting the eigenvalues of $U_N$ allows to check the numerical
      implementation and the unitarity of $U_N$;
      the picture shows for $N=200$ and $\kappa=1.5$ the eigenvalues 
      $\lambda_n$
      for the quantized standard map \eqref{eq:quantized-standard}. 
     }
     {fig:eigenvalues-on-unit-circle}

For the unitary operator one has to solve the eigenvalue problem
\begin{equation} \label{eq:quantum-map-eval-eq}
  U_N \psi_n = \lambda_n \psi_n \qquad \text{ with } n=0,\dots,N-1, \;\;
         \psi_n\in\C^N \;\;.
\end{equation}
Here $\lambda_n$ is the $n$--th eigenvalue and the corresponding
eigenvector $\psi_n$ consists of $N$ complex components,
where $N$ is the size of the unitary matrix $U_N$.
Because of the unitarity of $U_N$ the eigenvalues lie on
the unit circle, i.e.\ $|\lambda_n|=1$.

Let us discuss some of the numerical aspects relevant for 
finding the solutions of \eqref{eq:quantum-map-eval-eq}
without going into implementation specific details
(see the appendix and \cite{MyHomepage} for an 
implementation using {\tt Python}).

Computing the eigenvalues of \eqref{eq:quantum-map-eval-eq} consists
of two main steps
\begin{itemize}
  \item Setting up the matrix $U_N$:

        The computational effort increases proportional to $N^2$
(unless each matrix element requires further  loops)
        as we have to fill the $N^2$ matrix elements.

        The memory requirement to store $U_N$ is $16 \, N^2 $ Bytes
        (for a IEEE-compliant machine a double precision
         floating point number requires 8 Bytes;
        as we have both real and imaginary part we end
        up with 16 Bytes per matrix element).
  \item Computing the eigenvalues:

        The computational effort for the matrix diagonalization 
        (typically) scales like $N^3$. 
        
        Usually one will use a black-box routine
        such as one from the NAG-library \cite{naglib}
        or from  LAPACK \cite{lapack}.
        To my knowledge there are no routines
        which make use of the fact that the matrix $U_N$ is unitary
        so we may for example use the NAG routine {\tt F02GBF}
        or the LAPACK routine {\tt ZGEES} (or
        the more recent routine {\tt ZGEEV} which is
        faster for larger matrices, e.g.\ $N\ge 500$) which 
        compute all eigenvalues of a complex matrix.

For certain maps specific optimizations are possible,
see e.g.\ \cite{ProRob94} for the standard map.
For this type of mapping a 
different approach employing a combination
of fast Fourier transform and Lanczos method
reduces the computational effort to $N^2 \ln N$ \cite{KetKruGei99}.

\end{itemize}

After successful compilation and running of the program it is useful 
to see whether the eigenvalues really lie on the unit circle.
In fig.~\ref{fig:eigenvalues-on-unit-circle} 
this is illustrated for $N=200$  and the standard map
with $\kappa=1.5$.
For small $N$ the running times of the program for
setting up the matrix $U_N$ and its diagonalization
is just a matter of minutes. For example
on an Intel Pentium III processor with 666 MHz
one needs just 6 minutes to compute the eigenvalues of 
\eqref{eq:quantum-map-eval-eq} 
when $N=1000$. However, for $N=3000$ already 140 MB of RAM
are required to store $U_N$ and the computing time increases to 6 hours.
Depending on available memory, computing power, patience
and motivation one may use larger values of $N$.

Let us conclude this part with a more technical remark:
In addition to the choice of computer language, compiler, optimizations
and algorithm there is one very important component for 
achieving good performance
when doing numerical linear algebra computations: the BLAS
(Basic Linear Algebra Subprograms). Libraries such as
LAPACK defer all the basics tasks like 
adding vectors, vector--matrix 
or matrix--matrix multiplication to the BLAS such
that highly optimized (machine--specific) BLAS routines should be used.
Most hardware vendors provide these (of differing quality).
Recently the software system ATLAS 
(Automatically Tuned Linear Algebra Software) \cite{ATLAS} has 
been introduced which generates a machine dependent
optimized BLAS library.
For some computers ATLAS--based BLAS can be even faster
than the vendor supplied ones!

\FloatBarrier
\subsection{Eigenvalue statistics}

\BILD{bth}
     {
     \begin{center}
       \PSImagx{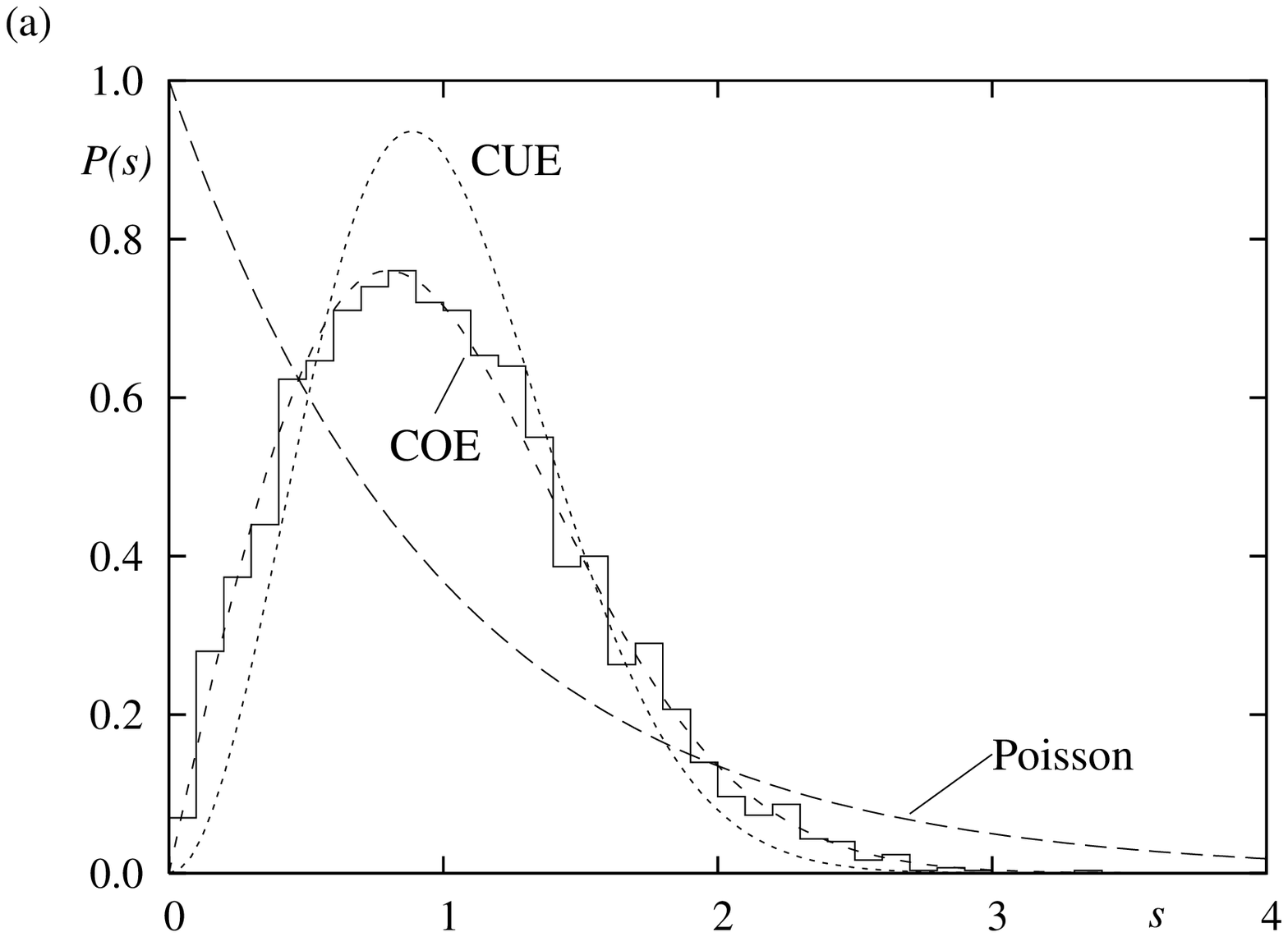}{14.5cm}

       \PSImagx{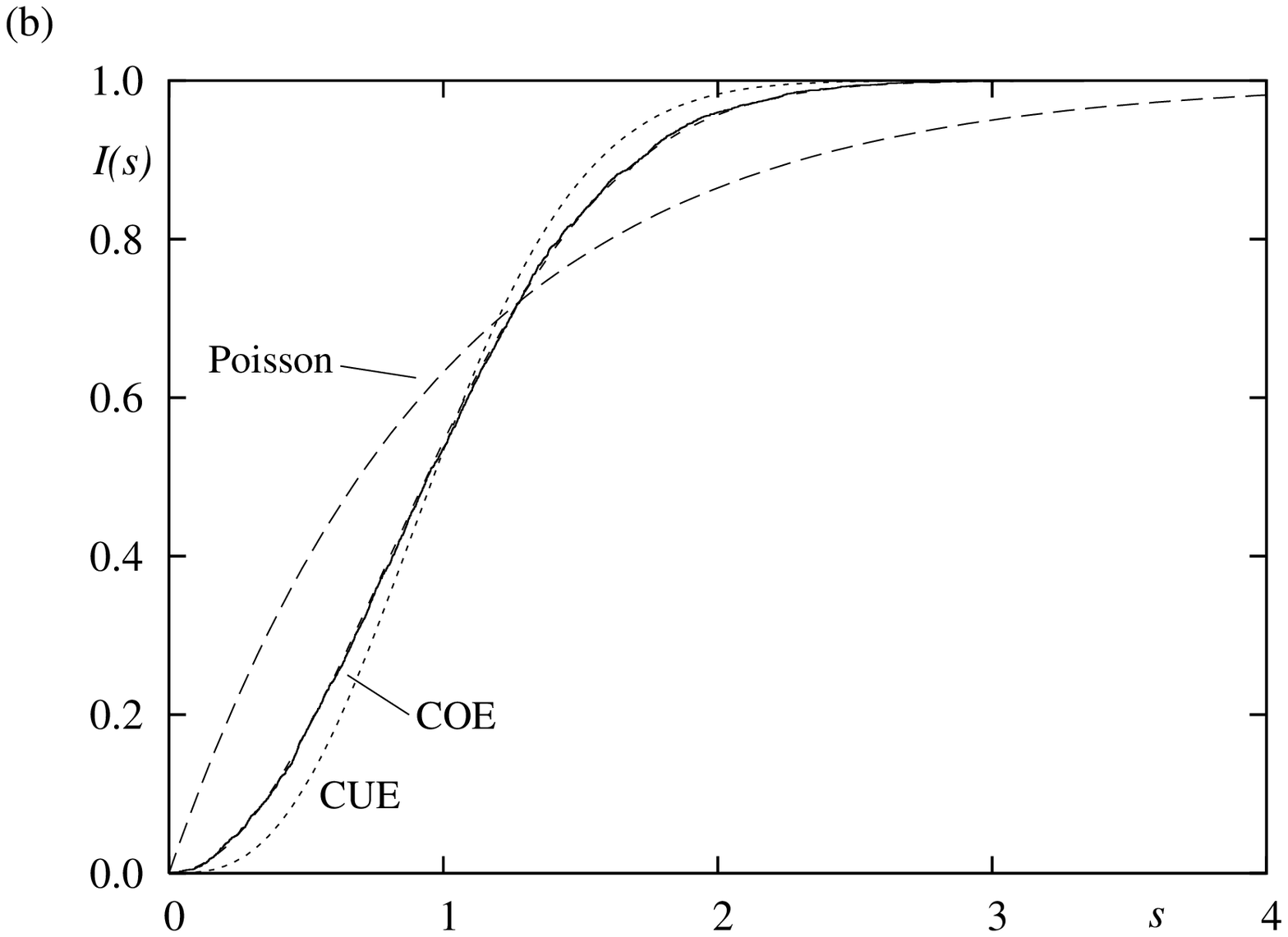}{14.5cm}
     \end{center}
     }
     {(a) Level spacing distribution $P(s)$ and (b) cumulative
      level spacing distribution $I(s)$ for the perturbed
      cat map \eqref{eq:p-cat} with $\kappa=0.3$ and $N=3001$.
     }
     {fig:spacing-statistics}

\BILD{bth}
     {
     \begin{center}
       \PSImagx{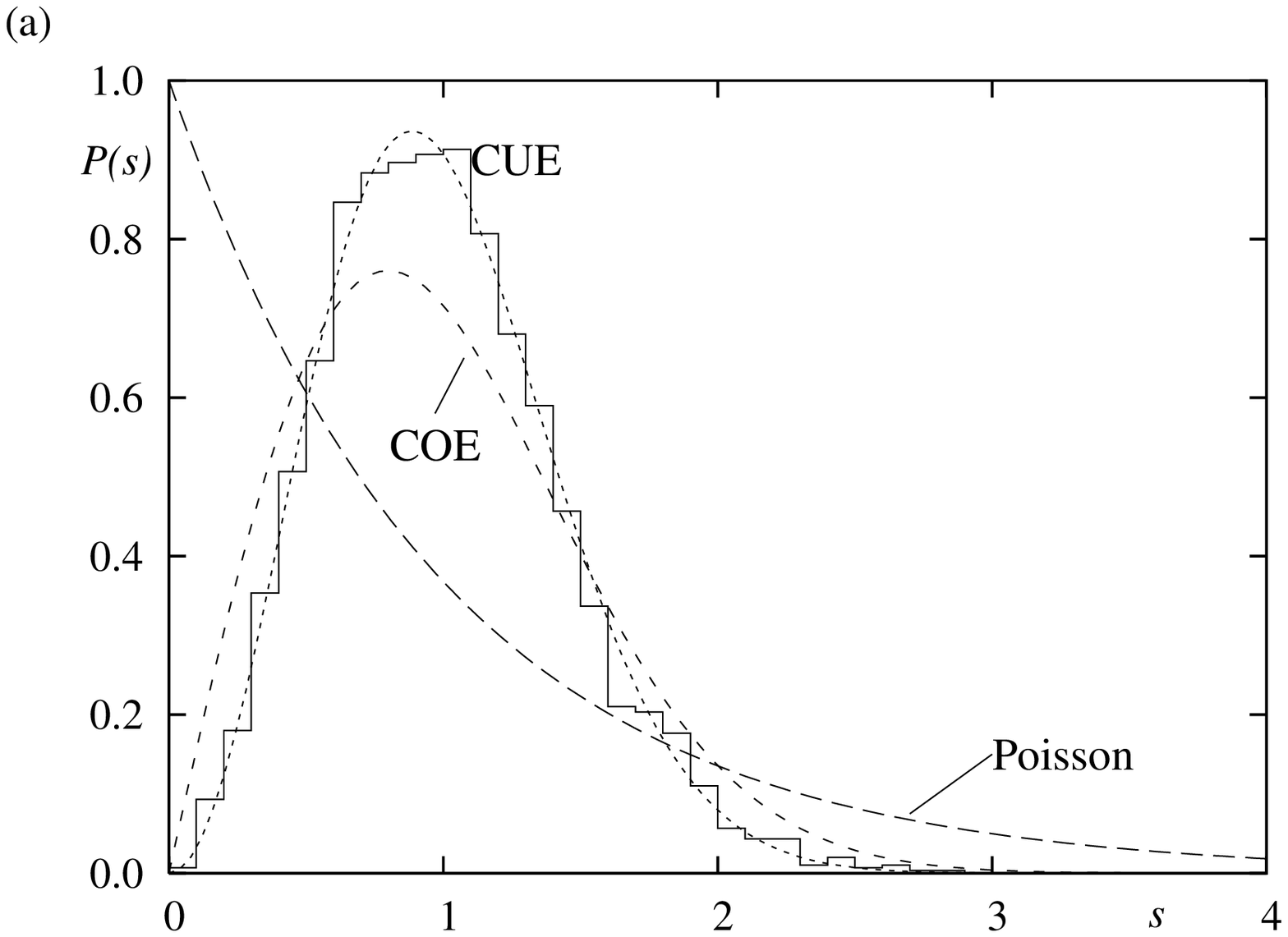}{14.5cm}

       \PSImagx{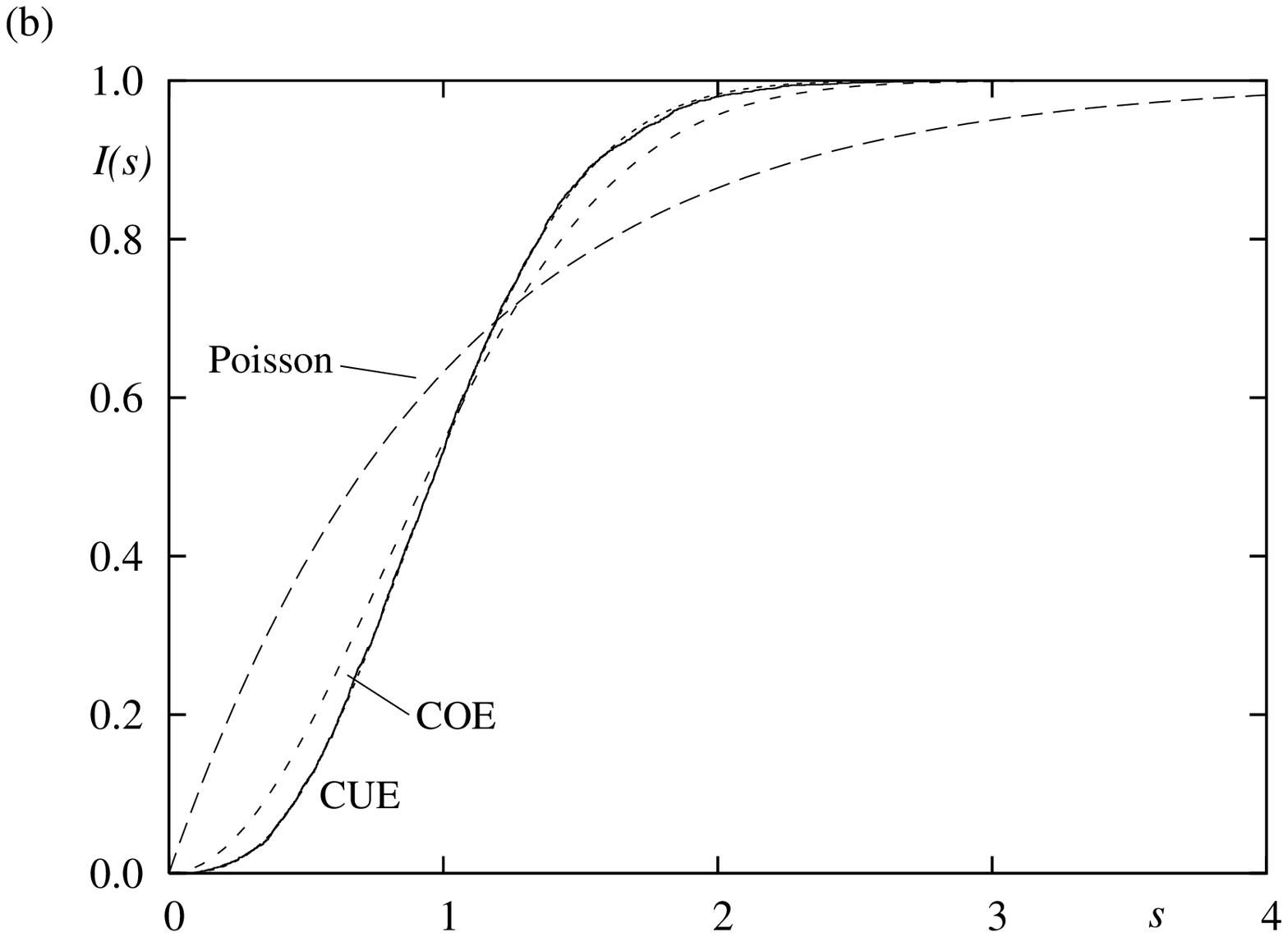}{14.5cm}
     \end{center}
     }
     {(a) Level spacing distribution $P(s)$ and (b) cumulative
      level spacing distribution $I(s)$ for the perturbed
      cat map \eqref{eq:CUE-map} with $\kappa_p=\kappa_q=0.012$ and $N=3001$.
     }
     {fig:spacing-statistics-pcat2}

One central research line in quantum chaos is the investigation
of spectral statistics. It has been conjectured \cite{BohGiaSch84}
that for generic chaotic systems the eigenvalue
statistics can be described by random matrix theory,
whereas generic integrable systems should follow Poissonian
statistics \cite{BerTab77}.
To study the eigenvalue statistics for quantum maps
one considers
the eigenphases $\varphi_n \in [0,2\pi [$, 
defined by $\lambda_n=\ue^{\ui\varphi_n}$
(in the following we will also call $\varphi_n$ levels
in analogy to the energy levels for the Schr\"odinger equation).
The simplest statistics is the nearest neighbour
level spacing distribution $P(s)$ which is
the distribution of the spacings 
\begin{equation*}
  s_n := \frac{N}{2\pi} (\varphi_{n+1} -\varphi_n)   \qquad  
           \text{with } n=0,\dots,N-1 
          \quad \text{ and }  \varphi_{N}:=\varphi_0 \;\;.
\end{equation*}
The factor $\frac{N}{2\pi}$ ensures that the average of all spacings 
$s_n$ is 1.
To compute the distribution practically one chooses
a division of the interval $[0,10]$ (usually this interval is sufficient,
but more precisely the upper limit is determined by the largest $s_n$)
into $b$ bins and determines the fraction of spacings $s_n$ falling
into the corresponding bins.
If $N$ is too small it is better to consider instead of $P(s)$
the corresponding cumulative distribution
\begin{equation}
  I(s) := \frac{\# \{ n \setsep s_n \le s \} }
               {N}
\end{equation}
which avoids the binning and results in a smoother curve.

Fig.~\ref{fig:spacing-statistics} shows for the 
perturbed cat map \eqref{eq:p-cat} with $\kappa=0.3$ the
level spacing distribution $P(s)$ and the cumulative
level spacing distribution $I(s)$ for $N=3001$.
For this parameter value $\kappa$ the map is still Anosov
so one expects that the correlations of the eigenphases
follow random matrix theory; in particular because the
perturbation should break up the number theoretical degeneracies
which lead to non-generic spectral statistics
for the cat maps at $\kappa=0$ \cite{Kea91b,Kea91c}.
In \cite{Mez1999:PhD,KeaMez2000}
it is shown that for all perturbations which are just a shear in position
one of the symmetries of the cat map survives, so that
the statistics are expected to be described by the 
circular orthogonal ensemble (COE).
In the limit $N\to\infty$ this is the same as the Gaussian orthogonal
ensemble (GOE). In fig.~\ref{fig:spacing-statistics} we
show the Wigner distribution $P_{\text{Wigner}}(s)$
which is very close to the COE distribution,
\begin{equation} \label{P(S)-GOE}
  P_{\text{COE}}(s) \approx P_{\text{Wigner}}(s) = 
      \frac{\pi}{2} s \exp\left(-\frac{\pi}{4}s^2\right) \;\;.
\end{equation}                                     
and for comparison the CUE distribution
\begin{equation}
  P_{\text{CUE}}(s)  \approx \frac{32}{\pi^2} s^2 
                      \exp\left(-\frac{4}{\pi}s^2\right)
\end{equation}
and the Poisson distribution (expected
for generic integrable systems)
\begin{equation} \label{P(S)-POIS}
  P_{\text{Poisson}}(s) = \ue^{-s} \;\; .
\end{equation}                        
The agreement with the expected COE distribution is very good.

A specific example, which breaks the above mentioned unitary symmetry and thus
leads to CUE statistics, uses two shears, one in position and one in momentum
\cite{Mez1999:PhD},  
\begin{equation} \label{eq:CUE-map}
  \twovec{q'}{p'} =   (A \circ P_q \circ P_p) 
  \twovec{q}{p}  \;\;,
\end{equation}
where
\begin{equation}
  A = \matII{ 12 & 7}{ 41 & 24 } 
\end{equation}
and 
$P_q(q,p) = (q+\kappa_q G(p),p)$,
$P_p(q,p) = (q,p+\kappa_p F(q))$
with $F(q)=\tfrac{1}{2\pi}(\sin(2\pi q)-\sin(4\pi q))$
and $G(p)=\tfrac{1}{2\pi}(\sin(4\pi q)-\sin(2\pi q))$.
For the corresponding quantum map with $\kappa_p=\kappa_q=0.012$
and $N=3001$ the level spacing distribution is shown in 
fig.~\ref{fig:spacing-statistics-pcat2}.
One observes very good agreement with the CUE distribution.

\FloatBarrier

\subsection{Eigenfunctions} 

Another interesting question concerns the statistical behaviour
of eigenfunctions, and more specifically for quantum
maps the eigenvector statistics and the properties
of phase space representations like the Husimi function.

\subsubsection{Eigenvector distributions}  
  \label{subsec:eigenvector-distributions-CUE}

Consider an eigenvector $\psi$ of a quantum map
given by the $N$ numbers $c_j\in \C$, $j=0,...,N-1$.
The distribution $P(\psi)$ is given (similarly to the level
spacing distribution)
by
\begin{equation}
  \frac{1}{N} \#\{ a \le |c_j|^2 \le b \} = \Int_a^b P(\psi) \ud \psi \;\;.
\end{equation}
Let us first discuss the corresponding random matrix results
(see e.g.~\cite{BroFloFreMelPanWon81,Haa2001}). 
For the COE the eigenvectors
can be chosen to be real and the coefficients $c_j$, $j=0,\dots,N-1$,
only have to obey the normalization condition
\begin{equation}
  \sum_{j=0}^{N-1} c_j^2 = 1 \qquad \text{ with $c_j\in\R$ }\;\;. 
\end{equation}
Thus the joint probability for an 
eigenvector $\BFc=(c_0,\ldots, c_{N-1})\in \R^N$ is 
\begin{equation}
  P_N^{\text{COE}}(\BFc) = \frac{\Gamma(N/2)}{\pi^{N/2}}  
         \delta\left(1-\sum_{j=0}^{N-1} c_j^2\right)
       \;\;,
\end{equation}
where the prefactor ensures normalization. So the probability
of one component to have a specific value $y$ is given by integrating
$P_N^{\text{COE}}(\BFc)$ over all other components,
\begin{equation}\
  P_N^{\text{COE}}(y)   = \Int \delta(y-c_1^2) P_N^{\text{COE}}(\BFc)
     \; \ud c_0 \cdots \ud c_{N-1}
            = \frac{1}{\sqrt{\pi y}} \frac{\Gamma(N/2)}{\Gamma( (N-1)/2)} 
          (1-y)^{(N-3)/2} \;\;.
\end{equation}

\BILD{th}
     {
     \begin{center}
   
     \vspace*{1.5cm}

      \PSImagx{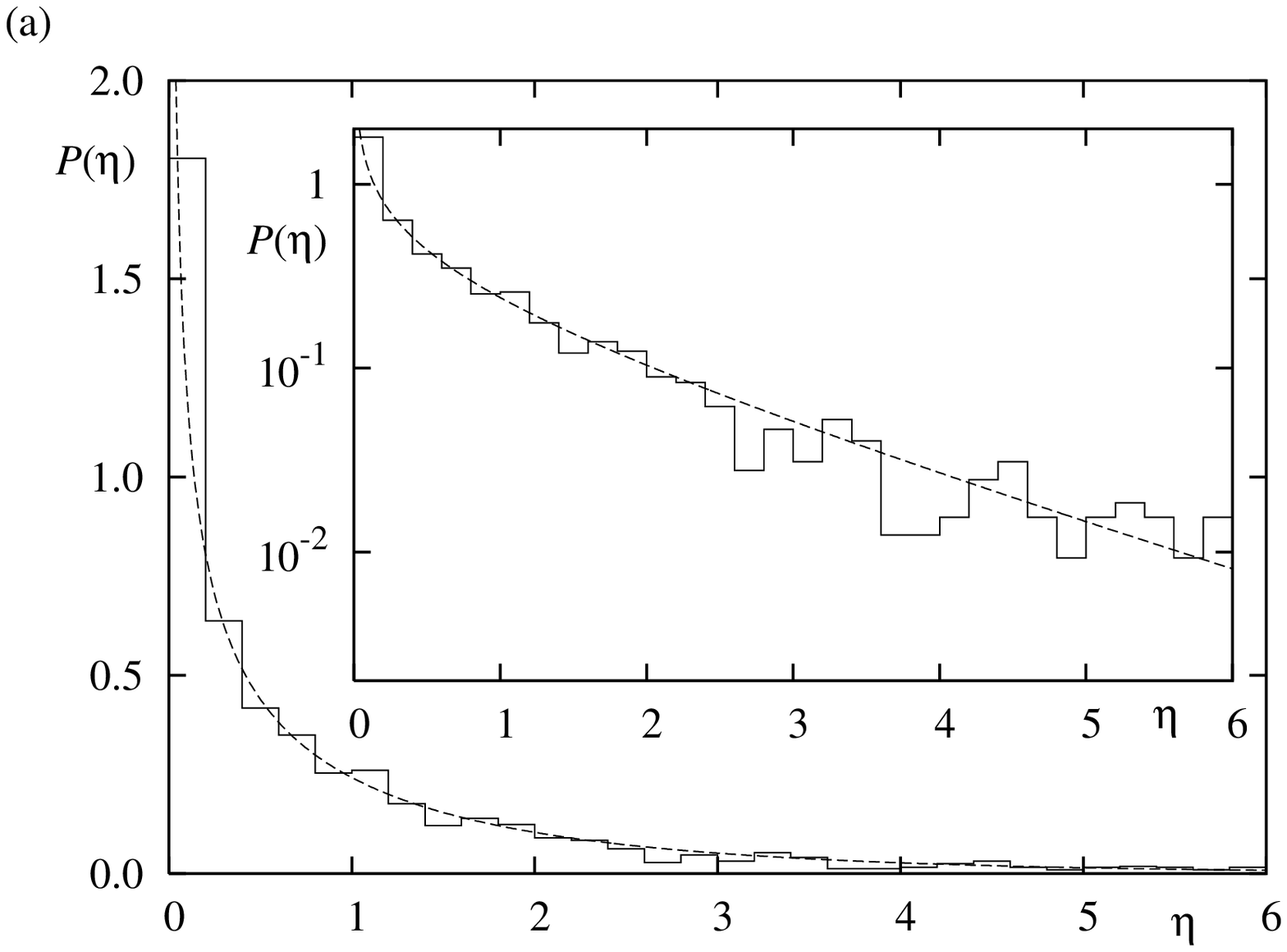}{13cm}
              
      \PSImagx{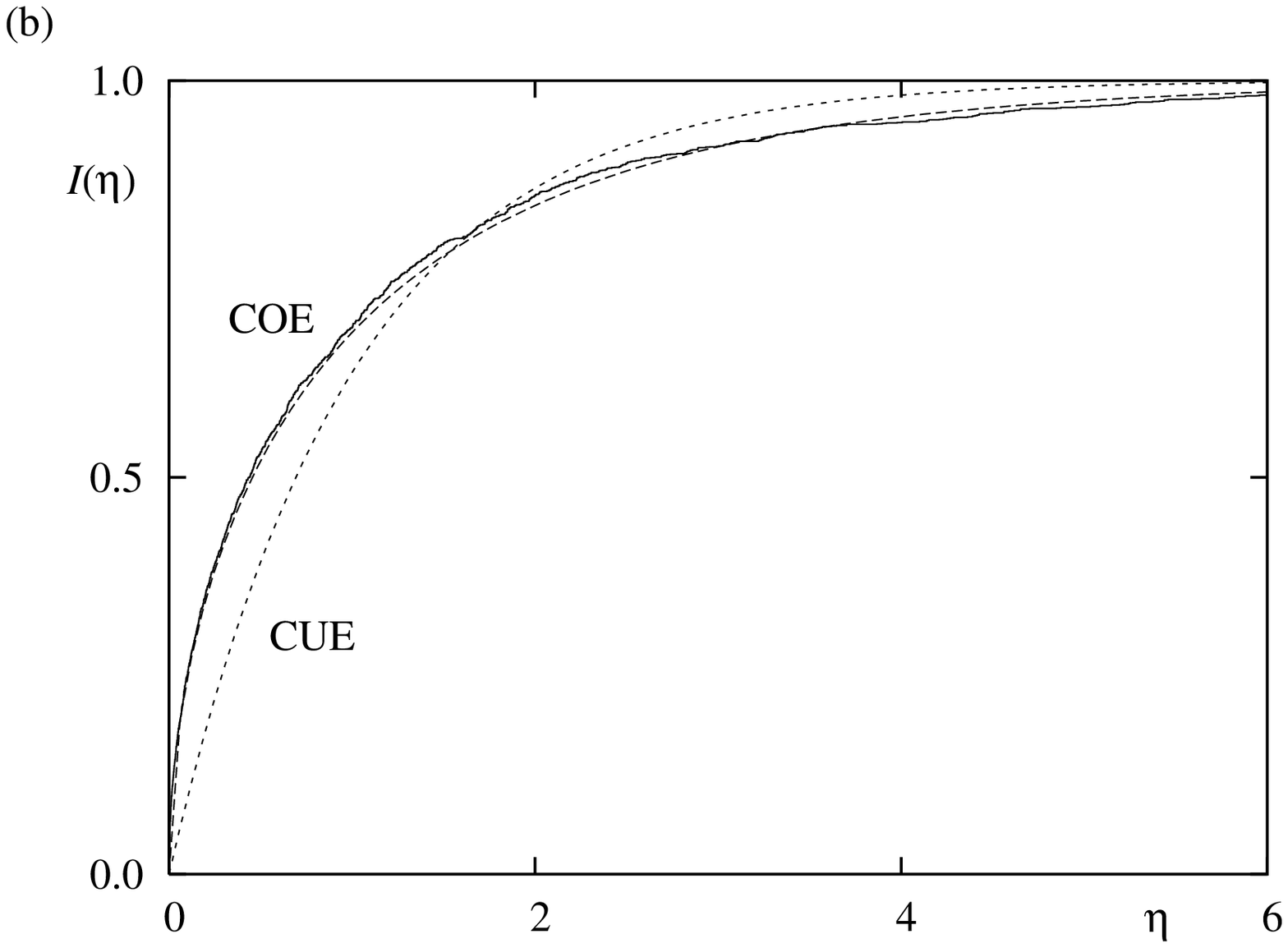}{13cm}

\vspace*{-7.0cm} \hspace*{3.5cm}
      \PSImagx{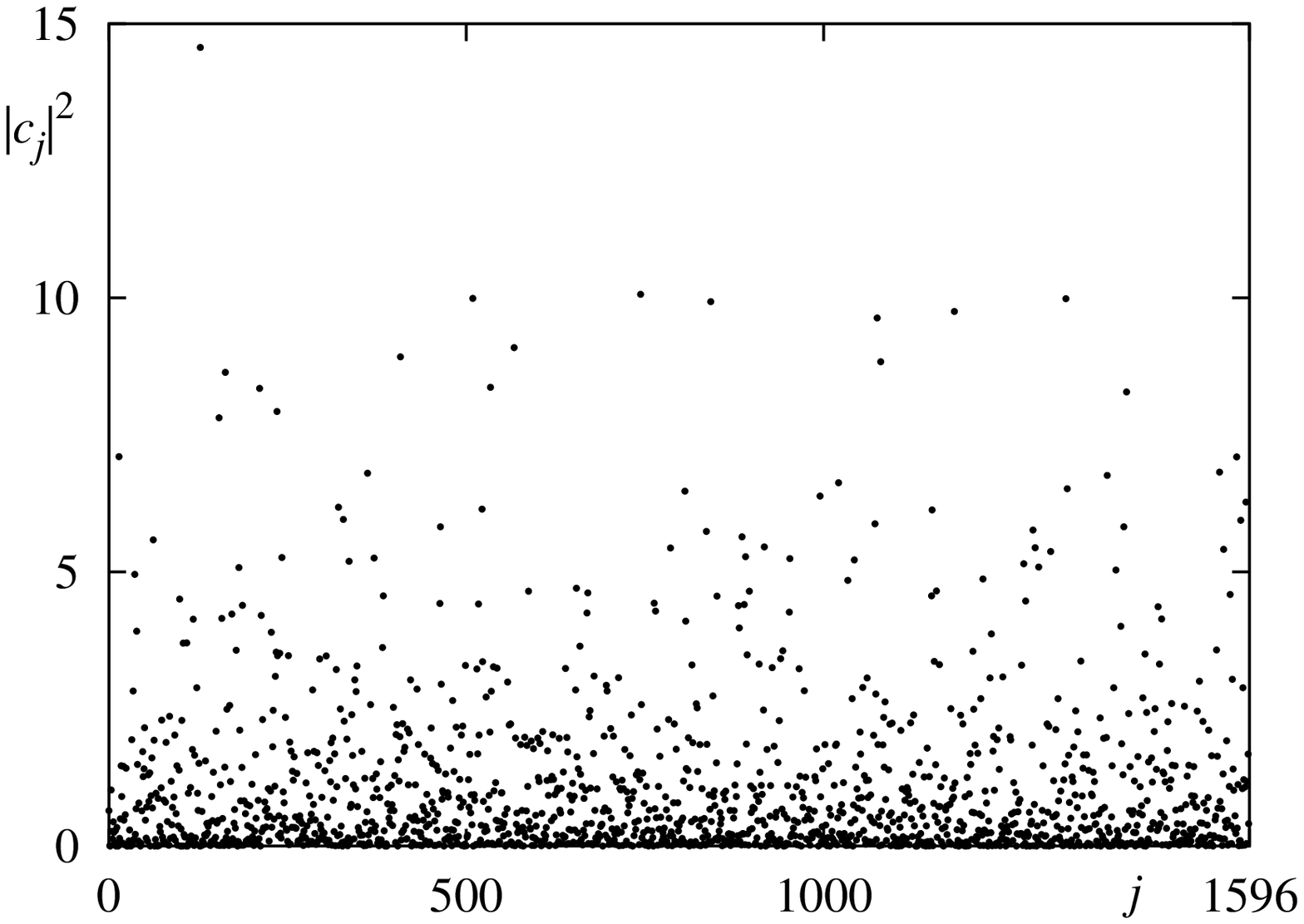}{8cm}

\vspace*{1cm}
     \end{center}
     }
     {(a) Eigenvector distribution for the perturbed cat map \eqref{eq:p-cat} 
      with $N=1597$ and $\kappa=0.3$.
     In comparison with the asymptotic COE distribution,
      eq.~\eqref{eq:distribution-COE}, dashed line.
      The inset shows the same curves in 
      a log--normal plot.
      In (b) the cumulative distribution is shown
      and in the inset a      
      plot of the absolute value of the components $c_j^{(n)}$, 
      $j=0,\dots,N-1$ of the corresponding eigenvector $\psi_{n=20}$ 
      is displayed.
     }
     {fig:evec-distribution}

\BILD{tbh}
     {

    \begin{center}
      \PSImagx{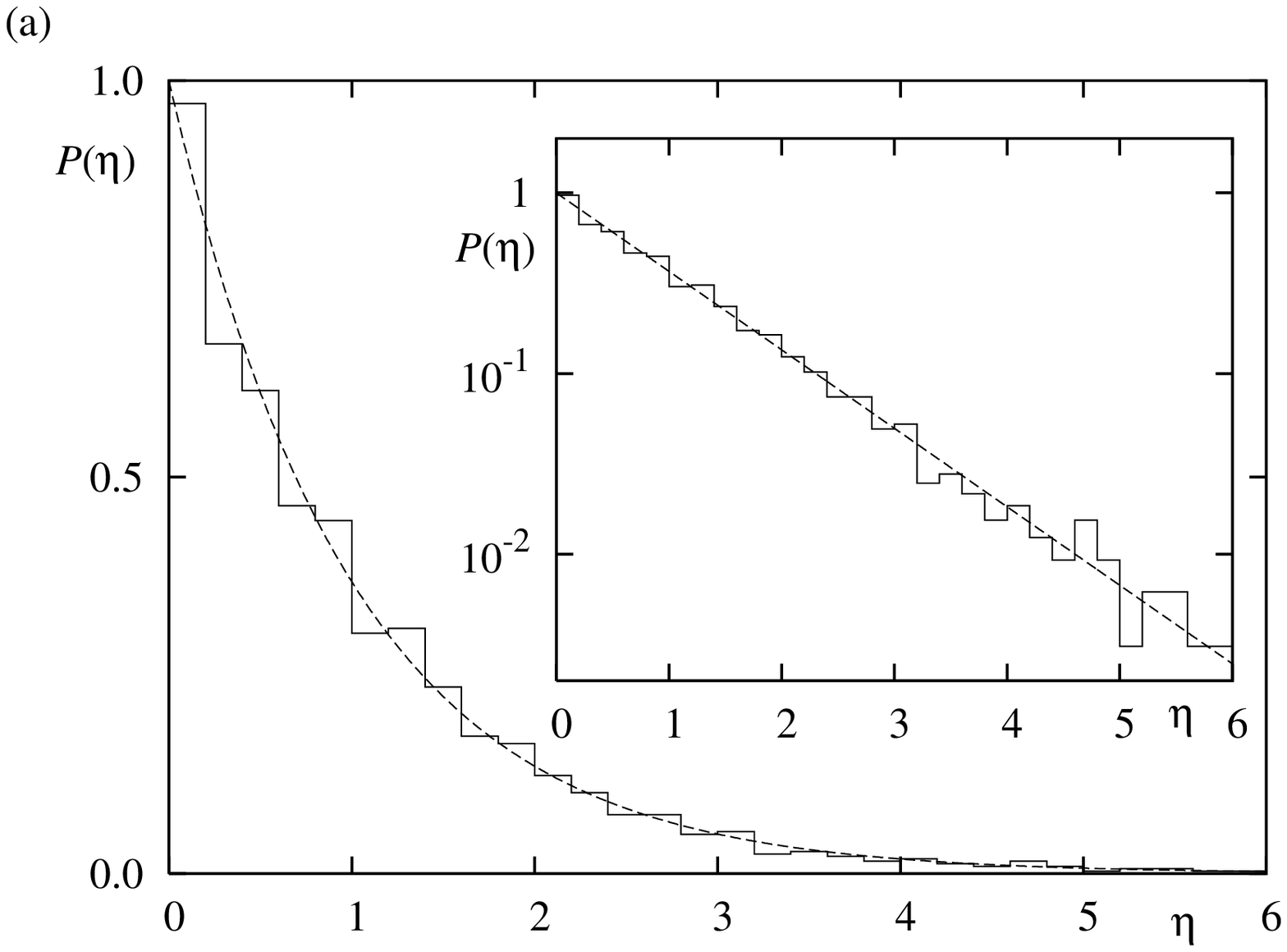}{13cm}

      \PSImagx{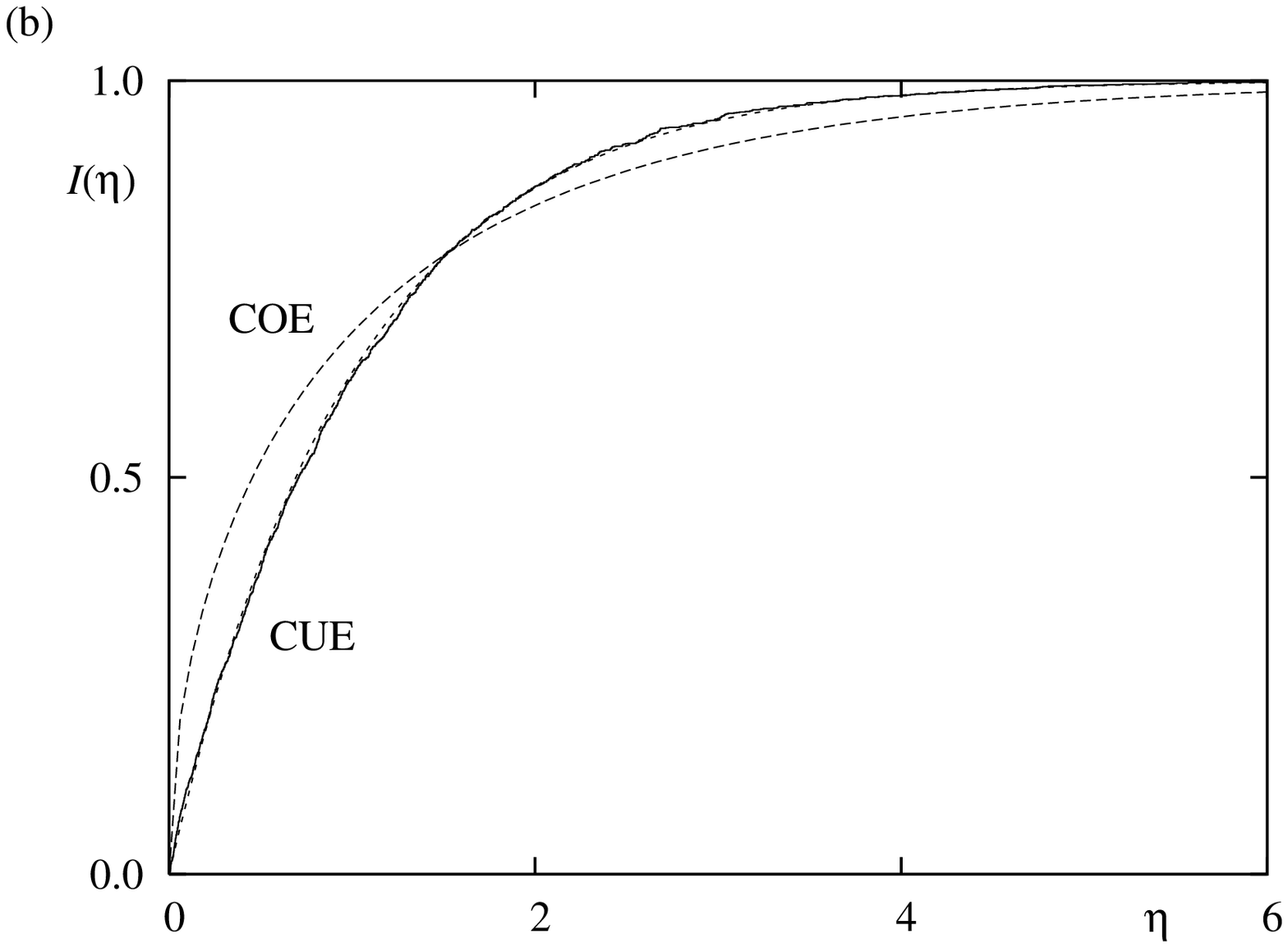}{13cm}

\vspace*{-7.0cm} \hspace*{3.5cm}
      \PSImagx{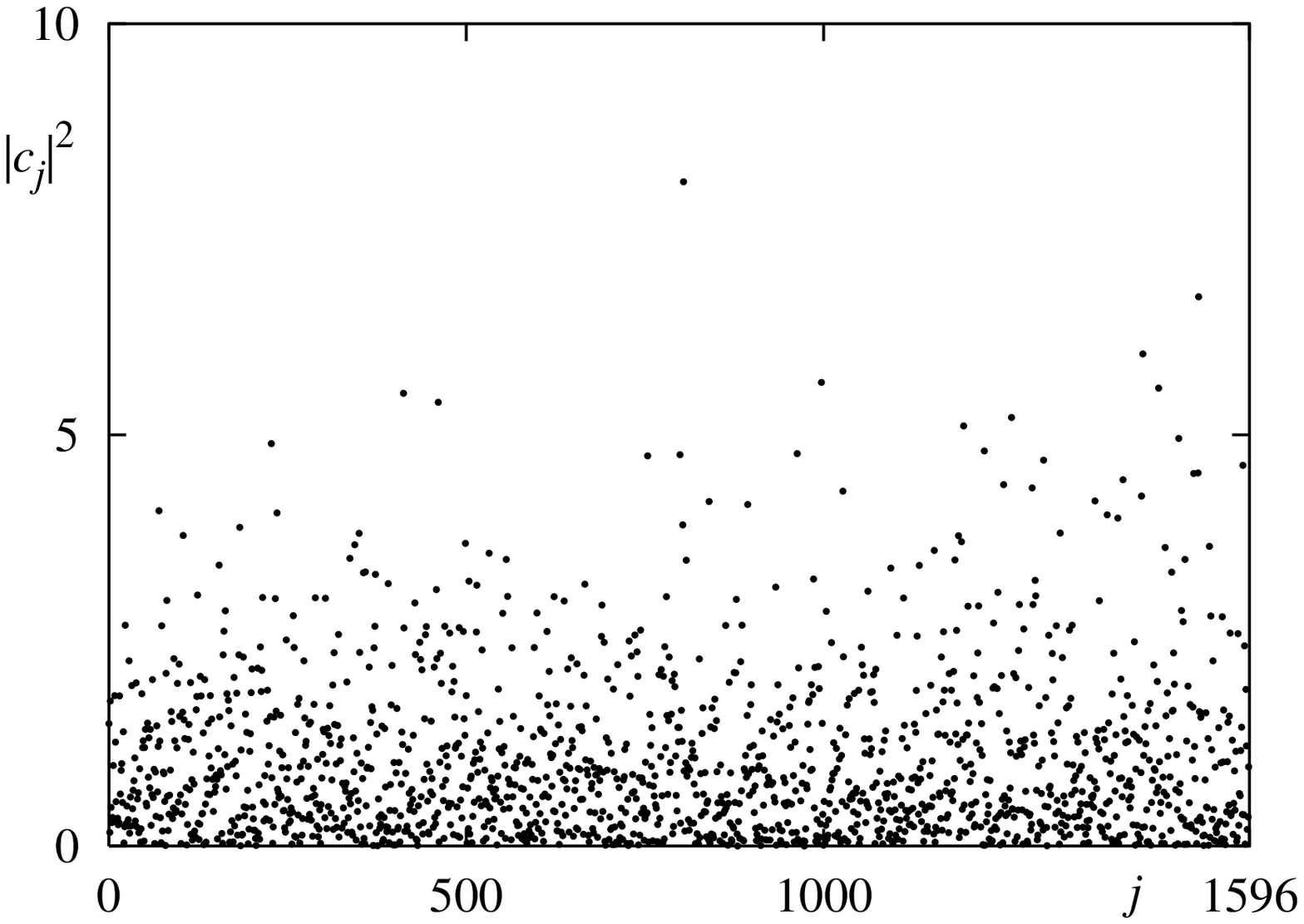}{8cm}

\vspace*{1cm}

     \end{center}
     }
     {
      (a) Eigenvector distribution of an eigenvector 
      for the perturbed cat map \eqref{eq:CUE-map} 
      with $N=1597$ and $\kappa_p=\kappa_q=0.012$ is shown 
      in comparison with the asymptotic CUE distribution,
      eq.~\eqref{eq:distribution-CUE}, dashed line.
      The inset shows the same curves in 
      a log--normal plot. In (b) the corresponding cumulative distributions
      are shown  and in the inset a      
      plot of the absolute value of the components $c_j^{(n)}$, 
      $j=0,\dots,N-1$ of the corresponding eigenvector $\psi_{n=2}$ 
      is displayed.
     }
     {fig:CUE-cat-evec-distribution}

\afterpage{\clearpage}

The mean of $P_N^{\text{COE}}(y)$ is 
$\int_0^1 y P_N^{\text{COE}}(y) \; \ud y = 1/N$.
So using the rescaling $\eta=y N$ gives
\begin{equation} \label{eq:distribution-COE}
  P_N^{\text{COE}}(\eta) =
 \frac{1}{\sqrt{\pi N \eta }} \frac{\Gamma(N/2)}{\Gamma( (N-1)/2)} 
          (1-\eta/N)^{(N-3)/2} \;\;.
\end{equation}
In the limit of large $N$ one gets
the so-called Porter-Thomas distribution \cite{PorTom56}
\begin{equation} \label{eq:distribution-N-limit-COE}
  P_N^{\text{COE}}(\eta)= \frac{1}{\sqrt{2\pi  \eta }} \exp(-\eta/2) \;\;,
\end{equation}
and the corresponding cumulative distribution
$I(y)=\Int_0^y P(y') \; \ud y'$   
reads
\begin{equation} 
     \label{eq:distribution-N-limit-COEc-cumulative} 
  I(\eta) = \erf\left(\sqrt{\eta/2}\right) \;\;.
\end{equation}
Fig.~\ref{fig:evec-distribution} shows an example 
for the eigenvector distribution of an eigenstate of
the perturbed cat map \eqref{eq:p-cat} with $\kappa=0.3$ and $N=1597$.
There is good agreement with
the expected COE distribution, eq.~\eqref{eq:distribution-N-limit-COE},
shown as dashed line.

Finally, let us consider again the map \eqref{eq:CUE-map}
which shows CUE level statistics. From this one would expect
that also the eigenvector statistics follows the CUE.
Similar to the case of the
COE one has the normalization
condition
\begin{equation}
  \sum_{j=0}^{N-1} |c_j|^2 = 1  \qquad \text{ with $c_j\in\C$ }\;\;.
\end{equation}
Thus the joint probability for an 
eigenvector $\BFc=(c_0,\ldots, c_{N-1})\in \C^N$ reads
\begin{equation}
  P_N^{\text{CUE}}(\BFc) = \frac{\Gamma(N)}{\pi^N} 
            \delta\left( 1- \sum_{j=0}^{N-1} |c_j|^2 \right) \;\;.
\end{equation}
The probability
of one component to have a specific value $y$ is given by integrating
$P_N^{\text{CUE}}(\BFc)$ over all other (complex) components,
\begin{equation}
  P_N^{\text{CUE}}(y)   = \Int \delta(y-|c_0|^2) P_N^{\text{COE}}(\BFc)
     \; \ud^2 c_0 \cdots \ud^2 c_{N-1} 
            =  (N-1) (1-y)^{N-2} \;\;.
\end{equation}
\BILD{b}
     {
     \begin{center}
       \PSImagx{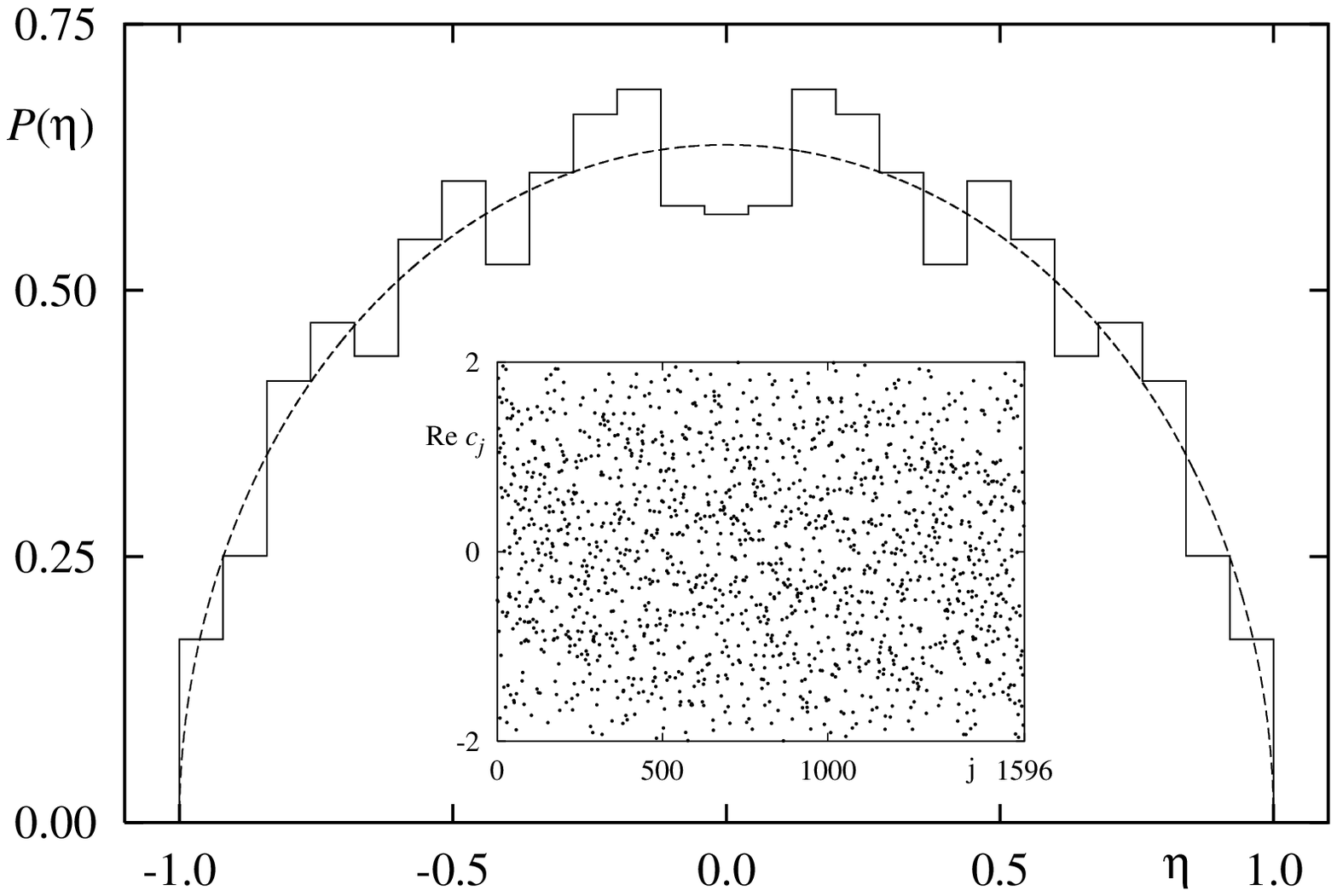}{13cm}
     \end{center}
     }
     {Eigenvector distribution of an eigenvector 
      for the unperturbed (i.e.\ $\kappa=0$) cat map \eqref{eq:p-cat} 
      with $N=1597$. This is compared with the asymptotic
      semicircle law, eq.~\eqref{eq:semicircle}.
      The inset shows the corresponding eigenvector
      (compare with the eigenvectors shown in the previous two figures).
     }
     {fig:semicircle-cat-evec-distribution}
Again as for the COE, the mean of $P_N^{\text{CUE}}(y)$ is $1/N$
and the rescaling $\eta := y N$ leads to
\begin{equation} \label{eq:evec-distrib-CUE}
  P_N^{\text{CUE}}(\eta) 
     = \frac{N-1}{N} \left(1-\frac{\eta}{N}\right)^{N-2}
\end{equation}
which has mean $1$.
In the large $N$ limit we have 
\begin{equation} \label{eq:distribution-CUE}
 P^{\text{CUE}}(\eta)  = \exp(-\eta) \;\;.
\end{equation}
and the cumulative distribution simply is
\begin{equation} \label{eq:distribution-N-limit-CUEc-cumulative}
  I^{\text{CUE}}(\eta) = 1-\exp(-\eta) \;\;.
\end{equation}
Fig.~\ref{fig:CUE-cat-evec-distribution} shows $P(\eta)$
for one eigenvector of the perturbed cat map \eqref{eq:CUE-map}.
There is good agreement with $P^{\text{CUE}}(\eta)$.

A different distribution is obtained for unperturbed cat maps:
for certain subsequences of prime
numbers (which depend on the map)
the distribution of $\eta=\tfrac{1}{2} \Real \,\psi$ tends to
the semicircle law,
\begin{equation} \label{eq:semicircle}
  P(\psi)= \begin{cases} 
             \frac{2}{\pi}\sqrt{1-\eta^2} & \qquad \text{for } \eta \le 1 \\
             0               & \qquad \text{for }\eta > 1  \;\;,
           \end{cases} 
\end{equation}
see \cite{KurRud2001} for details (see also \cite{Eck86}).
In fig.~\ref{fig:semicircle-cat-evec-distribution}
we show an example of an eigenstate with $N=1597$ for the quantum map
corresponding to the map \eqref{eq:p-cat} with $\kappa=0$.
For this $N$ the map fulfils the conditions of \cite{KurRud2001}
and one observes a nice semicircle distribution of the eigenvector.
However, it seems that the approach to the asymptotic distribution
is slower than for the case of the random matrix situations.

\FloatBarrier
\subsubsection{Husimi functions}

A different representation of eigenstates is to consider 
a  phase space representation,
like for example the Husimi function,
which allows for a more direct comparison with the structures
for the classical map.
Without going into the mathematical details, the Husimi
representation is obtained by projecting
the eigenstate onto a coherent state centred in a point $(q,p)\in\T^2$,
\begin{align} 
  H_n(q,p) &= \left|  \langle C_{q,p} | \psi_n  \rangle \right|^2 
           = \left|  \sum_{j=0}^{N-1} 
               \langle C_{q,p} | q_j  \rangle 
               \langle  q_j | \psi \rangle \right|^2 
           = \left|  \sum_{j=0}^{N-1}  
               \langle   C_{q,p} | q_j     \rangle 
                c_j  \right|^2  \nonumber \\ 
           &=  \left|  \sum_{j=0}^{N-1}    
            (2N)^{1/4}   
        \exp\left(-\pi N (q^2-\ui p q) \right)
        \exp(\pi N (-q_j^2+2(q-\ui p) q_j)) \right. \label{eq:Husimi}
                \\ \nonumber
         &\left.\qquad  \qquad \quad \vartheta_3\left.\left(
                   \ui \pi N \left( q_j -\frac{\ui \theta_1}{N} 
                                    - q+\ui p\right)
                        \right| \ui N \right)  c_j \right|^2  \;\;.
\end{align}
Here $q_j= \tfrac{1}{N}(\theta_2+j)$, $j=0,\ldots, N-1$
and
$\vartheta_3(Z | \tau)$ is the Jacobi-Theta function,
\begin{equation}
  \vartheta_3(Z | \tau) = \sum_{n\in\Z} \ue^{\ui \pi \tau n^2+2\ui n Z}\;\;,
    \qquad \text{ with } Z,\tau \in \C, \;\; \Imag(\tau) > 0\;\;.
\end{equation}
The coefficients $c_j$ are the components of the eigenvector $\psi_n$
in the position representation as obtained from the diagonalization
of $U_N$ (for other generating functions than the one used
in eq.~\eqref{eq:generating-function-quantization} one has to adapt
eq.~\eqref{eq:Husimi}).

\newcommand{\einpcatbild}[4]{
\begin{minipage}{#2}
   \hspace*{0.25cm} #4 \hspace*{2.75cm} #3

  \PSImagx{#1}{#2}
\end{minipage}
}

\BILD{t}
     {
     \begin{center}
\vspace*{-0.5cm}
      \einpcatbild{}{7.8cm}{$\kappa=0.3$}{a)}
      \einpcatbild{}{7.8cm}{$\kappa=6.5$}{b)}

\vspace*{-1.5cm}
     \end{center}
     }
     {In a) a Husimi function $H_n(q,p)$  of
      the perturbed cat map \eqref{eq:p-cat} with $\kappa=0.3$
      is plotted which shows
      the expected `uniform' distribution. Here
      black corresponds to large values of $H_n(q,p)$.
      In b) for $\kappa=6.5$ a state localizing on one of the 
      elliptic islands is shown
     (compare with fig.~\ref{fig:orbits-perturbed-cat}).}
     {fig:Husimi-p-cat}

\newcommand{\einstandardbild}[4]{
\begin{minipage}{#2}
   \hspace*{0.25cm} #4 \hspace*{2.75cm} #3

  \PSImagx{#1}{#2}
\end{minipage}
}

\BILD{b}
     {
     \begin{center}
\vspace*{-0.5cm}

        \einstandardbild{}{7.8cm}{}{a)}
        \einstandardbild{}{7.8cm}{}{b)}
         
\vspace*{-0.75cm}
         
        \einstandardbild{}{7.8cm}{}{c)}
        \einstandardbild{}{7.8cm}{}{d)}
     \end{center}

\vspace*{-1.5cm}
     }
     {Examples of Husimi functions for the standard map with $\kappa=3.0$
      and $N=1600$.
     }
     {fig:Husimi-standard}

\renewcommand{\einstandardbild}[4]{
\begin{minipage}{#2}
\vspace*{-0.35cm}

  \PSImagx{#1}{#2}
\end{minipage}
}

\BILD{h}
     {

     \begin{center}

        \einstandardbild{}{4.4cm}{}{}
        \hspace*{-0.6cm}
        \einstandardbild{}{4.4cm}{}{}
        \hspace*{-0.6cm}
        \einstandardbild{}{4.4cm}{}{}
        \hspace*{-0.6cm}
        \einstandardbild{}{4.4cm}{}{}

\vspace*{0.35cm}

        \einstandardbild{}{4.4cm}{}{}
        \hspace*{-0.6cm}                                       
        \einstandardbild{}{4.4cm}{}{}
        \hspace*{-0.6cm}                                       
        \einstandardbild{}{4.4cm}{}{}
        \hspace*{-0.6cm}                                       
        \einstandardbild{}{4.4cm}{}{}

\vspace*{0.35cm}

        \einstandardbild{}{4.4cm}{}{}
        \hspace*{-0.6cm}                                        
        \einstandardbild{}{4.4cm}{}{}
        \hspace*{-0.6cm}                                        
        \einstandardbild{}{4.4cm}{}{}
        \hspace*{-0.6cm}                                        
        \einstandardbild{}{4.4cm}{}{}

\vspace*{0.35cm}

        \einstandardbild{}{4.4cm}{}{}
        \hspace*{-0.6cm}                                        
        \einstandardbild{}{4.4cm}{}{}
        \hspace*{-0.6cm}                                        
        \einstandardbild{}{4.4cm}{}{}
        \hspace*{-0.6cm}                                        
        \einstandardbild{}{4.4cm}{}{}

\vspace*{0.35cm}

        \einstandardbild{}{4.4cm}{}{}
        \hspace*{-0.6cm}                                        
        \einstandardbild{}{4.4cm}{}{}
        \hspace*{-0.6cm}                                        
        \einstandardbild{}{4.4cm}{}{}
        \hspace*{-0.6cm}                                        
        \einstandardbild{}{4.4cm}{}{}
     \end{center}
     }
     {Examples of Husimi functions $H_n(q,p)$
      for the standard map with $\kappa=1.5$
      and $N=1600$ for $n=0,\dots,19$.
      (Compare with fig.~\ref{fig:orbits-standard-map}.)
     }
     {fig:Husimi-standard-1.5}

If one wants to compute a Husimi function on a grid of $N\times N$
points on $\T^2$ the computational effort grows with $N^3$.
So for computing all Husimi function 
of a quantum map for a given $N$ the computational effort grows
with $N^4$. Already for moderate $N$ this can be quite
time--consuming, but even more importantly,
usually one also wants to store all these Husimi functions on the hard-disk
which limits the accessible range of $N$.
Sometimes a smaller grid, e.g.\
of size $10\sqrt{N} \times 10\sqrt{N}$
can be sufficient which reduces the growth of the computational
effort to $N^2$ for a single Husimi function and to $N^3$ for
all Husimi functions at a given $N$.
Even then one still needs $800\, N^2$ Bytes to store these
on the hard-disk. For example for $N=1600$ this roughly leads to 
2 GB of data and for $N=3000$ one needs approximately 7 GB.
However, there are also cases where a finer grid, e.g.\ $2N \times 2N$
is necessary.

Theoretically one expects that for $N\to\infty$
the Husimi functions concentrate on those
regions in phase space which are invariant under the map
(this follows from the Egorov property).
So for ergodic systems the expectation is that (in the weak sense)
\begin{equation}
  H_n(q,p) \to 1  \quad \text{with $n=0,...,N-1$ 
                              as the matrix size $N\to\infty$} \;\;.
\end{equation}
The precise formulation of this statement is the
contents of the quantum ergodicity theorem for maps \cite{BouDeB96}
(see \cite{DeBDeg96} for the case of discontinuous maps).
The quantum ergodicity theorem only makes a statement about a subsequence
of density one (i.e.\ almost all states) which
for example leaves space for scars, i.e.\ eigenstates
localized on unstable periodic orbits.
For systems with mixed phase space one (asymptotically) expects localization
in the stochastic region(s) and on the tori in the elliptic regions.

In fig.~\ref{fig:Husimi-p-cat}a) we show for the perturbed
cat map with $\kappa=0.3$
the Husimi function for the same eigenstate as in 
fig.~\ref{fig:evec-distribution}. As expected it shows a quite uniform
distribution (of course with the usual fluctuations).
In contrast for $\kappa=6.5$ there are eigenstates
such as the one shown in fig.~\ref{fig:Husimi-p-cat}b) 
which localizes on the elliptic island 
(compare with fig.~\ref{fig:orbits-perturbed-cat}).

In some sense more interesting are the Husimi functions for
mixed systems as the classical dynamics shows
more structure.
In fig.~\ref{fig:Husimi-standard} we show some examples
for the standard map with $\kappa=3.0$
Fig.~\ref{fig:Husimi-standard}a) shows a Husimi function
which is spread out in the irregular component.
In contrast in b) the Husimi function localizes on
a torus around the elliptic fixpoint.
The Husimi function in c) shows quite strong localization
around the small elliptic island of a periodic orbit with period
4. This island is so small that it is not visible in 
fig.~\ref{fig:orbits-standard-map}.
Therefore, the Husimi function displayed in fig.~\ref{fig:Husimi-standard}d)
indicates that the region of `influence' of this island is
much larger than the area of the island.
This region is also visible in the Husimi function
in fig.~\ref{fig:Husimi-standard}a), as the irregular
state has a very small probability in the regions around these islands.
A longer sequence of Husimi functions for the standard map with $\kappa=1.5$
shown in fig.~\ref{fig:Husimi-standard-1.5} 
illustrates the different types of
localized states (compare with fig.~\ref{fig:orbits-standard-map}).

\FloatBarrier
\section{Billiards}\label{sec:}

\subsection{Classical billiards}

A two--dimensional Euclidean billiard is given by the
free motion of a point particle in some 
domain $\Omega \subset \R^2$ with elastic
reflections at the boundary $\partial \Omega$.
Depending on the boundary one obtains completely
different dynamical behaviour, see fig.~\ref{fig:billiard-dyn}
where this is illustrated by showing orbits of
billiards in a circle, a square
and an ellipse, which are all integrable giving
rise to regular motion.
In contrast the Sinai billiard (motion in a square with
a circular scatterer), the stadium billiard (two semicircles
joined by parallel straight lines) and the cardioid billiard
show strongly chaotic motion (they are all proven
to be hyperbolic, ergodic, mixing and $K$--systems).

\newcommand{\einorbitbild}[1]{\begin{minipage}{4cm}\PSImagx{#1.ps}{4cm}\end{minipage}}

\BILD{tbh}
     {

\vspace*{0.5cm}

\hspace*{1.0cm}{\bf Integrable systems} 

\begin{center}
  \einorbitbild{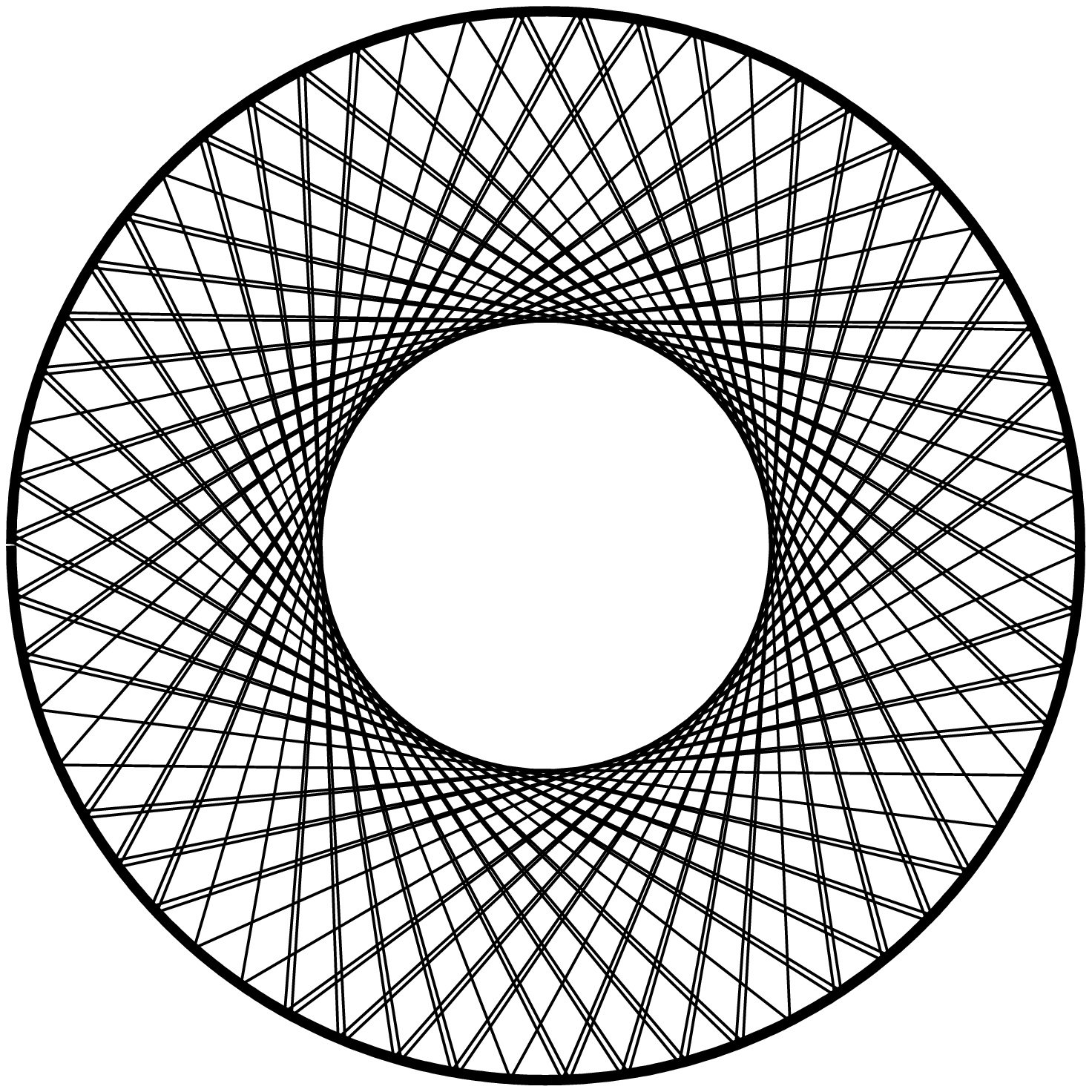}\hspace*{1.5cm}
  \einorbitbild{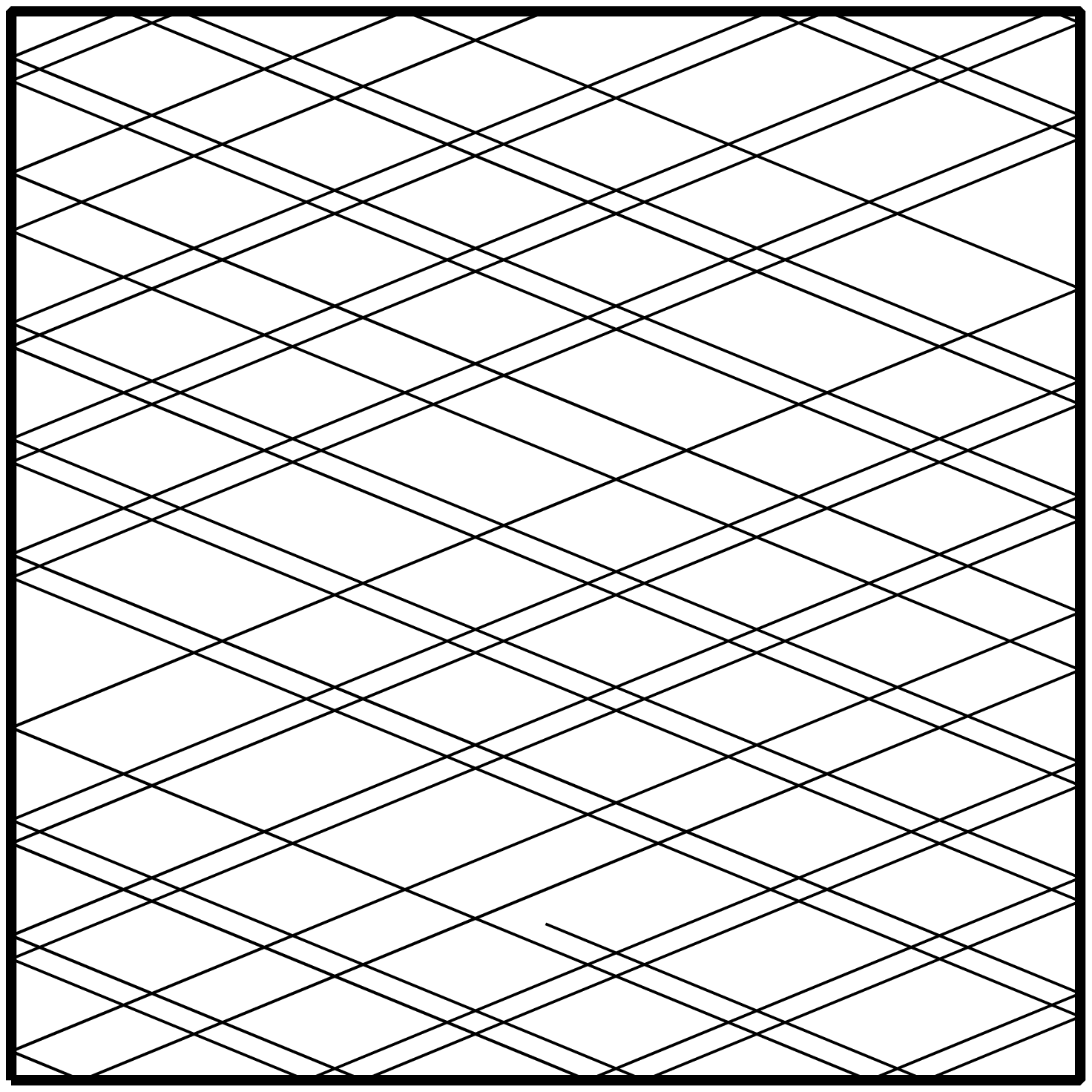}\hspace*{1.5cm}
  \einorbitbild{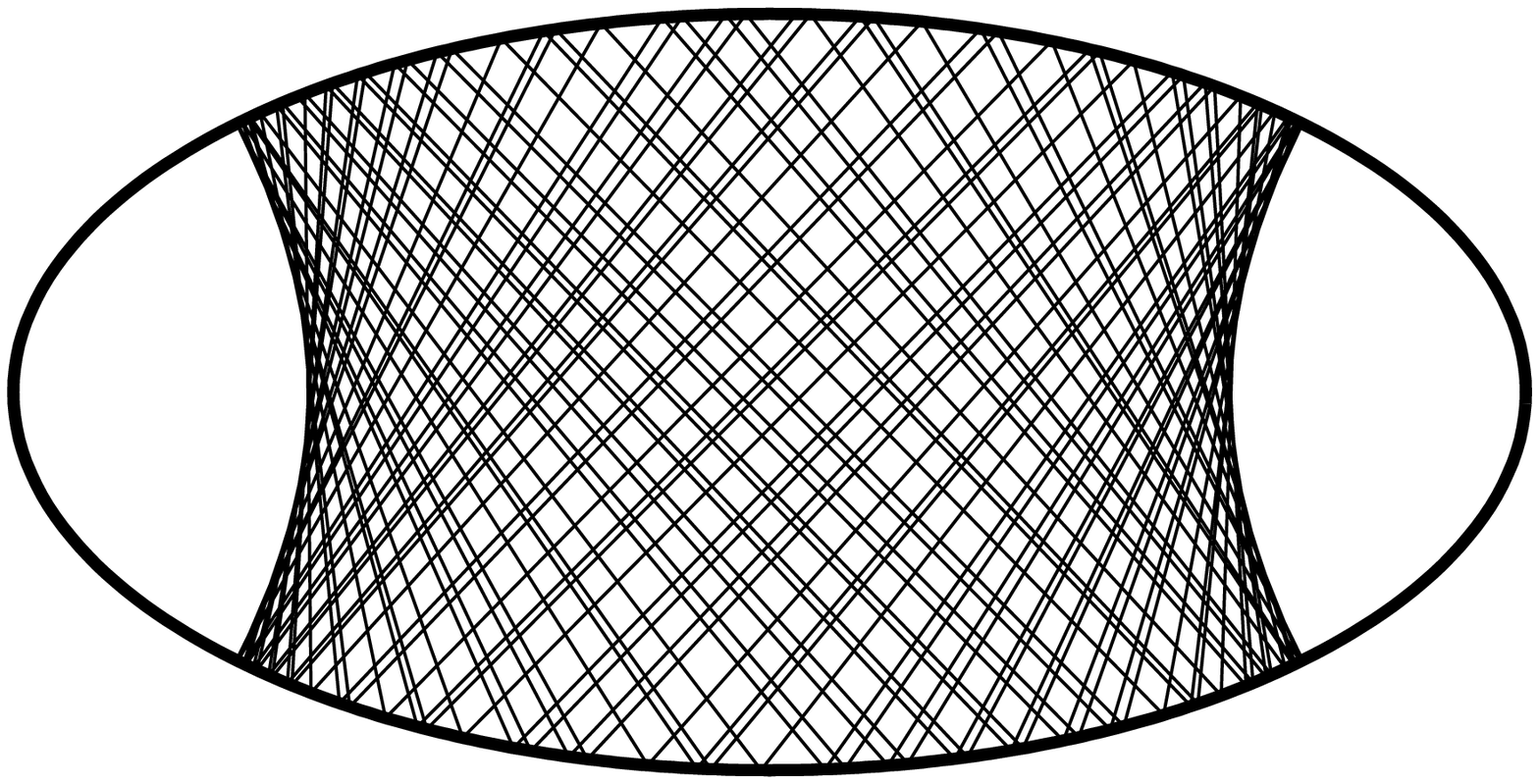}
\end{center}

\vspace*{0.5cm}

\hspace*{1.0cm} {\bf Chaotic systems}

\begin{center}
  \einorbitbild{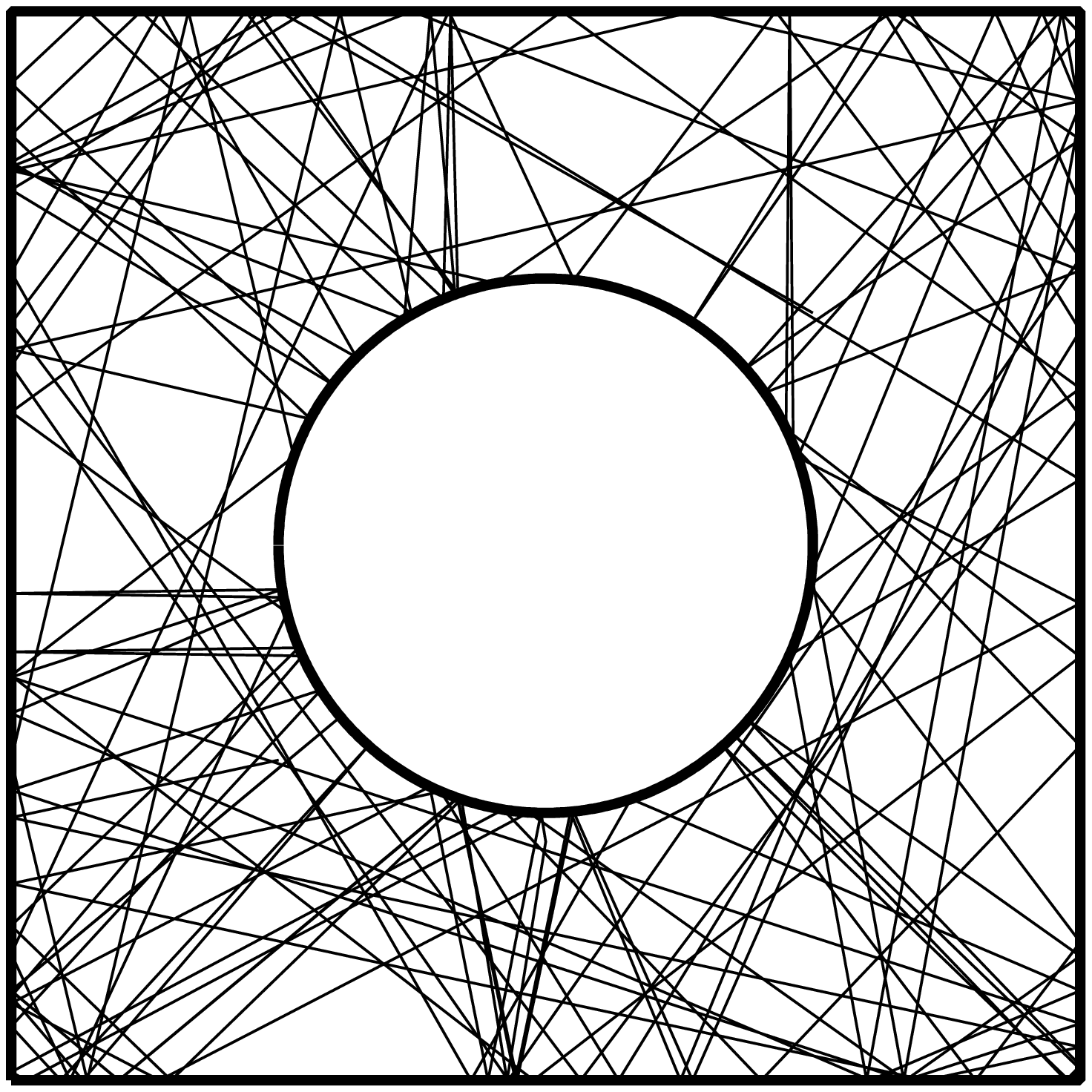}\hspace*{1.5cm}
  \einorbitbild{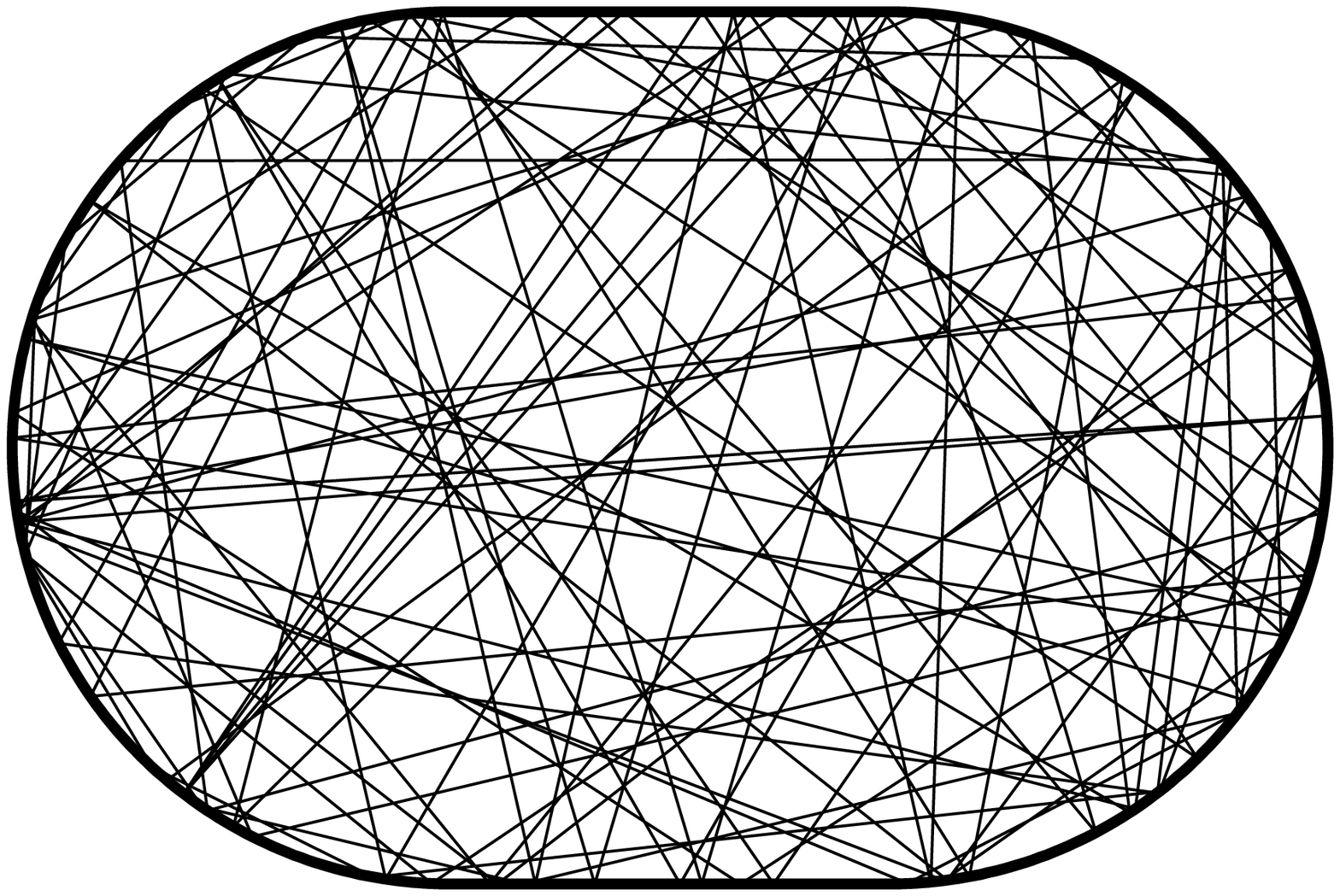}\hspace*{1.5cm}
  \einorbitbild{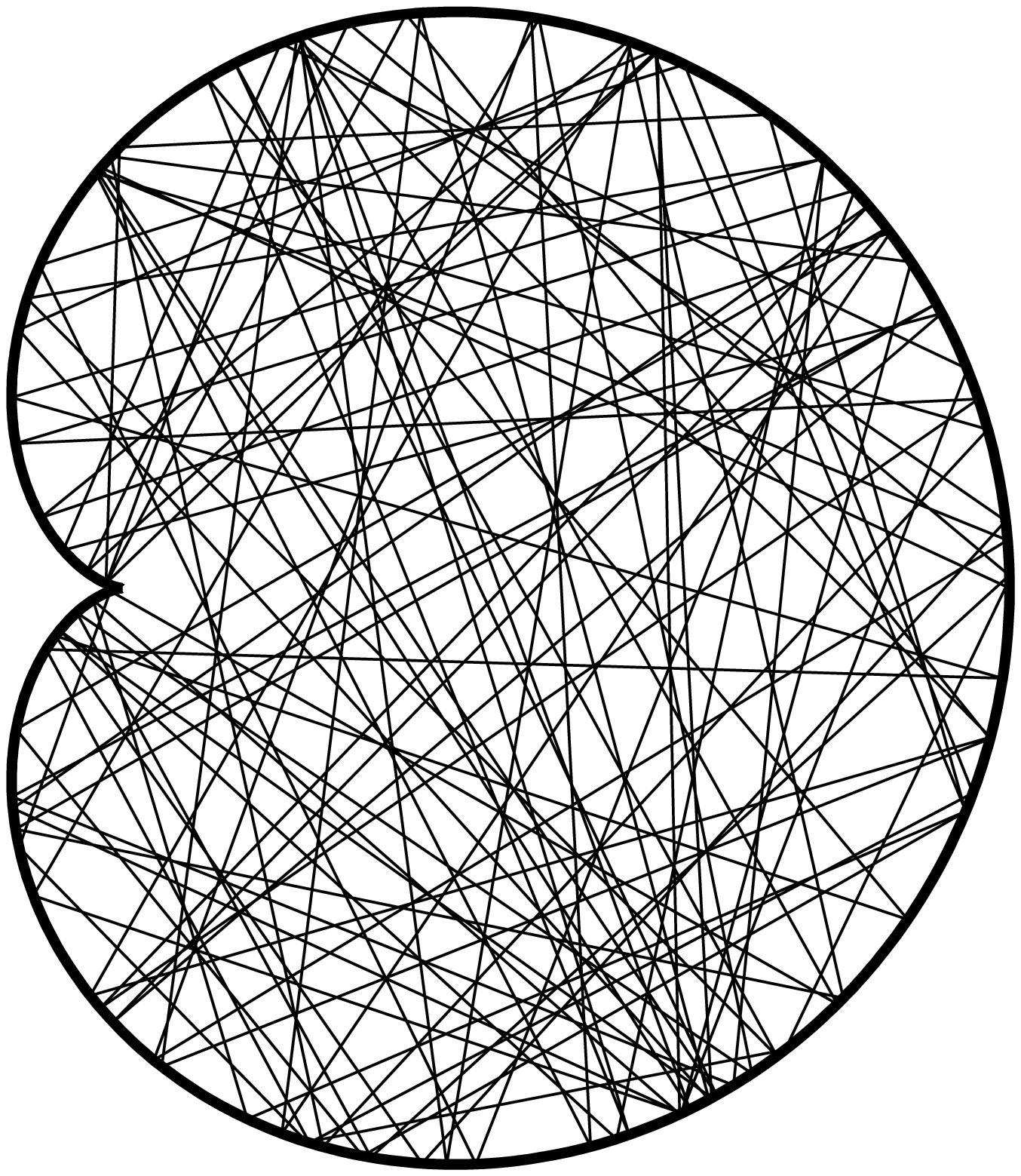}
\end{center}

     }
     {Billiard dynamics in integrable and chaotic systems.
     }
     {fig:billiard-dyn}

As the motion inside the billiard is on straight lines
it is convenient to use the boundary to define a Poincar\'e section,
\begin{equation}
   \cP:=\left\{ (s,p) \; | \; s\in[0,|\partial\Omega|], 
                            \;p\in[-1,1] \right\} \;\;.
\end{equation}
Here $s$ is the arclength along $\partial \Omega$
and $p=\langle\BFv,\BFT(s)\rangle$ is
the projection of the unit velocity vector $\BFv$
after the reflection on the unit tangent vector $\BFT(s)$
in the point $s\in\partial\Omega$.
The Poincar\'e map is then given by
\begin{align*}
  P:\cP&\to\cP\\
 \xi = (s,p) &\mapsto \xi' = (s',p') \;\;,
\end{align*}
i.e.\ for a given point $\xi = (s,p)$
one considers the ray starting at the point $\BFr(s)\in \partial\Omega$
in the direction specified by $p$ and determines the first
intersection with the boundary, leading to the new point $\xi' = (s',p')$.
Explicitly, the Cartesian components of the unit 
velocity $\BFv$ of a point particle starting on $\pO$ at $\BFr(s)$ are
determined by the angle $\beta\in[-\pi/2,\pi/2]$ 
measured with respect to the inward pointing normal 
$\BFN=(-T_y,T_x)$.
The velocity in the $\BFT,\BFN$ coordinate system is denoted by
$(p,n)=(\sin\beta,\cos\beta)$, so that in Cartesian coordinates
\begin{equation}
    \BFv= (v_x,v_y) = \matII{T_x & N_x}{T_y & N_y } (p,n) 
      = \left(T_x p + N_x \sqrt{1-p^2}, T_y p + N_y \sqrt{1-p^2}\right)  \;\;.
\end{equation}
Starting in the point $\BFr(s) \in\partial\Omega$ in the direction $\BFv$,
the ray $\BFr+t\BFv$ intersects $\partial \Omega$ at some point 
$\BFr'=(x',y')$. 
If the boundary is determined by the implicit equation
\begin{equation} \label{eq:boundary-implicit}
  F(x,y)=0 \;\;,
\end{equation}
the new point $\BFr'$ can be determined by solving 
\begin{equation} \label{eq:billiard-map-via-implicit}
        F(x+t v_x,y+t v_y) = 0 \;\;.
\end{equation}
For non--convex billiards there
are points $\xi=(s,p)\in\cP$ 
for which there is more than one solution (apart from $t=0$);
obviously the one with the smallest $t>0$ has to be chosen.
The condition \eqref{eq:boundary-implicit} can be used to remove
the $t=0$ solution analytically from \eqref{eq:billiard-map-via-implicit}.
If $F$ is a polynomial in $x$ and $y$ this allows
to reduce the order of \eqref{eq:billiard-map-via-implicit}
by one. This approach has for example been used for
the cardioid billiard leading to a cubic equation for $t$,
see \cite{BaeDul97} for details.
From the solution $t$ 
one gets the coordinates $(x',y')=(x,y)+t\BFv$ which have to be converted
(in a system dependent way) to the arclength coordinate $s'$
(in many practical applications there is a more suitable internal
variable, for example the polar angle etc.).
The corresponding new projection of the momentum is given
by $p'=-\langle\BFv,\BFT(s')\rangle$.

\subsection{Quantum billiards} \label{sec:quantum-billiards}

For a classical billiard system the associated
quantum billiard is given 
by the stationary Schr\"odinger equation (in units $\hbar=2m=1$)
\begin{equation}
\label{Schroedinger}
  -\Delta \psi_n({\bfq}) = E_n  \psi_n({\bfq})\;\;, \quad \bfq\in \Omega
\end{equation}                                            
with (for example) Dirichlet boundary conditions, 
i.e.\ $\psi_n({\bfq})=0$ for $\bfq\in \partial\Omega$.
Here
$\Delta$ denotes the Laplace operator, which reads in two dimensions
\begin{equation}
  \Delta=  \left(
                  \frac{\partial^2}{\partial q_1^2} 
                + \frac{\partial^2}{\partial q_2^2} 
              \right)\;\;.
\end{equation}

In the Schr\"odinger representation 
the state of a particle is described in configuration space by
a wave function $\psi\in L^2(\Omega)$,
where $L^2(\Omega)$ is the Hilbert space of square integrable
functions on $\Omega$.
The interpretation of $\psi$ is that $\int_D |\psi(\BFq)|^2 \;\ud^2 \noBFq$ 
is the probability of finding the particle inside the domain $D\subset\Omega$.

Due to the compactness of $\Omega$, 
the quantal energy spectrum $\{E_n\}$ is purely discrete
and can be ordered as $0<E_1\le E_2 \le E_3 \le \ldots$.
The eigenfunctions can be chosen to be real and to form
an orthonormal basis of $L^2(\Omega)$,
\begin{equation*}
   \langle \psi_n | \psi_m \rangle :=
\Int_{\Omega}   \psi_n({\bfq}) \psi_m({\bfq}) \; \ud^2q 
                     = \delta_{mn} \;\; .
\end{equation*}

The mathematical problem defined by eq.~\eqref{Schroedinger}
is the well--known eigenvalue problem of
the Helmholtz equation, which for example also describes
a vibrating membrane or flat microwave cavities.
For some simple domains $\Omega$ 
it is possible to solve eq.~\eqref{Schroedinger}
analytically. For example for the billiard in a rectangle
with sides $a$ and $b$ the (non--normalized) eigenfunctions are given by
$\psi_{n_1,n_2}(\BFq)=\sin(\pi n_1 q_1/a) \sin(\pi n_2 q_2/b)$
with corresponding eigenvalues $E_{n_1,n_2} = \pi^2 (n_1^2/a^2+n_2^2/b^2)$
and $(n_1,n_2)\in{\N\,}^2$.
For the billiard in a circle the  eigenfunctions
are given in polar--coordinates by 
$\psi_{mn}(r,\varphi)=J_m(j_{mn}r) \exp(\ui m \varphi)$,
where $j_{mn}$ is the $n$--th zero of the Bessel function $J_m(x)$
and $m\in \Z$, $n\in \N$.
However, in general no analytical solutions of eq.~\eqref{Schroedinger}
exist so that numerical methods have to be used to compute
eigenvalues and eigenfunctions.

\BILD{tbh}
     {
      \begin{center}
      \hspace*{1.25cm}\PSImagxy{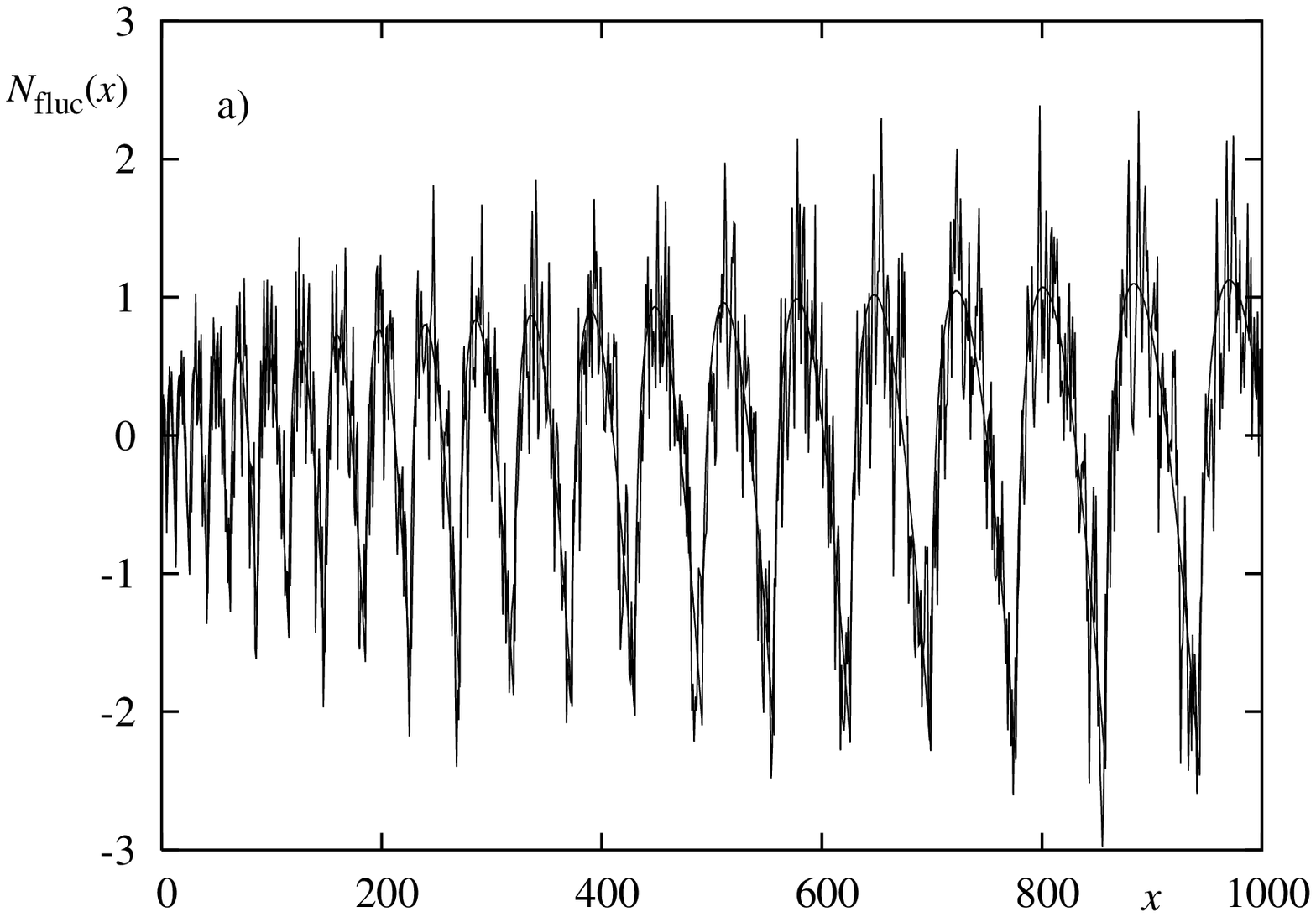}{14cm}{10.0cm}
          \hspace*{0.25cm}

      \hspace*{1.25cm}\PSImagxy{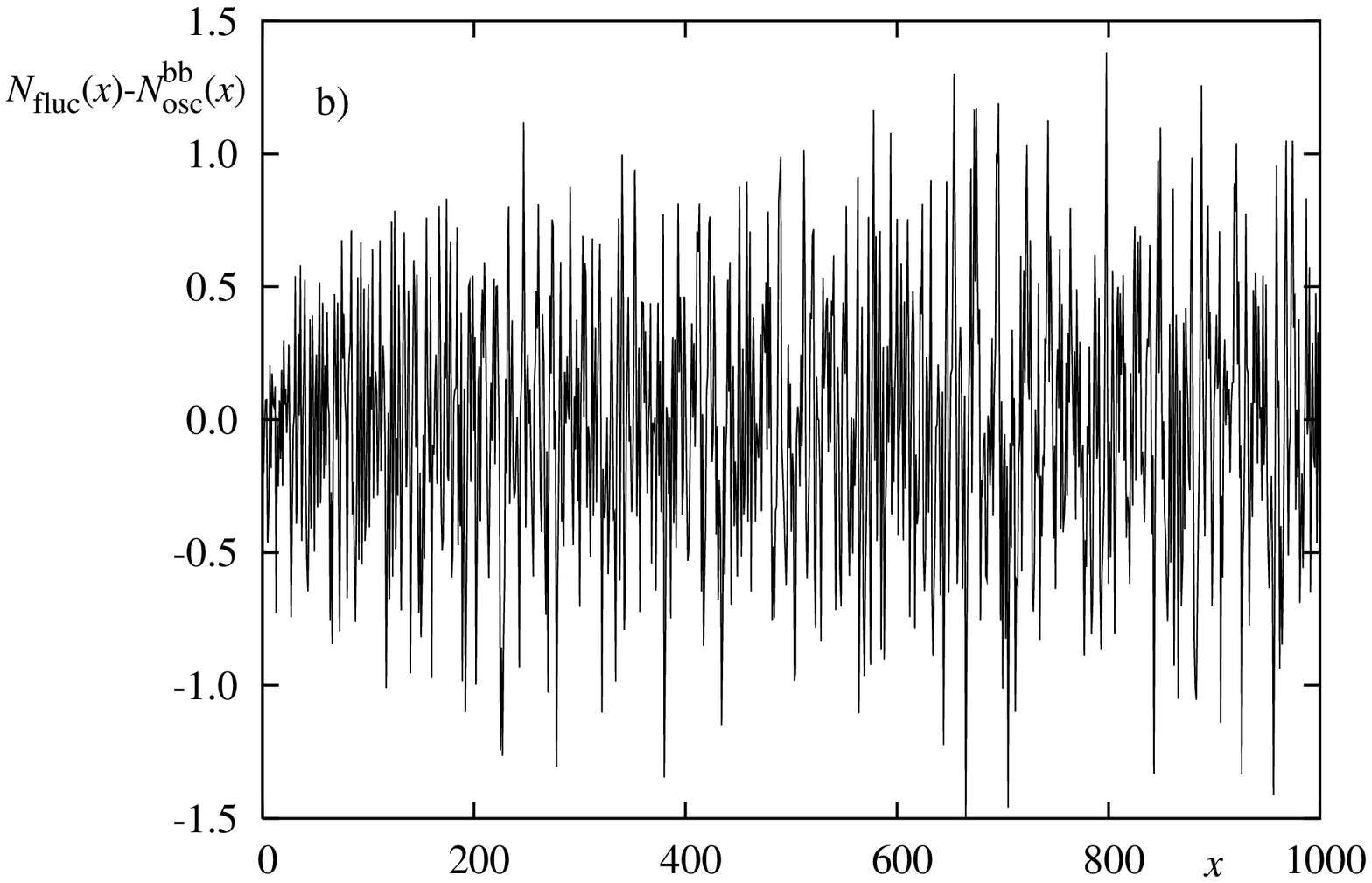}{14cm}{10.0cm}
          \hspace*{0.25cm}
      \end{center}
     }
     {Plot of $N_{\text{fluc}}(x)$ 
      for the stadium billiard ($a=1.8$, odd-odd symmetry) 
      together with the contribution
      from the bouncing ball orbits, dashed line, 
      see eq.~\eqref{eq:bb-stadium}.
      In b) 
      the fluctuating part after subtraction of the contribution
      of the bouncing ball orbits is shown.}
     {fig:delta-n-stadium}

\BILD{tbh}
     {
      \begin{center}
      \hspace*{1.25cm}\PSImagxy{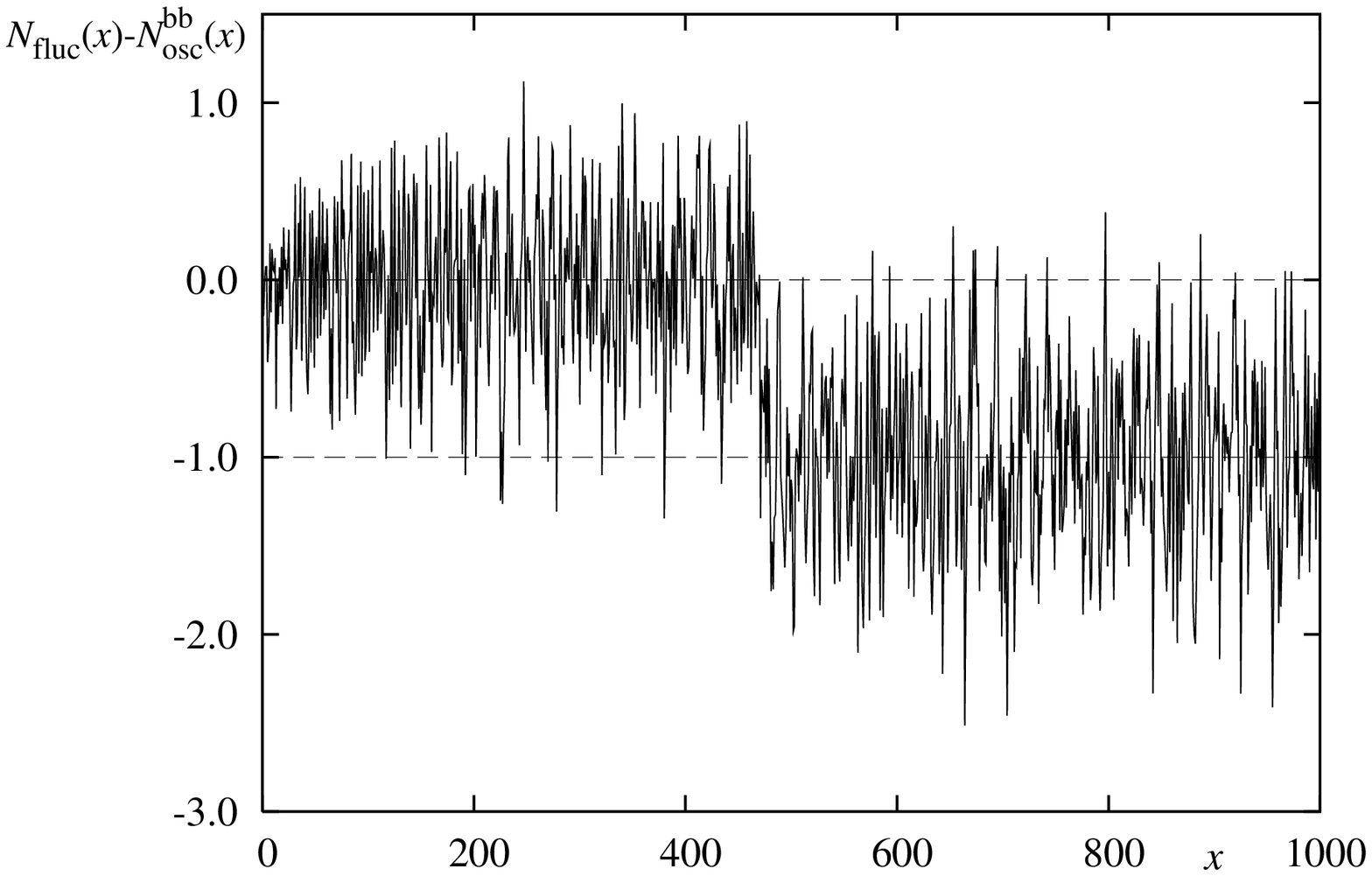}{14cm}{10.0cm}
          \hspace*{0.25cm}

      \end{center}
     \vspace*{-0.5cm}
     }
     {Detection of missing levels using the $\delta_n$--statistics.
      Shown is $N_{\text{fluc}}(x)-N^{\text{bb}}_{\text{osc}}(x)$
      for the case where one level has been removed from the spectrum.
      One clearly sees the jump in the fluctuations which allows
      to roughly locate the place where the level is missing.
     }
     {fig:delta-n-with-missing-eval}

The spectral staircase function $N(E)$
(integrated level density)
\begin{equation}
  N(E) := \#\{n \; | \; E_n \le E\} \;\;
\end{equation}
counts the number of energy levels $E_n$ below a given energy $E$.
$N(E)$ can be separated into a mean smooth part $\overline{N}(E)$
and a fluctuating part
\begin{equation}
  N(E) =  \overline{N}(E) + N_{\text{fluc}}(E) \;\;.
\end{equation}
For two-dimensional billiards, $\overline{N}(E)$ is given by the
generalized Weyl formula \cite{BalHil76}
\begin{equation} \label{Weyl}
  \overline{N}(E) =  \frac{\CA}{4\pi} E 
                      - \frac{\CL}{4\pi} \sqrt{E} 
                      + \CC  + \ldots\;\;,
\end{equation}
where $\CA$ denotes the area of the billiard, and
$\CL:=\CL^- - \CL^+$, where $\CL^-$ and $\CL^+$ 
are the lengths of the pieces of the boundary $\partial \Omega$
with Dirichlet and Neumann boundary conditions,
respectively.  
The constant $\CC$ takes curvature and corner corrections into account.

The simplest quantity is the
$\delta_n$--statistics, which is obtained from
the fluctuating part of the spectral
staircase evaluated at the unfolded energy eigenvalues 
$x_n:=\overline{N}(E_n)$
\begin{equation}
\label{delta_n}
  \delta_n := N_0(E_n)-\overline{N}(E_n)
            = n - \frac{1}{2}-x_n \;\; ,
\end{equation}                                    
where 
\begin{equation} 
  N_0(E):=\lim_{\epsilon\to 0} \frac{N(E+\epsilon)+N(E-\epsilon)}{2}  \;\;.
\end{equation}
The quantity $\delta_n$ is a good measure for the completeness
of a given energy spectrum.
For a complete spectrum $\delta_n$, or equivalently
$N_{\text{fluc}}(x)$, should fluctuate around zero. 
Fig.~\ref{fig:delta-n-stadium}a) shows $N_{\text{fluc}}(x)$ 
for the stadium billiard,
which indeed fluctuates around zero. In addition
there is an overall modulation of $N_{\text{fluc}}(x)$
which is caused by the bouncing ball orbits. They
form a one parameter family of periodic orbits
having perpendicular reflections at the two parallel walls
(of length $a$, see fig.~\ref{fig:stadium-geometry}) of the stadium.
The contribution of these orbits to the spectral staircase  function
reads 
        \cite{SieSmiCreLit93}
        \begin{align}
           \label{eq:bb-stadium}
           N^{\text{bb}}_{\text{fluc}}(E) &= \frac{a}{\pi} \sum_{n=1}^\infty
                 \sqrt{E-E_n^{\text{bb}}} \; \;
                  \Theta\left(\sqrt{E}-\sqrt{E_n^{\text{bb}}}\right)
            - \left(\frac{a}{4\pi} E - \frac{1}{2\pi} \sqrt{E} \right) \\
             &=\frac{a}{2\sqrt{\pi^3 }}\,E^{\frac{1}{4}}\,\sum_{n=1}^\infty
\frac{1}{n^{\frac{3}{2}}}\,\cos\left(2an \sqrt{E}-\frac{3\pi}{4}\right)
    \;\;, 
         \label{eq:bb-stadium-}
        \end{align}
        where $E^{\text{bb}}_n= \pi^2 n^2$ are 
        the eigenvalues of a particle in a one-dimensional box of length $1$,
        and $\Theta$ is the Heaviside step function.
Subtracting $N^{\text{bb}}_{\text{osc}}(x)$ from $N_{\text{fluc}}(x)$
removes the additional oscillation, see fig.~\ref{fig:delta-n-stadium}b).
If an eigenvalue is missing 
this is clearly visible by a `jump' of $\delta_n$ in comparison to 
points fluctuating around $0$,
see fig.~\ref{fig:delta-n-with-missing-eval}
for an example where one eigenvalue has been removed `by-hand'.
Clearly, the energy interval in which a level is missing can
be estimated from the plot.

In the same way as for quantum maps one
can study the level spacing distribution
and more complicated statistics, like the number variance,
$n$--point correlation functions etc.,
see for example \cite{BaeSteSti95,BaeSte2001}
for some further examples for the cardioid billiard.

\FloatBarrier
\subsection{Computing eigenvalues and eigenfunctions for quantum billiards}

There exist several numerical methods to solve the Helmholtz
equation 
inside a domain $\Omega \subset \R^2$,
\begin{equation} \label{eq:Helmholtz}
  \Delta \psi(\bfq) + k^2 \psi(\bfq) = 0 \;\; , 
            \qquad \bfq \in \Omega \backslash \partial \Omega \;\;,
\end{equation}                                                         
with Dirichlet boundary conditions 
\begin{equation} 
  \label{eq:DC-bc}
  \psi(\bfq) = 0  \;\; , \qquad \bfq \in \partial \Omega.
\end{equation}
For a good review on the determination of the
eigenvalues of \eqref{eq:Helmholtz} see \cite{KutSig84},
which however does not cover finite element methods
or boundary integral methods.
Additionally, in the context of quantum chaos the plane
wave decomposition \cite{Hel91} (see also \cite{LiRob94} for a detailed
description of the method),
the scattering approach, see e.g.\ \cite{DorSmi92,DieSmi93,SchSmi95}, 
and more recently the scaling method \cite{VerSar95},
are commonly used.

Here I will give a sketch of the derivation
of the boundary integral method and discuss in more detail the numerical
implementation. 
The boundary integral method reduces the problem of
solving the two-dimensional Helmholtz equation \eqref{eq:Helmholtz} 
to a one-dimensional integral equation, see e.g.\
\cite{BurMil71,KleRoa74,Rid79,Rid79b,Mar82,BerWil84,SieSte90b,%
BisJai90,Boa92,Boa94,AurSte93,Pis96,KosSch97,LiRobHu98,Sie98,HorSmi2000}
and references therein.
Of course, the general approach also applies 
to higher dimensions 
but we will only discuss the two-dimensional case.
For studies of three-dimensional systems by various
methods see e.g.\ \cite{AurMar96,PriSmi95,Ste96,Pro97a,Pro97b}.
Boundary integral methods are also used in many other areas so
that it is impossible to give a full account.
For example they are also commonly used in acoustics,
see e.g.\ \cite{CisBre1991} and the detailed list of references therein.
Finally, the boundary integral method provides
a starting point to derive the Gutzwiller trace formula,
see e.g.~\cite{Bog92,HarShu92,Bur95,SiePavSch97,Sie98}.

\subsubsection{Boundary integral equation}

Let $G(\bfq,\bfq')$ be a Green function of the inhomogeneous equation,
i.e.\
\begin{equation} \label{eq:Inhomogene-Helmholtz-Gleichung}
  (\Delta + k^2) G_k(\bfq, \bfq') 
       = \delta(\bfq-\bfq') \;\;.
\end{equation}                            
Considering the integral over $\Omega$ of the difference 
$\psi(\bfq') \cdot$\eqref{eq:Inhomogene-Helmholtz-Gleichung}$-
G_k(\bfq, \bfq')\cdot$\eqref{eq:Helmholtz}        
one obtains
\begin{equation}
  \int\limits_{\Omega}  \left[
     \psi(\bfq') \Delta' G_k(\bfq, \bfq')  
    - G_k(\bfq, \bfq') \Delta' \psi(\bfq')
    \right] \;\ud^2\bfq'
  = \int\limits_{\Omega}  
     \psi(\bfq') \delta(\bfq - \bfq') \;\ud^2\bfq' \;\; .
\end{equation}
Using the second Green theorem gives the Helmholtz representation 
\begin{equation} \label{eq:Integral-Gleichung}
  \oint\limits_{\partial \Omega}  \left[
     \psi(\bfq') 
        \frac{\partial G_k}{\partial n'} (\bfq, \bfq')  
    - G_k(\bfq, \bfq') 
        \frac{\partial \psi}{\partial n'}(\bfq')
    \right] \;\ud s' 
  = \left\{ \begin{array}{cl}
              \psi(\bfq) & 
                    ; \quad \bfq \in 
                   {\Omega} \setminus {\partial \Omega} \\[0.3ex]
              \halb \psi(\bfq) & 
                    ; \quad \bfq \in {\partial \Omega} \\[0.3ex]
               0  & ; \quad \mbox{else}
            \end{array}   
    \right.       \;\;.
\end{equation}         
Here $\bfq'\equiv \bfq(s')$ and 
$\tfrac{\partial}{\partial n'} = \bfn(s') \nabla$
with $\bfn(s)=(q_2'(s),-q_1'(s))$ denoting the outward
pointing normal vector,
where $(q_1(s),q_2(s))$ is a parametrization
of the billiard boundary $\partial\Omega$ in terms
of the arclength $s$, oriented counterclockwise.
Special care has to be taken to obtain the result 
for $\BFq\in\partial\Omega$, see e.g.~\cite{KleRoa74,Bur95}.
(When $\bfq$ is in a corner of the billiard the
factor $\halb$ has to be replaced by $\tfrac{\theta}{2\pi}$,
where $\theta$ is the (inner) angle of the corner.)
For Dirichlet boundary conditions one obtains
\begin{equation} \label{eq:int-darstellung}
 \oint\limits_{\partial\Omega}
         u(s')
         G_k(\bfq,\bfq')
        \; \ud s' = 0 \;\;, \qquad \bfq \in \partial \Omega
         \;\;,
\end{equation}          
where 
\begin{equation}
 u(s):=\frac{\partial}{\partial n} \psi(\bfq(s)) := 
   \bfn(s) \nabla\psi(\bfq(s))  := 
            \bfn(s)  \lim_{\substack{\bfq'\to \bfq(s)\\ 
                  \BFq'\in \Omega\backslash\partial\Omega}} 
                 \nabla\psi(\bfq') 
\end{equation}
is the normal derivative function of $\psi$.

In two dimensions a Green function for a free particle is given
by the Hankel function of first kind
\begin{equation} \label{eq:Green-fct-two-dim}
  G_k(\bfq,\bfq') = - \frac{\ui}{4} \;H_0^{(1)} \left(k \,|\bfq-\bfq'|\right) 
                =  -\frac{\ui}{4} \;\left[    J_0 \left(k \,|\bfq-\bfq'|\right)
                                     +\ui \;Y_0  \left(k \,|\bfq-\bfq'|\right)
                               \right] \;\;.
\end{equation}                                                           
Since $H_0^{(1)}(z) \sim \tfrac{\ui}{\pi} \ln z$ for $z\to 0$,
the Green function $G_k(\bfq,\bfq')$ diverges logarithmically
such that it is more convenient to derive
an integral equation whose kernel is free of this singularity.
To that end one (formally)
applies the normal derivative $\frac{\partial}{\partial n}$
to eq.~\eqref{eq:Integral-Gleichung}.
More carefully one has to consider a jump relation for
the normal derivative function, see e.g.~\cite{KleRoa74,Bur95}.
The result is
\begin{equation} \label{eq:pre-Fredholm}
  u(s) = -2 \oint\limits_{\partial\Omega} \frac{\partial}{\partial n}
            G_k\left(\bfq(s),\bfq(s')\right)\,u(s') \;\ud s' \;\;.
\end{equation}                               
For the derivative of the Green function one obtains
\begin{equation}
  \frac{\partial}{\partial n} G_k\left(\bfq(s),\bfq(s')\right)
  = \frac{\ui k }{4} \cos(\phi(s,s')) \; H_1^{(1)}\left(k \,\tau(s,s')\right)
  \;\;,
\end{equation}   
where $\tau(s,s')=|\bfq(s)-\bfq(s')|$ is the Euclidean distance
between the two points on the boundary and
\begin{equation}
  \cos \phi(s,s') = \frac{\bfn(s) \left(\bfq(s)-\bfq(s')\right)}{\tau(s,s')}
  \;\;.
\end{equation}   
This gives the integral equation for the normal derivative $u(s)$
\begin{equation} \label{eq:Fredholm}
    u(s) = \oInt_{\partial\Omega} Q_k(s,s') \; u(s') \; \ud s' \;\;,  
\end{equation}         
with integral kernel
\begin{equation}
   Q_k(s,s') = - \frac{\ui k}{2} \cos \phi(s,s')  
             \; H_1^{(1)}\left(k \,\tau(s,s')\right) \;\;.
\end{equation}                                              
Eq.~\eqref{eq:Fredholm} is a Fredholm equation of second
kind which has non--trivial solutions if the determinant
\begin{equation}
  D(k):=\det(\mathbbm{1}-\widehat{Q}_k)
\end{equation}
vanishes. Here $\widehat{Q}_k$ is the integral operator on $\partial\Omega$
defined by 
\begin{equation}
  \widehat{Q}_k(u(s)) =  \oInt_{\partial \Omega} Q_k(s,s') \; u(s') \; \ud s' \;\;.
\end{equation}

For eigenvalues $E_n$ of the Helmholtz equation with Dirichlet
boundary conditions one has $D(k)=0$ for $k=\sqrt{E_n}$,
see e.g.\ \cite{Bur95} for a detailed proof.
However, for $\Imag\, k<0$ there can be further zeros of
$D(k)$ which (for the interior Dirichlet problem)
correspond to the outside scattering problem
with Neumann boundary conditions \cite{BurSte95:p,TasHarShu97,EckPil97}
(see also \cite{KleRoa74}).
Explicitly this can be seen from the factorization
$D(k)=D(0)D_{\text{int}}(k) D_{\text{ext}}(k)$,
where the factors can be written exclusively
in terms of the interior and exterior problem.
More aspects concerning the additional spurious solutions will be discussed
in sec.~\ref{sec:spurious}.

Before turning to the numerical implementation,
let us discuss the behaviour of the integral kernel
for small arguments.
The Hankel function $H_1^{(1)} (x)$ reads for small arguments
\begin{equation}
  H_1^{(1)}\left(k \,\tau(s,s')\right) \sim - \frac{2 \ui}{\pi k |s-s'|}
  \;\;, \qquad \text{ for } s-s' \to 0 \;\;.
\end{equation}                                        
This singularity is compensated by the behaviour of
\begin{equation}
  \cos \phi(s,s') \sim - \frac{1}{2} \kappa(s) \, |s-s'| 
  \;\;, \qquad \text{ for } s' \to s \;\;,
\end{equation}                                        
where $\kappa(s)$ is the curvature of the boundary in the point $s$.
Here the curvature is defined by $\kappa(s)=q_1'(s)q_2''(s)-q_2'(s)q_1''(s)$
such that for example $\kappa(s)=1$
for a circle of radius one.
Thus for the integral kernel we obtain 
\begin{equation} \label{eq:kernal-at-s-s}
  Q_k(s,s') \to \frac{1}{2\pi} \kappa(s) 
  \;\;, \qquad \text{ for } s-s' \to 0 \;\;.  
\end{equation}

\subsubsection{Desymmetrization}

\BILD{t}
     {
     \begin{center}
       \PSImagx{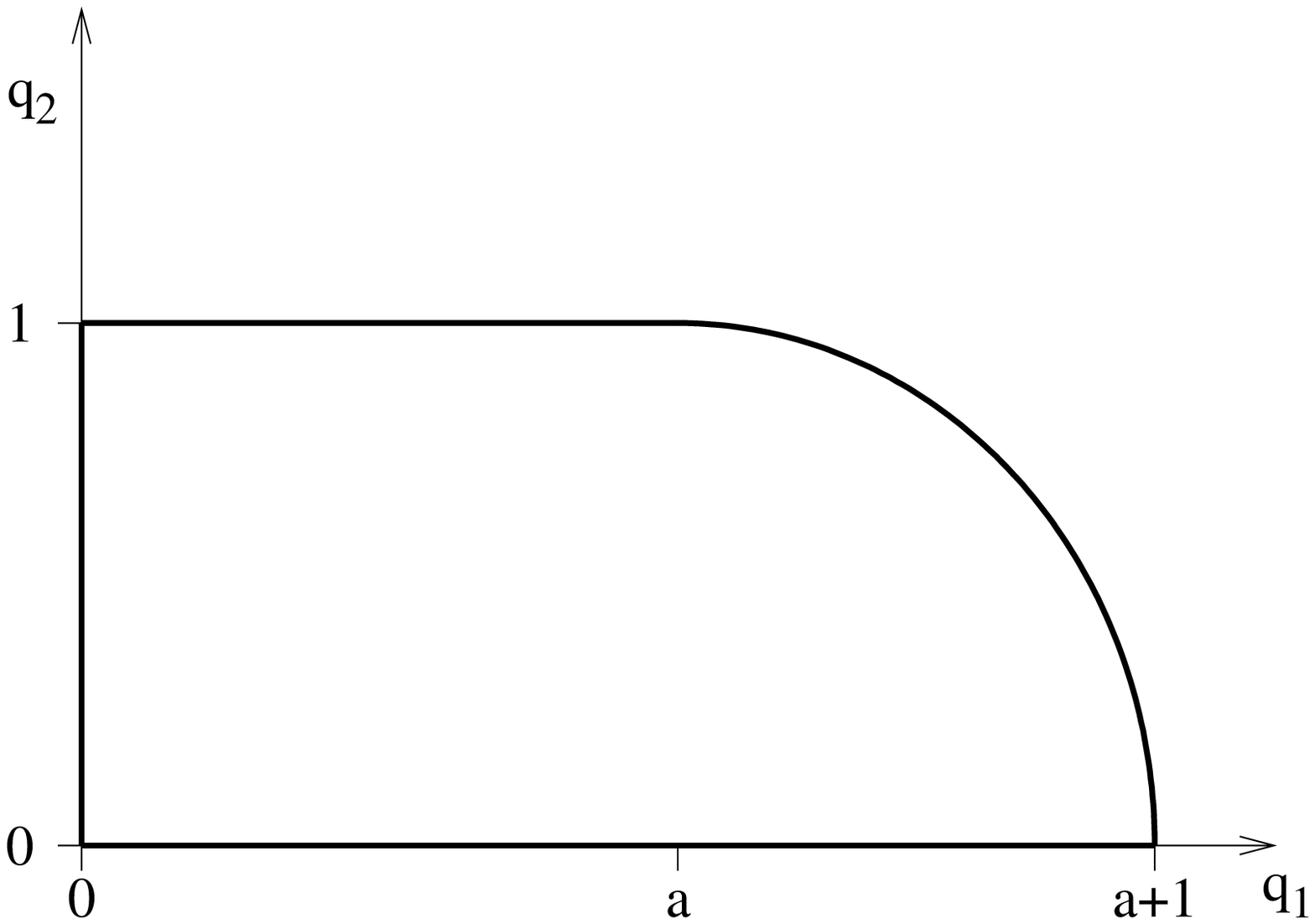}{10cm}
     \end{center}
     }
     {Geometry of the desymmetrized stadium billiard.
     }
     {fig:stadium-geometry}

For systems with symmetries the numerical effort can
be reduced by considering instead of the full system
the symmetry reduced system
with the corresponding Green function, see e.g.\ \cite{SieSte90b}.
For a reflection symmetry 
with respect to the $q_1$--axis one has
\begin{equation} \label{eq:desymm-x}
   G_k^\pm(\bfq,\bfq') := 
    G_k(|\bfq-\bfq'|) \pm    
    G_k(|\bfq-(q_1',-q_2')|) \;\; ,
\end{equation}
where $+$ applies to the case of  even eigenfunctions (i.e.\ Neumann
boundary conditions on the symmetry axis) and
$-$ to odd eigenfunctions  (i.e.\ Dirichlet 
boundary conditions on the symmetry axis).

For a two-fold reflection symmetry (as in the case of the stadium billiard,
see fig.~\ref{fig:stadium-geometry} for a sketch of the geometry
and notations)
one has altogether four different subspectra,
corresponding to DD, DN, ND and DD boundary conditions on the symmetry axes
$q_1$ and $q_2$, respectively. For example for Dirichlet-Dirichlet 
boundary conditions on the $q_1$-- and $q_2$--axes the Green function
reads
\begin{equation} \label{eq:desymm-xy-DD}
 G^{\text{DD}}_k = G_k(|\bfq-\bfq'|)- G_k(|\bfq-(q_1',-q_2')|)
                                   + G_k(|\bfq-(-q_1',-q_2')|)
                                   - G_k(|\bfq-(-q_1',q_2')|)  \;\;.
\end{equation}
For Neuman boundary conditions on these two axes one gets
\begin{equation}\label{eq:desymm-xy-NN}
 G^{\text{NN}}_k = G_k(|\bfq-\bfq'|)+ G_k(|\bfq-(q_1',-q_2')|)
                                   + G_k(|\bfq-(-q_1',-q_2')|)
                                   + G_k(|\bfq-(-q_1',q_2')|)   \;\;.
\end{equation}
The advantage of exploiting the symmetries of
the system is two-fold: firstly, we can separate
the eigenvalues and eigenfunctions for the different
symmetry classes, which is necessary for the investigation
of the spectral statistics. Secondly, the numerical
effort is reduced, since the integral over the whole boundary 
$\partial \Omega$ is reduced to an integral over a part of the boundary,
which in the above examples is half or a quarter
of the original boundary. The boundary 
along the symmetry axes need not be discretized as
the boundary condition is already fulfilled by construction.
Of course, for other geometries different choices for $G$ can
be more appropriate.

\subsubsection{Finding the eigenvalues}

\BILD{tbh}
     {
      \PSImagxy{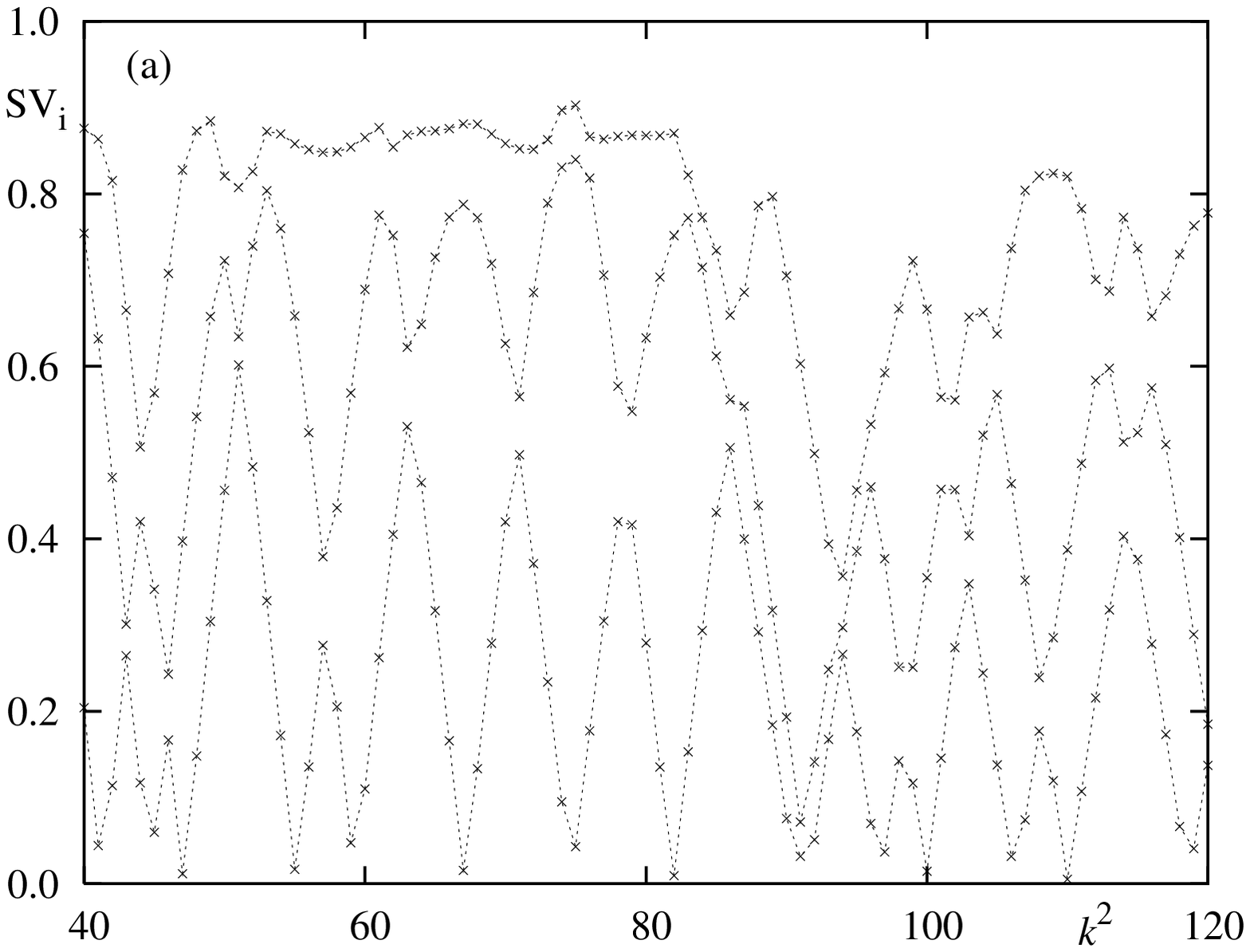}{16cm}{9.5cm}
      
      \vspace*{0.25cm}  

      \PSImagxy{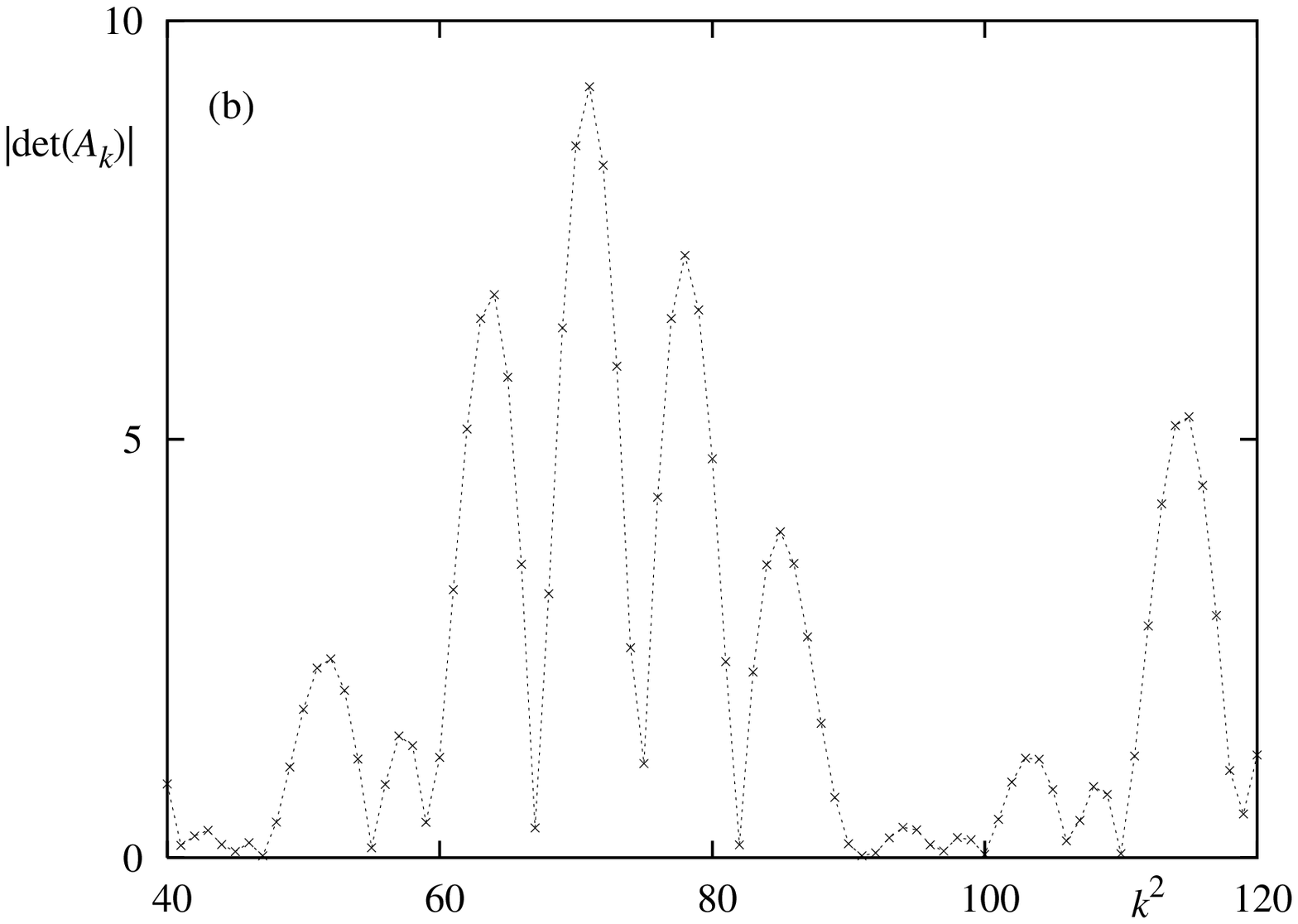}{16cm}{9.5cm}
     }
     {In a) the three smallest singular
      values are shown as a function of the energy $E=k^2$ for the
      stadium billiard with $a=1.8$ and odd-odd symmetry.
      The eigenvalues are located at the minima of the first
      singular value. The second and third singular values
      allow to locate places with near degeneracies
      as next to $k^2=90$, which can be resolved by
      magnification of the corresponding region, see 
      fig.~\ref{fig:svd-magnification}.
       In b) $|\det (A_k)|$ is shown. The minima tend
      to be not as pronounced as
       those of the singular values.}
     {fig:singular-value-example}

\BILD{t}
     {
      \begin{center}
         \PSImagx{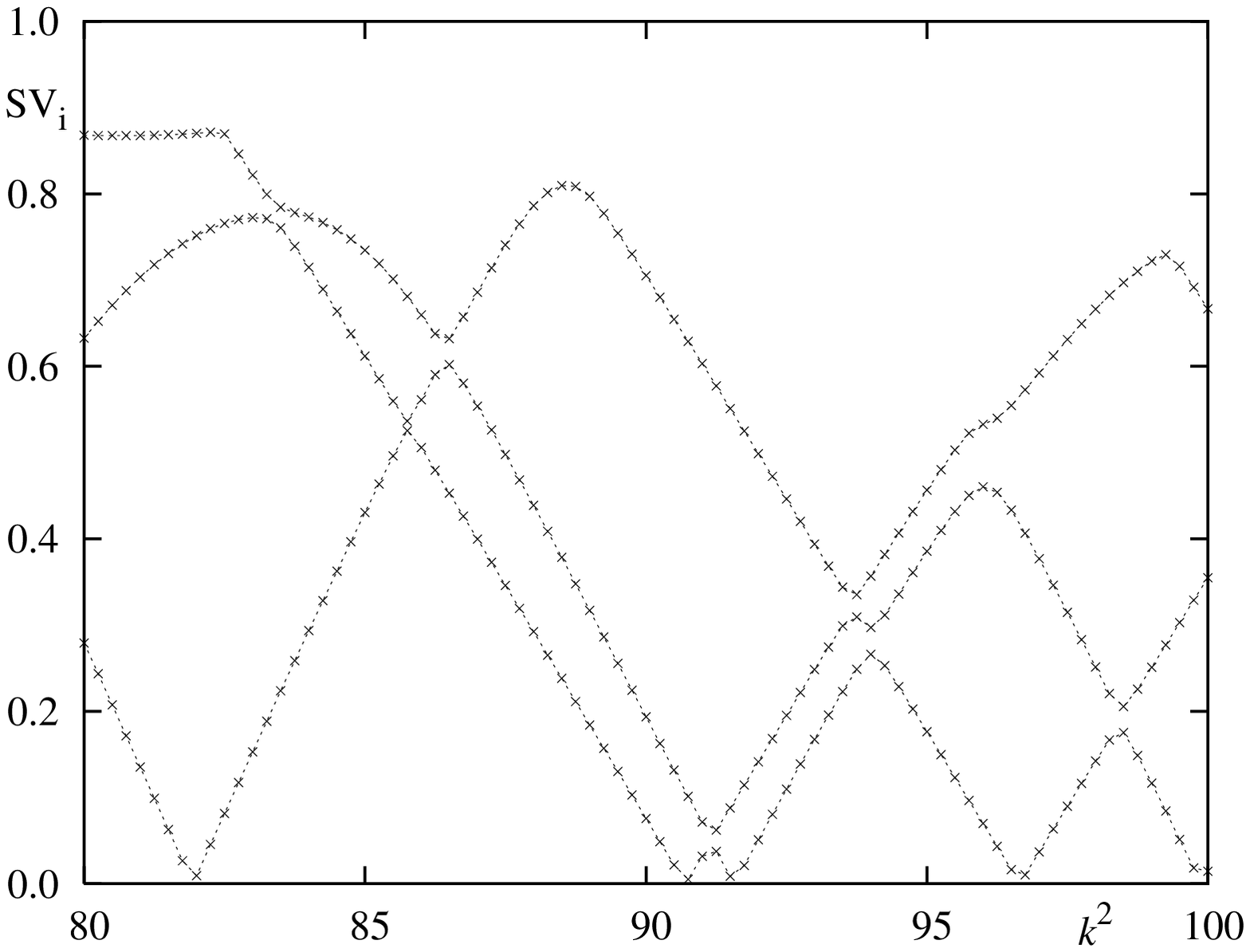}{13cm}
      \end{center}
     }
     {A magnification of fig.~\ref{fig:singular-value-example}
      shows that the singular value decomposition method
       easily  allows to locate nearly degenerate energy levels.
     }
     {fig:svd-magnification}

\BILD{b}
     {
      \begin{center}
       \PSImagx{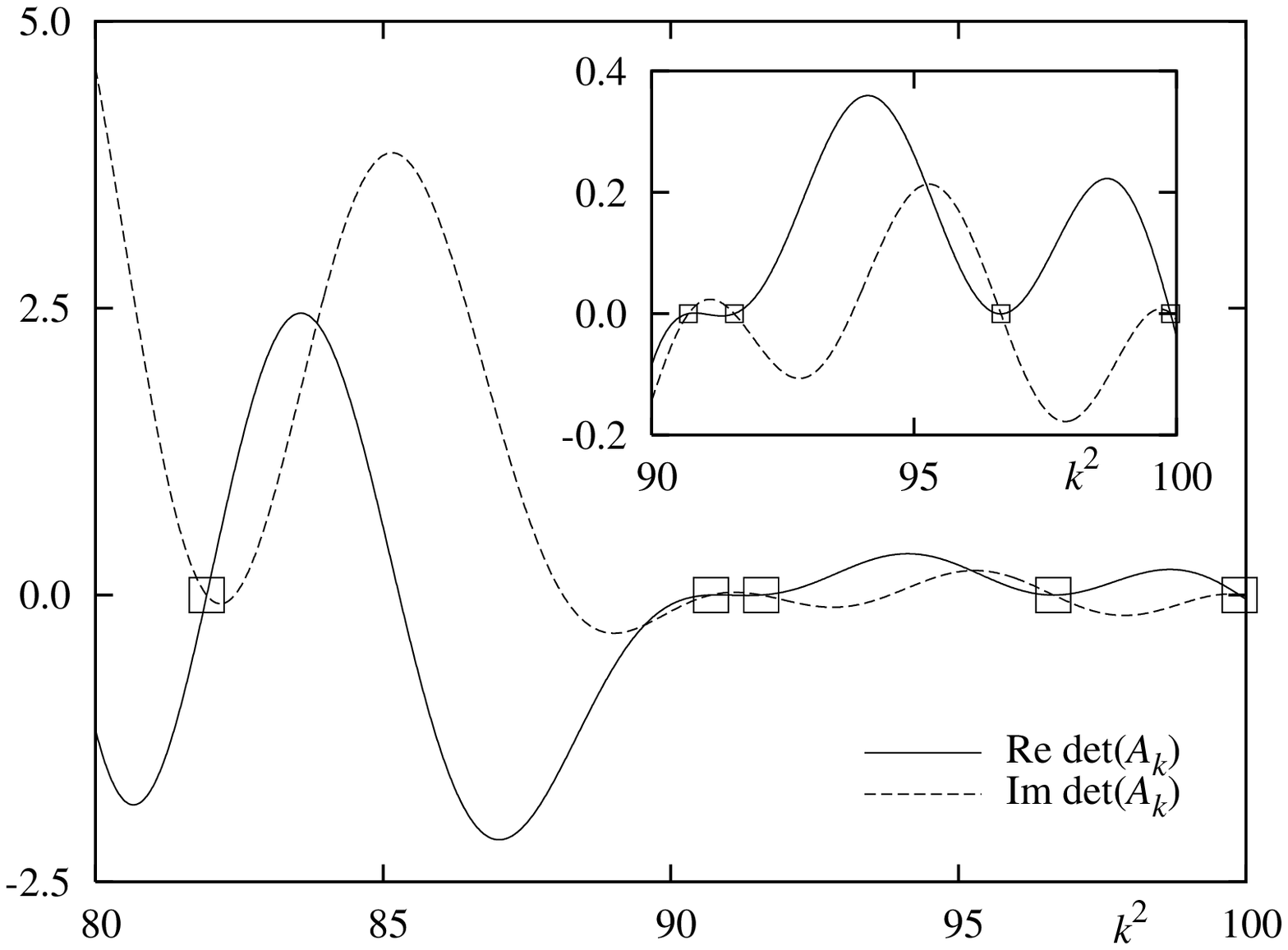}{13cm}
      \end{center}
     }
     {Plot of real and imaginary part of $\det (A_k)$
      as a function of $k$; the evaluation was done
      for 10 times as many points in $k^2$
      than for fig.~\ref{fig:svd-magnification}.
       Approximately simultaneous zeros
      correspond to minima of  $|\det (A_k)|$.      
      The locations of the eigenvalues are marked by squares.
     }
     {fig:real-imag-det-plot}

\afterpage{\clearpage}

In the numerical computations the integral over the boundary
is replaced by a Riemann sum. (There also exist more 
refined methods using polynomial approximations
combined with Gau{\ss}-Legendre integration, see e.g.~\cite{AurSte93}, 
which allow for a less fine discretization.)
Let $\Delta s = \CL / N $ be the discretization length 
of the boundary of length $\CL$ into $N$ pieces.
Then we have
\begin{equation} \label{eq:discrete-integral-eq}
   u(s_i) = \Delta s \sum_{j=0}^{N-1} Q_k(s_i,s_j) \,u(s_j) \;\;,
\end{equation}                                                           
where $s_i = (i+1/2) \Delta s$, $i=0,\ldots,N-1$.
Equation \eqref{eq:discrete-integral-eq}
can be written in matrix form as
\begin{equation}
  A_k \bfu = 0 \;, \qquad \text{ with }\quad 
     A_{ij} = \delta_{ij} - \Delta s \, Q_k(s_i,s_j)    
  \;\;.
\end{equation}   
Recall that for $s_i=s_j$ the kernel $Q_k(s_i,s_j)$ reduces
to the result given in eq.~\eqref{eq:kernal-at-s-s}.
The solutions of this linear equation provide approximations
to the eigenvalues $k_n^2$ and eigenvectors $u_n$. This 
leads to the problem of finding the real zeroes
of the determinant 
\begin{equation} 
  \label{eq:determinant-condition}
  \det(A_k)=0
\end{equation}
as a function of $k=\sqrt{E}$,
where $A_k$ is a dense, complex non-Hermitean matrix.
Due to the discretization of the integral the determinant $\det(A_k)$
will not become zero but only close to zero
(actually, the discretization shifts the zeros
slightly away from the real axis, see \cite{Boa92,Boa94}).

In the numerical computations it is very useful \cite{AurSte93}
to compute the singular values of the matrix $A$ instead of its determinant.
The singular value decomposition of a complex matrix is given
by the product of an unitary matrix $U$,
a diagonal matrix $S$ and a second unitary matrix $V$
\begin{equation}
   A = U S V^\dagger \;\;.
\end{equation}  
The diagonal matrix $S$ contains as entries SV$_i$
the singular values of $A$
and we have $|\det A| = |\prod  \text{SV}_i|$.
Since the original integral equation has been discretized,
the smallest singular value in general never gets zero,
but just very small, see fig.~\ref{fig:singular-value-example}.
Thus the minima of the smallest singular value provide approximations
to the eigenvalues of the integral equation.
For the numerical computation of the singular value decomposition
one may for example use the NAG routine {\tt F02XEF}
or the LAPACK routines {\tt ZGESVD} or {\tt ZGESDD}.
It turns out that the (more recent) routine {\tt ZGESDD}
is significantly faster 
(factor 3-5, at the expense of a higher memory consumption), 
in particular when also singular vectors are computed.

The advantage of the singular value decomposition
in comparison to locating the zeros of the determinant is that
degeneracies of eigenvalues can be detected
by looking at the second singular value, which also gets 
small when there are two nearby eigenvalues
(similarly higher degeneracies can be found by looking
at the next singular values).
In  fig.~\ref{fig:singular-value-example}a) an example of the behaviour 
of the three smallest singular values is shown in the case 
of the  stadium billiard ($a=1.8$) with Dirichlet boundary conditions.
\BILD{b}
     {
     \begin{center}
       \PSImagx{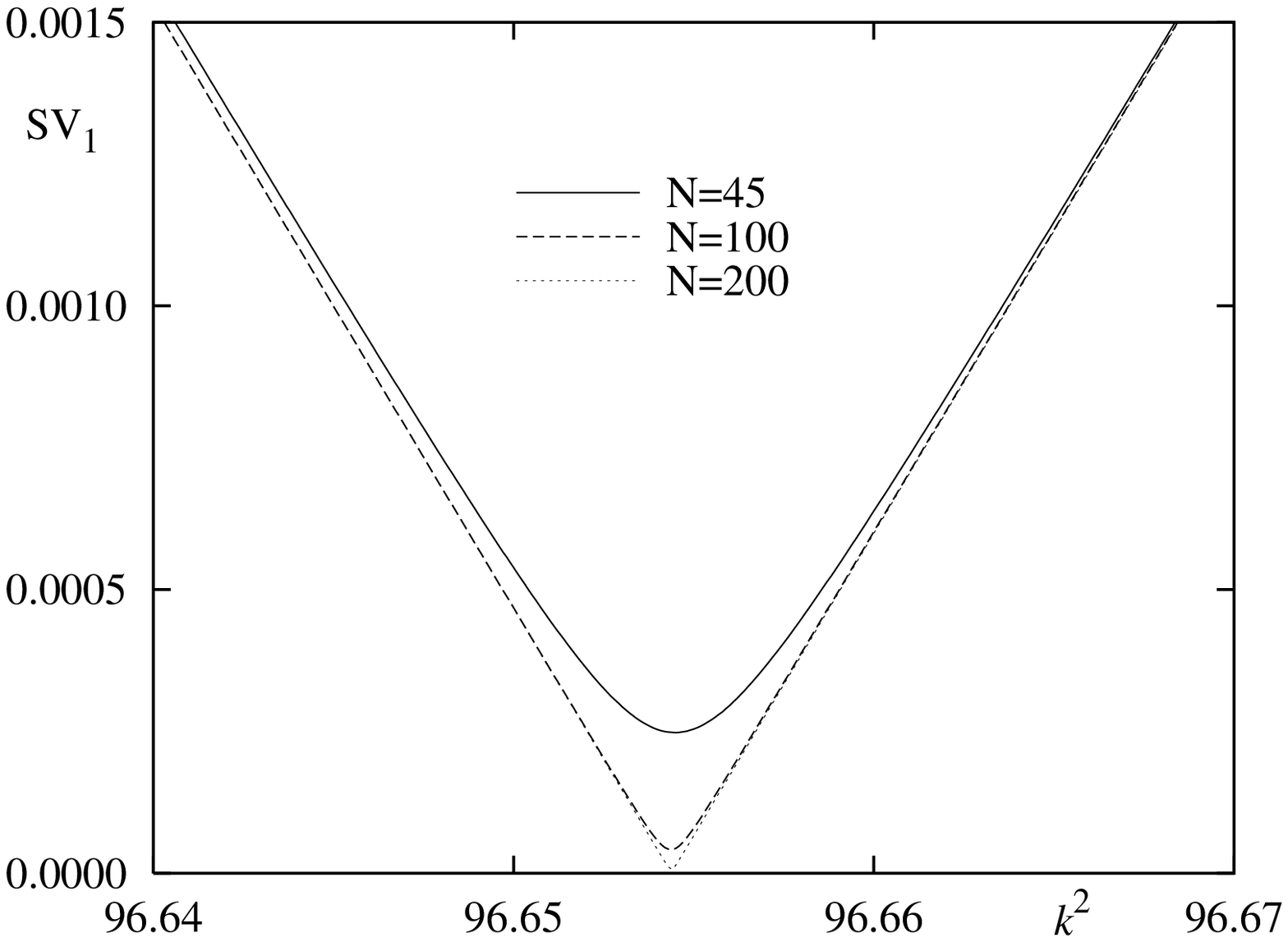}{12cm}
     \end{center}
     }
     {Magnification of fig.~\ref{fig:svd-magnification}
      around the minimum with $k^2\approx 96.5$ for different
      matrix sizes $N$. One nicely sees the pronounced
      parabolic structure for $N=45$    which gets
      smaller for larger $N$.
     }
     {fig:broad-minimum}
For comparison a plot of $|\det(A_k)|$ is shown in 
fig.~\ref{fig:singular-value-example}b).
One clearly sees that the singular value decomposition
provides more information.
For example, next to $k^2=90$
the minimum of $|\det (A_k)|$ looks slightly broader than the
others, however, this does not give a clear indication that
there might be more than one eigenvalue.
In contrast, the singular value decomposition method 
allows to resolve such kind of near-degeneracies
efficiently, see  fig.~\ref{fig:svd-magnification}.
Of course, this information is also available
via $\det (A_k)$, see fig.~\ref{fig:real-imag-det-plot}
where its real and imaginary part are plotted separately.
Here (approximately) simultaneous zeros correspond
to minima of $|\det (A_k)|$. However, notice that
compared to the singular value decomposition approach
much more discretization points in $E=k^2$ are necessary.

To determine all  energy levels in a given energy
interval $[E_1,E_2]$ one proceeds in the following way:
first one computes the singular values at equidistantly
chosen points $k^2\in[E_1,E_2]$; the energy is chosen as variable
because for two-dimensional billiards the mean distance
between two energy levels is approximately constant and 
according to the generalized Weyl formula 
\eqref{Weyl}
given by $\frac{4\pi}{\cA}$.
The finer the step size is chosen the easier the minima can be resolved,
however, at the same time the computing time to cover
a given energy range increases correspondingly.
The actual step size is a compromise between these two aspects;
good results have been achieved by using a step size of the
order of $\frac{1}{5}\,\frac{4\pi}{\cA}$
(for systems with many near level degeneracies,
e.g.\ integrable or near--integrable systems, a smaller step
size can be helpful).

The matrix size $N$ is chosen according to
$N= b \frac{\CL}{\lambda}=b\frac{\CL k}{2\pi}$, such that one 
obtains $b$ discretization points per units of the inverse of 
the de Broglie wave length $\lambda=\tfrac{2\pi}{k}$
along the boundary $\CL$. Typical choices for $b$
are between 5 and 12 depending on the system and
the wanted accuracy.

From the first scan one locates all minima of the smallest singular value.
If also the second singular value has a minimum next to a minimum of the
first one, one has to use a refined discretization in $E$
around the minimum
(the numerical implementation is a bit more sophisticated, in order
to account for several special situations, so that only a minimal
number of additional points need to be computed).
Once an isolated minimum is found, an approximation to
the eigenvalue can be computed by different methods.
Either one can perform a refined computation around the minimum,
which can be quite time-consuming,
or one can use a local approximation by a parabola \cite{Aur:privcomm}.
A linear interpolation also gives good results:
From the three points 1:$(k_1^2,SV_1(k_1^2))$, 2:$(k_2^2,SV_1(k_2^2))$, 
3:$(k_3^2,SV_1(k_3^2))$, characterizing a minimum of the first singular
value, one has two different lines  $\overline{12}$ and $\overline{23}$
with different slopes, of which the line with the larger slope has to be 
chosen. The intersection of this line with the zero axis gives
a good approximation to the eigenvalue,
which one can refine if necessary.
By repeatedly applying this for all minima, all energy levels in a given
interval can be found.
In fact, it is possible to develop a computer program which
takes care of all this such that all levels can be found
automatically.

A good check of the completeness is provided by considering
the $\delta_n$ statistics, see the example
in sec.~\ref{sec:quantum-billiards}.
The accuracy of the computed eigenvalues can be estimated 
from the bracket of the minimum given by the three points 1,2,3
if the matrix dimension $N$ is large enough.
For $N$ too small (for a given resolution in $E$) 
 one does not obtain a peaked, but a broad
minimum. This is illustrated in fig.~\ref{fig:broad-minimum}
by magnifying fig.~\ref{fig:svd-magnification}
around the minimum with $k^2\approx 96.5$ for different $N$.
One clearly sees the parabolic structure around the minimum for smaller $N$
and for larger $N$ one recovers the essentially linear
behaviour of the smallest singular value.

Tests of the accuracy of the method can be obtained
by considering a system where the eigenvalues are
known. For example for the circular billiard the eigenvalues can be computed
with arbitrary accuracy.
Also billiards where the eigenvalues can be computed
by other methods (e.g.\ conformal mapping method \cite{Rob84,ProRob93})
allow a determination of the accuracy of the method.
For a study of the scaling of the error for various billiards
see \cite{LiRobHu98}.
In addition  computations of the normal derivative function $u_n(s)$
and the eigenfunction (both inside and outside of $\Omega$)
allow to check the quality
of the numerical method and program.

\FloatBarrier
\subsubsection{Computing eigenfunctions}

\newcommand{\nwfandefct}[1]{
\begin{minipage}{7.6cm}
 \PSImagx{stad_nwf_#1.ps}{7.5cm}
\end{minipage}
\begin{minipage}{7.6cm}
  \PSImagx{stad_wavb_#1.ps}{7.5cm}
\end{minipage}
}

\BILD{t}
     {
     \vspace*{-0.25cm}
      \begin{center}
\nwfandefct{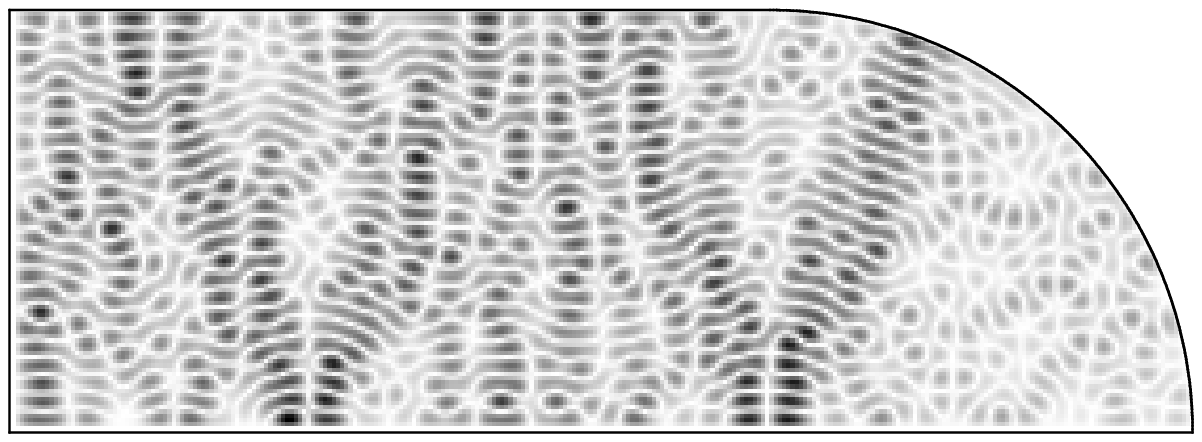}

\nwfandefct{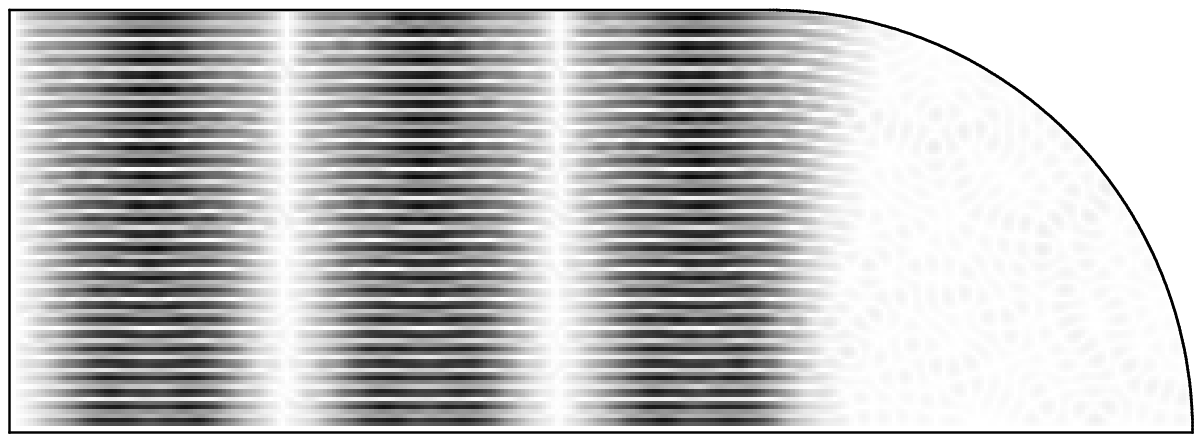}

\nwfandefct{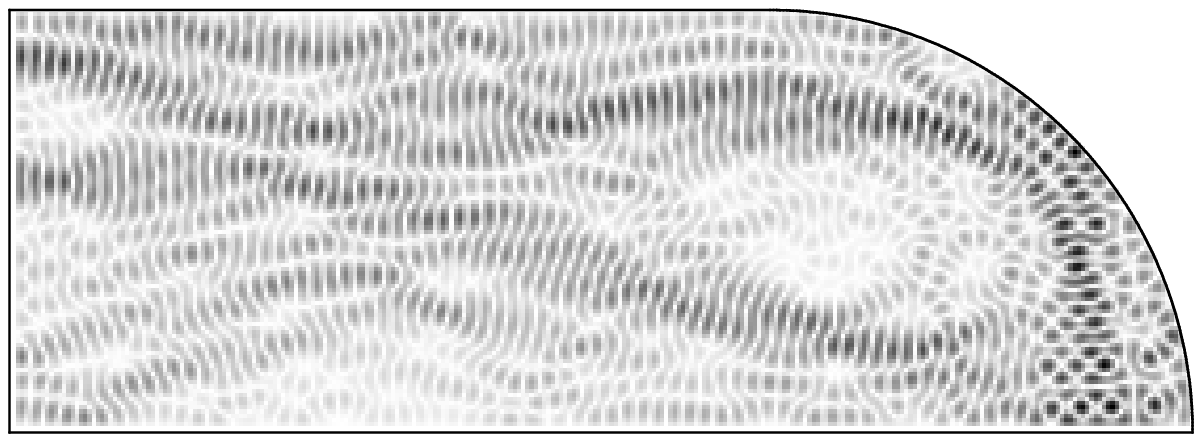}

\nwfandefct{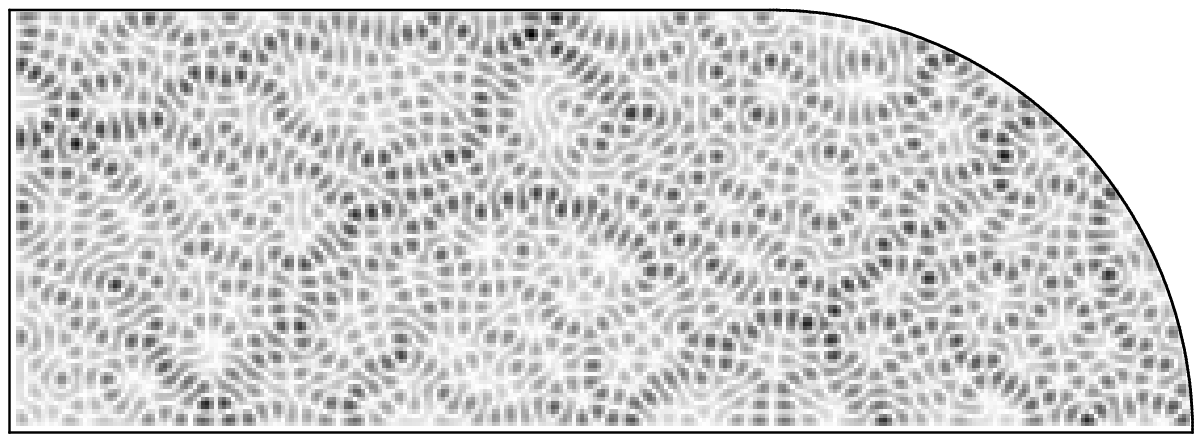}

      \end{center}
     \vspace*{-0.5cm}
     }
     {Examples of normal derivative functions $u_n(s)$ and
      the corresponding eigenfunctions in the stadium billiard
      (odd-odd symmetry, $a=1.8$). Here black corresponds to high intensity
      of $|\psi_n(\BFq)|^2$.
     }
     {fig:nwf-efct-stadium}

From a minimum of the smallest singular value we obtain
an approximation of the eigenvalue and at the same time
the corresponding singular vector $\bfu$
gives an approximation to the normal derivative function $u(s)$.
The NAG routine {\tt F02XEF} scales the singular
vector such that its first component is real.
Thus for a correct solution also the other
components should be essentially real, which provides
another check for the implementation of the method
and the accuracy of the eigenvalues.

The eigenfunction in the interior of the domain $\Omega$ can now
be calculated from the normal derivative function,
\begin{equation}
  \label{eq:wave-function-H0}
  \psi(\bfq) = - \frac{\ui}{4} \oint\limits_{\partial\Omega}  
            H_0^{(1)}\left(k \,|\bfq-\bfq(s)| \right) \, u(s) \; \ud s
            \;\;, \qquad \text{ for } 
             \bfq \in \Omega \backslash \partial \Omega \;\;.
\end{equation}                                                         
The computation of the eigenfunction can be simplified
by taking into account that
\begin{equation} \label{eq:omit-J_0}
        \oint\limits_{\partial\Omega}  
            J_0\left(k \,|\bfq-\bfq(s)| \right) \, u(s) \; \ud s = 0 \;\;,
\end{equation}
because the $J_0$--part of $G_k(\bfq,\bfq')$
is a solution of the homogeneous equation corresponding
to eq.~\eqref{eq:Inhomogene-Helmholtz-Gleichung}.
Thus \eqref{eq:wave-function-H0} is equivalent to 
\begin{equation}
  \label{eq:wave-function-Y0}
  \psi(\bfq) = \frac{1}{4} \oint\limits_{\partial\Omega}  
            Y_0\left(k \,|\bfq-\bfq(s)| \right) \, u(s) \; \ud s
            \;\;, \qquad \text{ for } 
             \bfq \in \Omega \backslash \partial \Omega \;\;.
\end{equation}
If one uses a desymmetrization, such as eq.~\eqref{eq:desymm-x}, 
\eqref{eq:desymm-xy-DD} or 
\eqref{eq:desymm-xy-NN}, the above formula
\eqref{eq:wave-function-Y0} has to be modified accordingly.

In fig.~\ref{fig:nwf-efct-stadium} we
show some examples of normal derivatives $u_n(s)$
and the corresponding eigenfunctions of the billiard, computed
via eq.~\eqref{eq:wave-function-Y0}.
The imaginary part of $u_n(s)$ is typically 5 or more orders
of magnitude smaller than the real part.
It is interesting to see that part of the structure
of the eigenfunctions is also reflected in $u_n(s)$. For example
for eigenstates with small probability in the region
of the quarter circle also the normal derivative is small
for $s< \pi/2$.

\subsubsection{Spurious solutions I: Real Green function approach} 
\label{sec:spurious}

\BILD{b}
     {
     \begin{center}
       \PSImagx{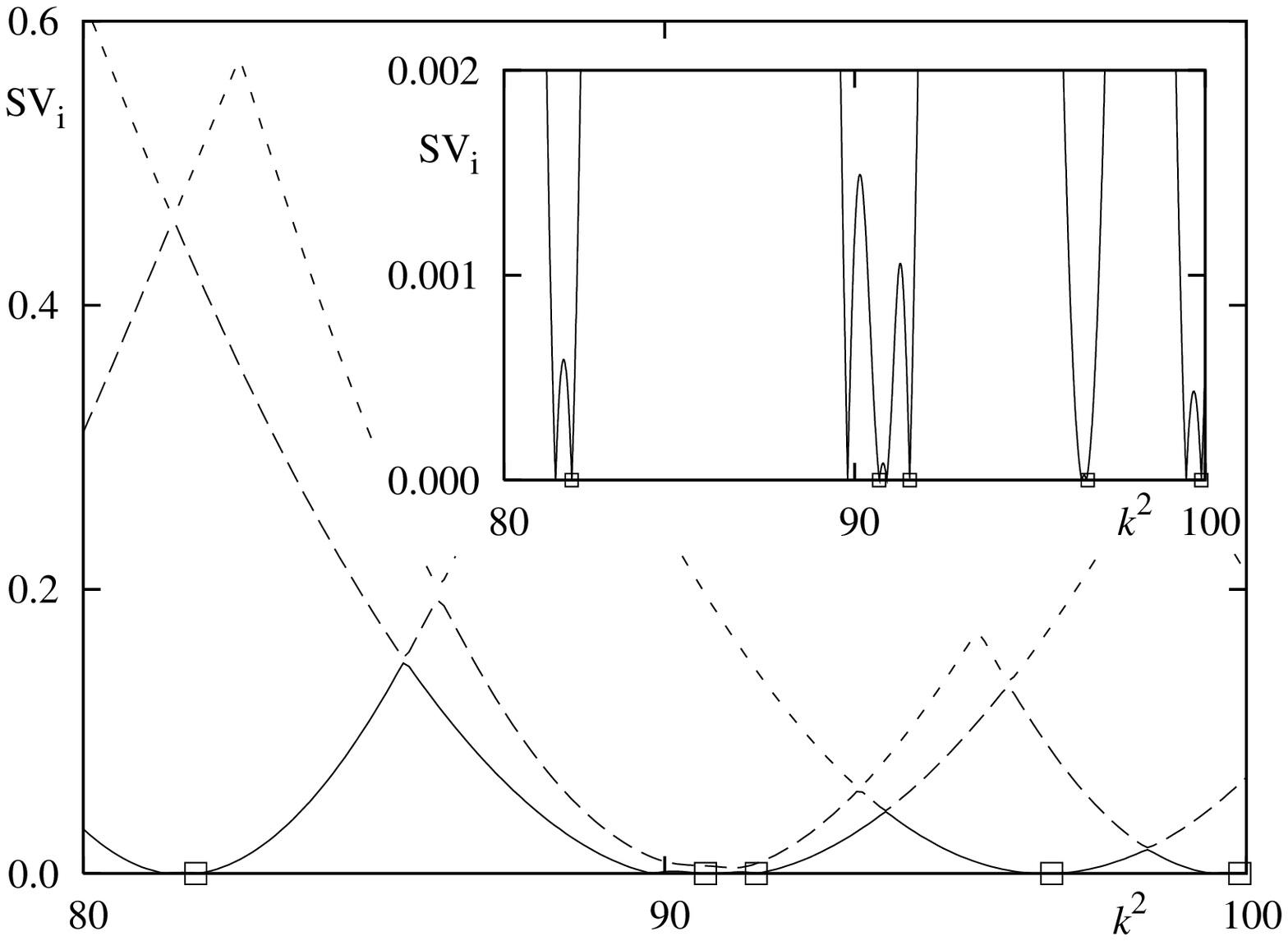}{13cm}
     \end{center}
     }
     {Using the real Green function \eqref{eq:just-real}
      leads to spurious solutions (see the inset)
      in addition to the correct eigenvalues marked by squares
      (compare with fig.~\ref{fig:svd-magnification}).
      For each true solution there is an additional spurious one
      (hardly visible at $k^2\approx 91$ and $k^2\approx 96$).
     }
     {fig:just-real-matrix}

\BILD{t}
     {
     \vspace*{-0.25cm}
      \begin{center}
         \PSImagx{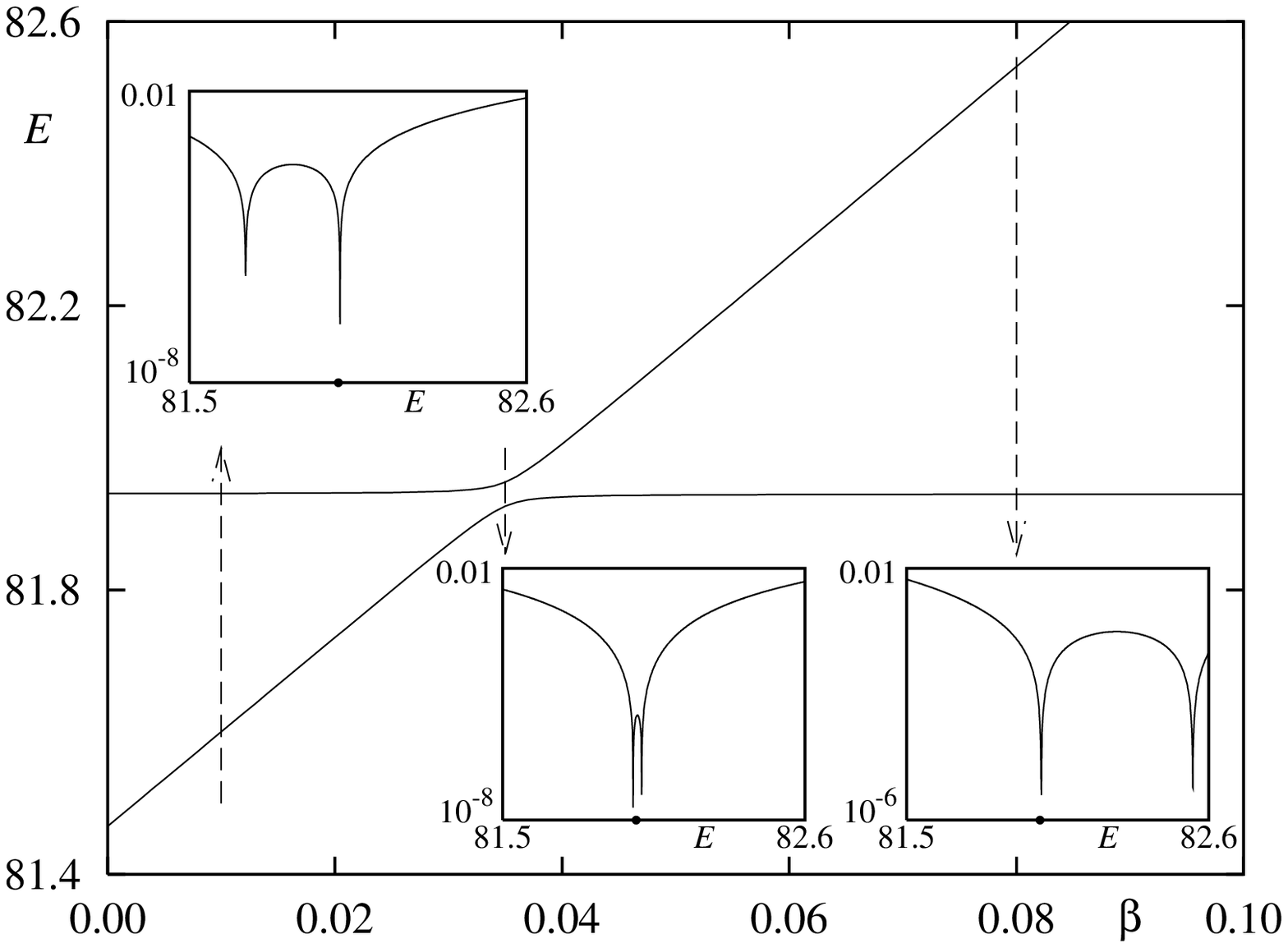}{14cm}
      \end{center}
     }
     {Plot of the minima of the singular values around   
      the eigenvalue $k^2=81.93\dots$
     with varying $\beta$ using the parametrized Green function
     \eqref{eq:Green-fct-two-dim-beta}.
     The insets show the corresponding structure of the first 
     singular value with a logarithmic vertical scale 
     (matrix size for this computation: $N=200$).
     }
     {fig:beta-var}

In certain situations and for some numerical methods it
may happen that one obtains in addition to the
true solutions of the Helmholtz equation \eqref{eq:Helmholtz} 
further so--called {\it spurious solutions}.
This question 
is discussed in some of the papers on the boundary integral method, 
in particular see \cite{BurMil71,Rid79,Rid79b}
and \cite{BerWil84,Boa92}.
There are essentially two different situations in which they are encountered.
The first is that one uses for the Green function
instead of the Hankel function, see eq.~\eqref{eq:Green-fct-two-dim},
just the real part, i.e.\
\begin{equation} \label{eq:just-real}
   G_k(\bfq,\bfq') =  \frac{1}{4} Y_0  \left(k \,|\bfq-\bfq'|\right) 
   \;\;.
\end{equation}
This seems reasonable as according to \eqref{eq:omit-J_0}
the $J_0$--Bessel function does not contribute to the eigenfunction.
Moreover then one can work with an entirely real matrix
for which the singular value decomposition can be computed
much faster.
However, when using this approach, there appear
additional zeros (for each correct one there is one additional one)
and the singular values loose their nice
linear structure, see fig.~\ref{fig:just-real-matrix}.
To overcome the problem of these additional zeros
a parametrized Green function
\begin{equation} \label{eq:Green-fct-two-dim-beta}
  G_k^{(\beta)}(\bfq,\bfq') =  
 \frac{1}{4} \;\left[ \beta   J_0 \left(k \,|\bfq-\bfq'|\right)
                                     + Y_0  \left(k \,|\bfq-\bfq'|\right)
                               \right] \;\;
\end{equation}                                                           
is used in \cite{Boa92}.
Thus for $\beta=0$ we obtain eq.~\eqref{eq:just-real}
and for $\beta=-\ui$ we get
eq.~\eqref{eq:Green-fct-two-dim}.
So using a purely real Green function
means to vary $\beta\in\R$ which  
changes the location of the spurious solutions
but not those of the true ones. This is illustrated 
in fig.~\ref{fig:beta-var} around the eigenvalue $k^2=81.93\dots$
with $\beta\in[0,0.1]$. Clearly on this scale the true solution
does not change under variation of $\beta$
(apart from the region of the avoided crossing 
which is due to the finite matrix size and gets smaller for larger $N$)
whereas the spurious solution strongly varies with $\beta$.
For $\beta=-\gamma \ui$ with increasing real $\gamma$
the additional zeros move away from the
real axis and it seems that for $\beta=-\ui$
they do not have any significant influence on the real axis.
Still there could be cases where also for $\beta=-\ui$
such a solution becomes relevant, but for convex geometries
we have not encountered
this situation. For an example of a non--convex geometry see 
section \ref{sec:non-convex}.

As an explicit example for the influence of parameterized
Green function \eqref{eq:Green-fct-two-dim-beta}
let us consider the circular billiard with radius 1,
where the Fredholm determinant reads (see e.g.\ \cite{Boa92,Bur95})
\begin{equation} \label{eq:Dk-circle}
  D(k) =  \prod_{l=-\infty}^{\infty} 
        \left[-\ui\pi k H_l^{(1)'}(k) J_l(k) \right] \;\;.
\end{equation}
As this product converges absolutely in the whole
complex $k$--plane (apart from a cut along the negative real axis)
zeros of $D(k)$ occur when one of the factors in the product
vanishes \cite{Bur95}. Clearly, the real zeros of $D(k)$ correspond to the
eigenvalues $j_{ml}$ of the circular billiard with radius $1$
and Dirichlet boundary conditions.
The further zeros stem
from the functions $H_l^{(1)'}(z)$ which 
have only zeros with $\Imag \, z <0$ \cite{AbrSte84},
and do not correspond to physical solutions of the interior problem.
However, they can be related to resonances of the exterior scattering
problem, but with Neumann boundary conditions \cite{BurSte95:p,TasHarShu97}.
Because of the radial symmetry
the $S$--matrix is diagonal in angular momentum space
\begin{equation}
  S_{l'l} =  -\frac{H_l^{(2)'}(k)}{H_l^{(1)'}(k)} \delta_{l'l}
\end{equation}
and therefore the resonances are at those complex $k$ for which
\begin{equation}
   H_l^{(1)'}(k)=0 \;\;,
\end{equation}
i.e.\ the same condition as implied by \eqref{eq:Dk-circle}.

If one uses the parametrized Green function   
\eqref{eq:Green-fct-two-dim-beta}
one can show (analogous to the derivation of eq.~\eqref{eq:Dk-circle})
that for the circular billiard
\begin{equation} \label{eq:Dk-circle-beta}
  D^{(\beta)}(k) =  \prod_{l=-\infty}^{\infty} 
        \left[\pi k  \left(\beta J_l'(k)+ Y_l'(k)\right) 
                   J_l(k) \right] \;\;.
\end{equation}
For $\beta=0$, which corresponds to the 
real Green function \eqref{eq:just-real},
we get additional zeros 
of $D^{(0)}(k)$ when
$Y_l'(k)=0$.
Varying $\beta$ from zero to $-\ui$ these real zeros turn complex.
At first sight one might think that these are connected to the
places with $H_l^{(1)'}(k)=0$, however
numerical computations show that (for all studied cases)
these move away from the real axis with a positive imaginary
part and for $\beta=-\ui$ one has $H_l^{(1)'}(k)=0$ only for
$\Imag\, k<0$.
Thus the spurious solutions for the real Green function
are not related to resonances of the 
scattering problem with Neumann boundary conditions. 

These examples suggest to use the full
complex Green function \eqref{eq:Green-fct-two-dim}
instead of the real variant \eqref{eq:just-real}.
Even though the numerical computation is more time--consuming for the complex
case their advantages over choosing \eqref{eq:just-real} are obvious
as the variation of $\beta$ is time--consuming as well
(and non--trivial to implement in an automatic way).

\FloatBarrier
\subsubsection{Spurious solutions II: Non--convex geometries} 
\label{sec:non-convex}

\BILD{t}
     {
\vspace*{0.5cm}
     \begin{center}
       \PSImagx{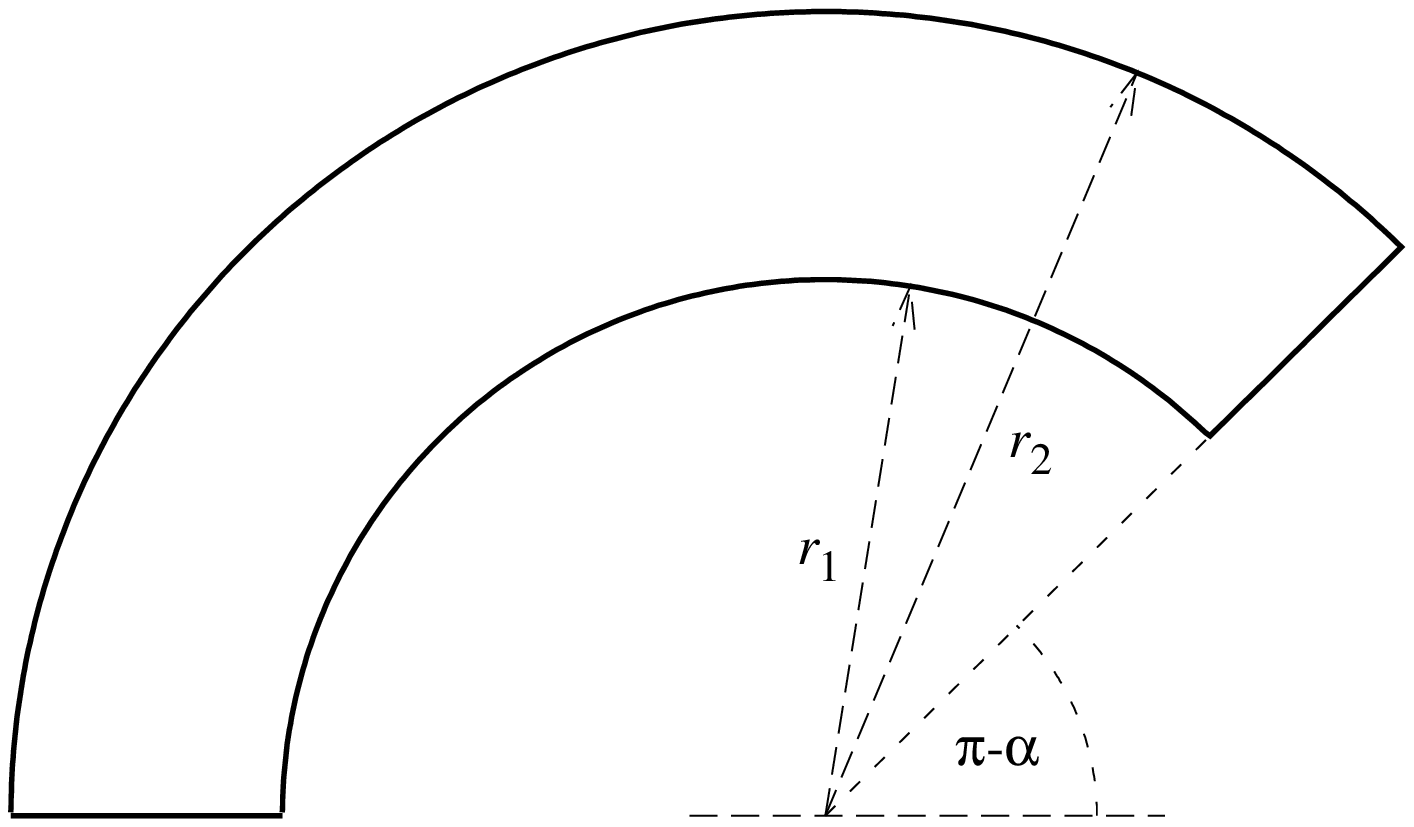}{11cm}
     \end{center}
     }
     {Boundary of the annular sector billiard for 
      $\alpha=\tfrac{7}{8} \pi$
      and $r_1=0.4$ and $r_2=0.6$.
     }
     {fig:kreisringsektor}

Even when choosing the complex Green function \eqref{eq:Green-fct-two-dim}
it is possible to encounter spurious solutions:
For the circular the additional complex 
zeros of $D(k)$ are sufficiently far
away from the real axis, i.e.\  $\Imag \, k \ll 0$ 
so that they do not
lead to problems with the application of the boundary integral
method. However, when one considers different geometries
the resonances of the corresponding scattering system
could be closer to the real axis. 
This can be nicely studied for the annular sector
billiard, see fig.~\ref{fig:kreisringsektor}, as the
eigenvalues and eigenfunctions can be determined numerically
with arbitrary accuracy.
Using the ansatz \cite[\S 25]{Som1984} 
\begin{equation}
 \psi(r,\phi) = \left[ J_\nu(kr) + c Y_\nu(kr) \right] \sin(\nu \phi)
\end{equation}
with $\nu=m\tfrac{\pi}{\alpha}$, $m=0,1,2,...$
and requiring $\psi(r_1,\phi)=0$ and $\psi(r_2,\phi)=0$ 
gives the (implicit) eigenvalue equation
\begin{equation} \label{eq:exact-eval-annular-sector}
  J_\nu(k r_1) Y_\nu(k r_2) - Y_\nu(k r_1) J_\nu(k r_2) = 0\;\;.
\end{equation}
For each $m$ one gets a sequence of zeros $k_{mn}=\sqrt{E_{mn}}$.

Fig.~\ref{fig:kreisringsektor-svd1} shows for the annular
sector billiard with $\alpha=\tfrac{49}{50}\pi$
the first
three singular values as a function of $k^2$.
The solutions of \eqref{eq:exact-eval-annular-sector}
are marked by triangles. Clearly, there are additional
minima, which can be associated with resonances
of the dual scattering problem
(for further details and examples of this association for the
annular sector billiard see \cite{Hes97:PhD}).
In the limit of $\alpha\to \pi$
these resonances  are given by the eigenvalues of 
the circular billiard of radius $r_1$ with Neumann boundary conditions.
For this billiard 
the ansatz $\psi(r,\phi)= J_m(k r)$ together
with $\left.\tfrac{\partial \psi(r,\phi)}{\partial r}\right|_{r=r_1}=0$
gives the eigenvalue equation
\begin{equation}\label{eq:exact-eval-Neumann-circle}
  m J_m(k r_1) - k r_1 J_{m+1}(k r_1)  = 0  \;\;.
\end{equation}
The circles shown in 
fig.~\ref{fig:kreisringsektor-svd1} correspond to the 
solutions of \eqref{eq:exact-eval-Neumann-circle}
and provide a very good description of the additional
minima.

Thus the question arises how to detect and distinguish
these additional solutions. First, of course
their existence and relevance strongly depends on
the system one is studying. In many situations (for example convex geometries)
there appear to be no complex solutions
coming close enough to the real axis.
Intuitively this seems reasonable as long as there 
are no trapped orbits outside of the billiard
as these should give rise to resonances
with small imaginary part.

However, if such additional solutions exist they will show
up in the $\delta_n$ statistics by an offset of $+1$ at
each additional eigenvalue (unless one by chance misses
the same number of `correct' eigenvalues).
If one has a system with such additional solutions
one approach is to plot the corresponding normal derivative
function $u(s)$ and the eigenfunction.
Usually they will behave quite differently for
a correct eigenvalue and for a spurious solution.
For example for the case of the annular sector billiard
the normal derivative function for a spurious solution
is discontinuous along
the boundary and the corresponding eigenfunction
also has contributions outside of the billiard,
see fig.~\ref{fig:nwf-kreisringsektor}.
Another test would be to use the normalization condition
\eqref{eq:normalization-of-psi} for the normal derivative and
compute the norm of the eigenfunction
in the interior of the billiard.
These two are the same
for proper eigenfunctions whereas
for spurious solutions they will disagree.
Unfortunately, this is a highly inefficient 
method as the computation of the eigenfunction
in $\Omega$ is quite time--consuming.
Instead of computing the normalization for the full
billiard one could restrict to smaller subregions,
e.g.\ for the annular sector billiard
one could integrate over the region of the circle with radius $r_1$
and check if it is different from zero indicating
a spurious solution.
For the annular sector billiard the additional
zeros of the Fredholm determinant $D(k)$
are complex as long as $\alpha<\pi$. Thus for $N\to\infty$
these minima will stay bounded away from
zero in contrast to the minima corresponding to the eigenvalues. 
However, in practice it is not possible to check
this as one has to make $N$ too large to distinguish
these from the correct eigenvalues.

\BILDD{t}
     {
\vspace*{-0.5cm}
     \begin{center}
       \PSImagx{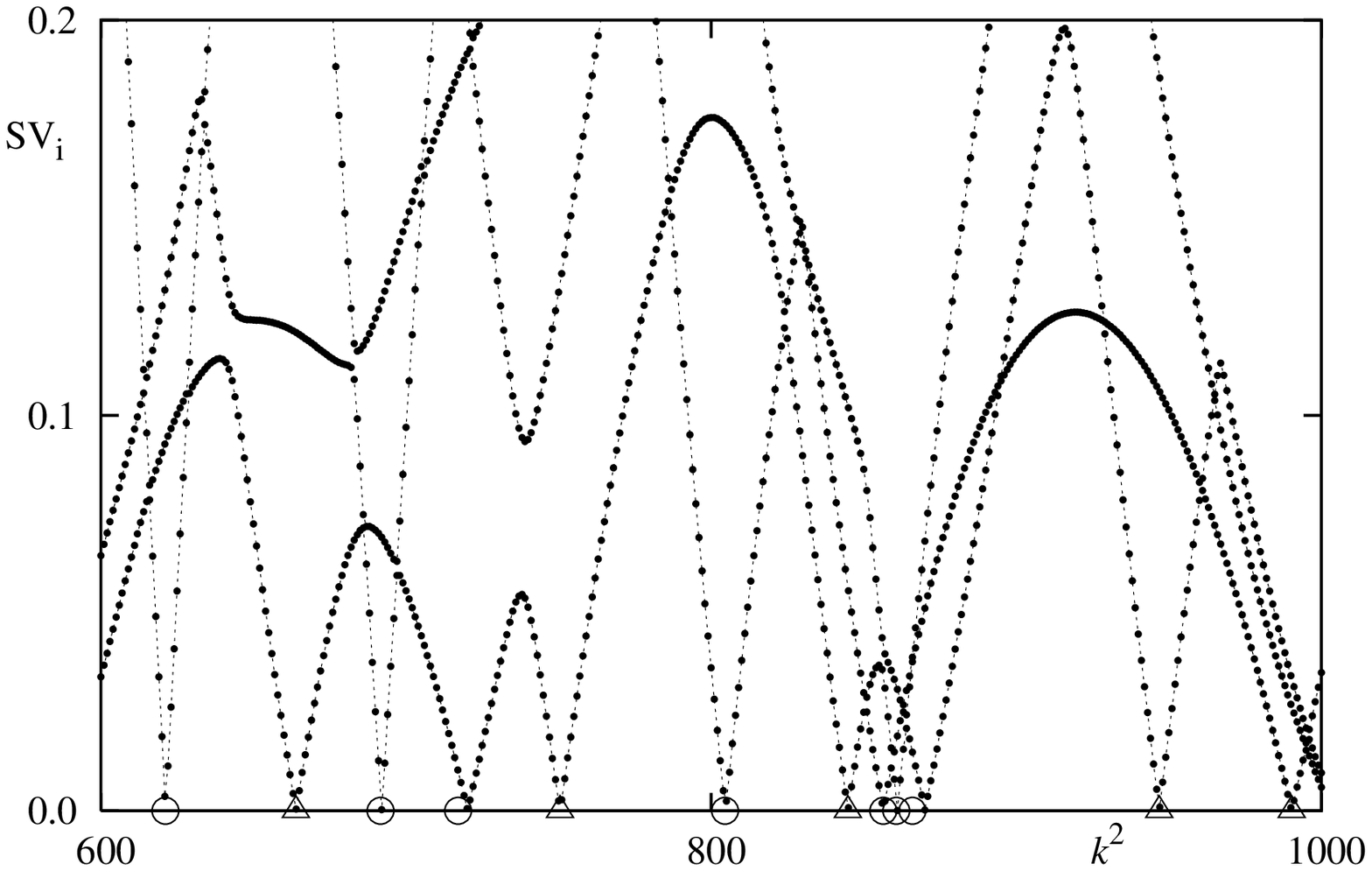}{15.25cm}
\vspace*{-0.75cm}
     \end{center}
     }
     {First three singular values as a function of $E=k^2$
      of the annular sector billiard for 
      $\alpha=\tfrac{49}{50} \pi$ and $r_1=0.4$ and $r_2=0.6$.
       The triangles correspond to the exact eigenvalues
       for the annular sector billiard, computed from 
       eq.~\eqref{eq:exact-eval-annular-sector} and 
       the circles correspond to the eigenvalues
       of the circular billiard with radius $r_1$ and Neumann boundary
       condition, determined via eq.~\eqref{eq:exact-eval-Neumann-circle}.
     }
     {fig:kreisringsektor-svd1}

\BILDD{b}
     {
\vspace*{-1cm}
     \begin{center}
       \PSImagx{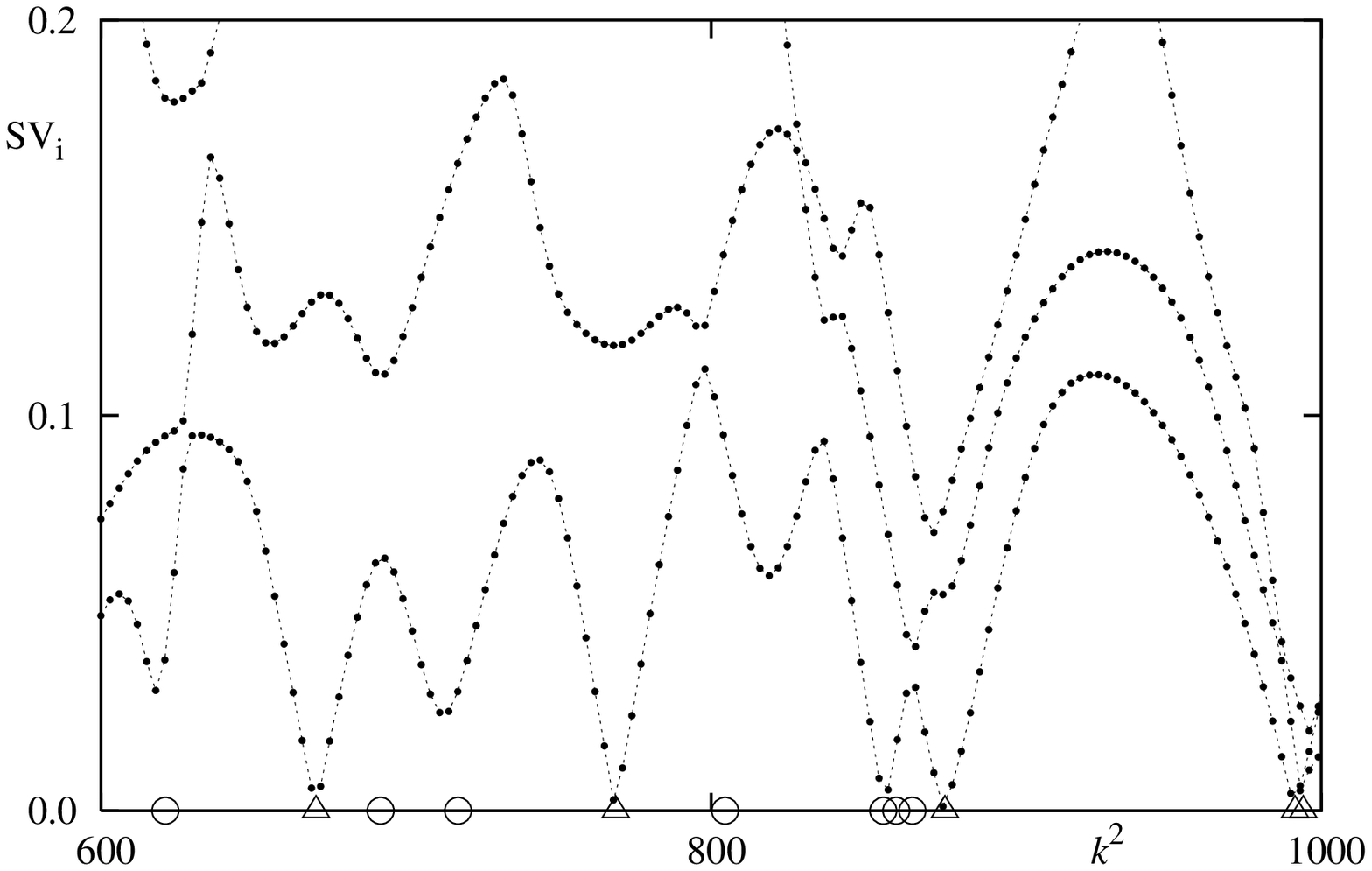}{15.25cm}
     \end{center}
\vspace*{-0.75cm}
     }
     {First three singular values as a function of $E=k^2$
      of the annular sector billiard for 
      $\alpha=\tfrac{7}{8} \pi$ and $r_1=0.4$ and $r_2=0.6$.
       The triangles correspond to the exact eigenvalues
       for the annular sector billiard, computed 
       from eq.~\eqref{eq:exact-eval-annular-sector} and the circles 
       correspond to the eigenvalues
       of the circular billiard with radius $r_1$ and Neumann boundary
       condition, determined via eq.~\eqref{eq:exact-eval-Neumann-circle}
     }
     {fig:kreisringsektor-svd2}

\BILD{tbh}
     {

\vspace*{0.5cm}

     \begin{center}
       \PSImagx{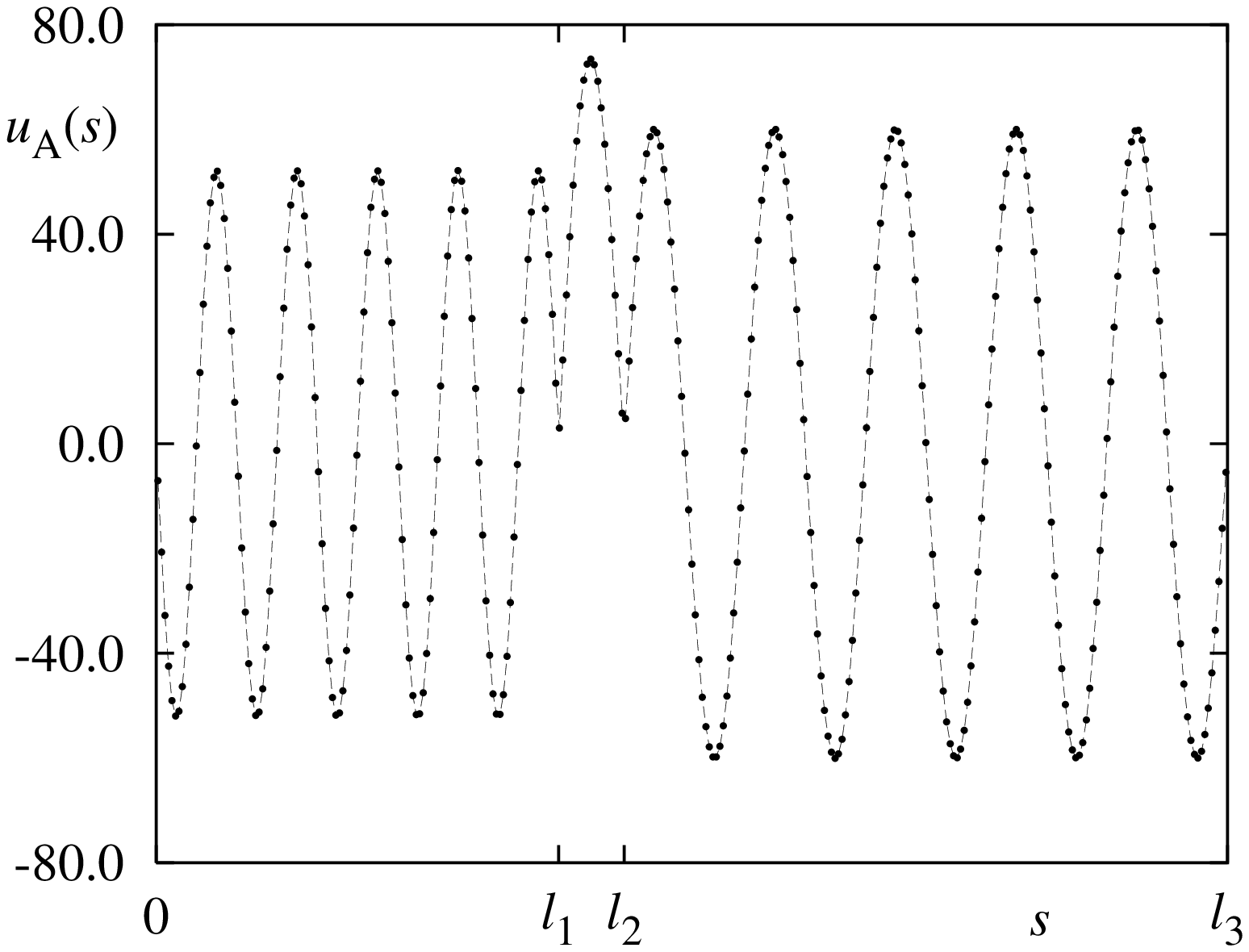}{8cm}
\hspace*{0.5cm}
       \PSImagx{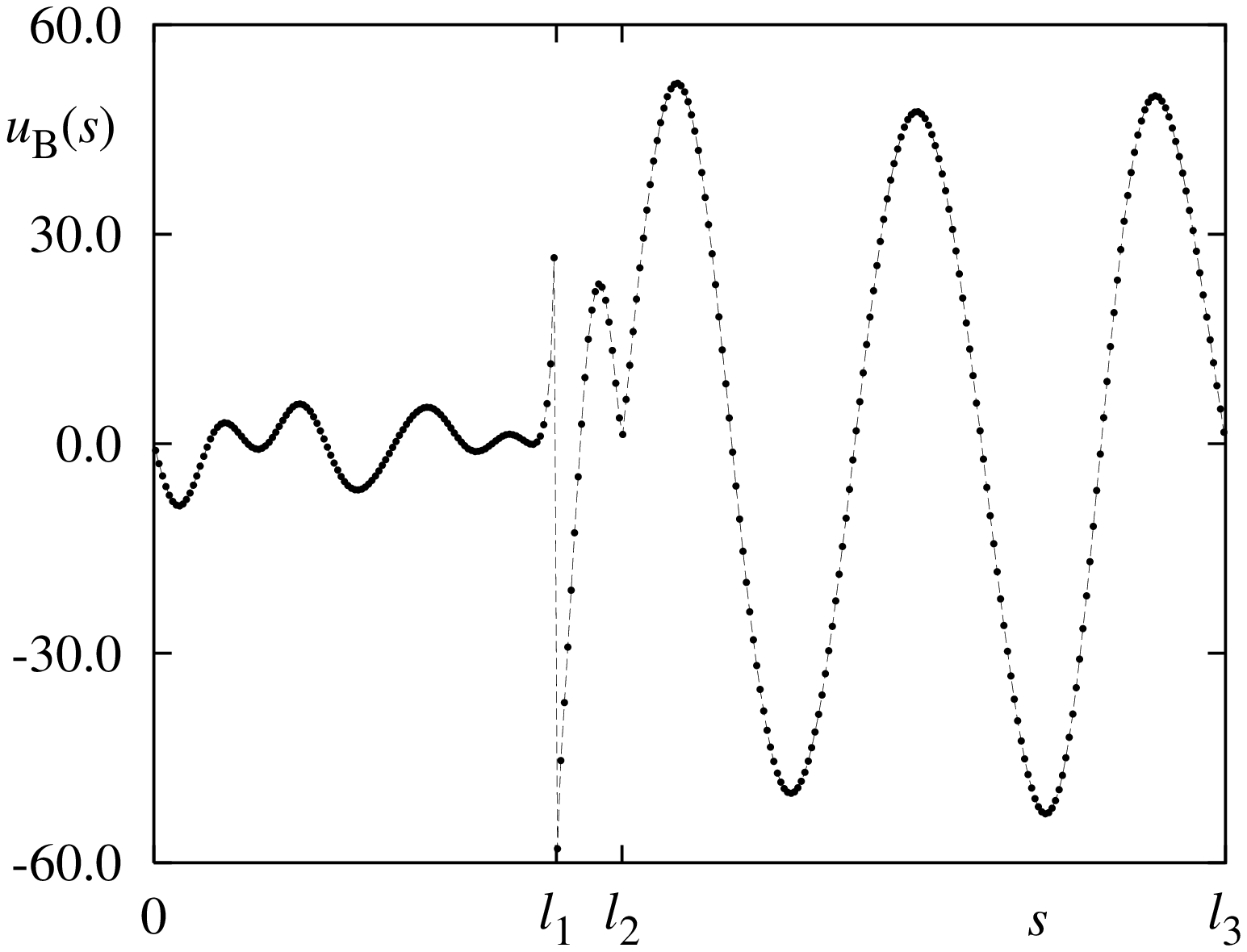}{8cm}

\vspace*{0.75cm}

\PSImagx{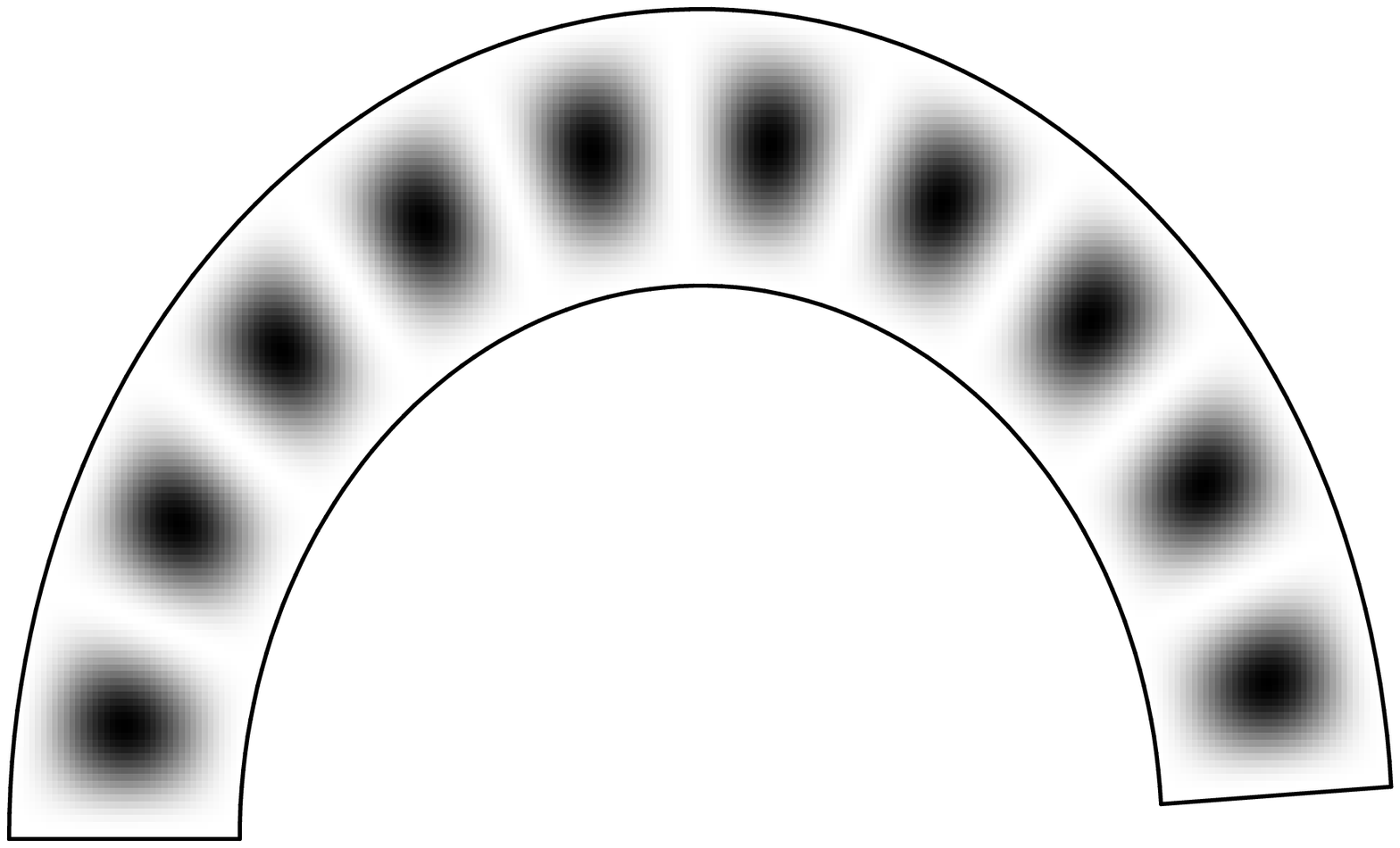}{8cm}
\hspace*{0.5cm}
\PSImagx{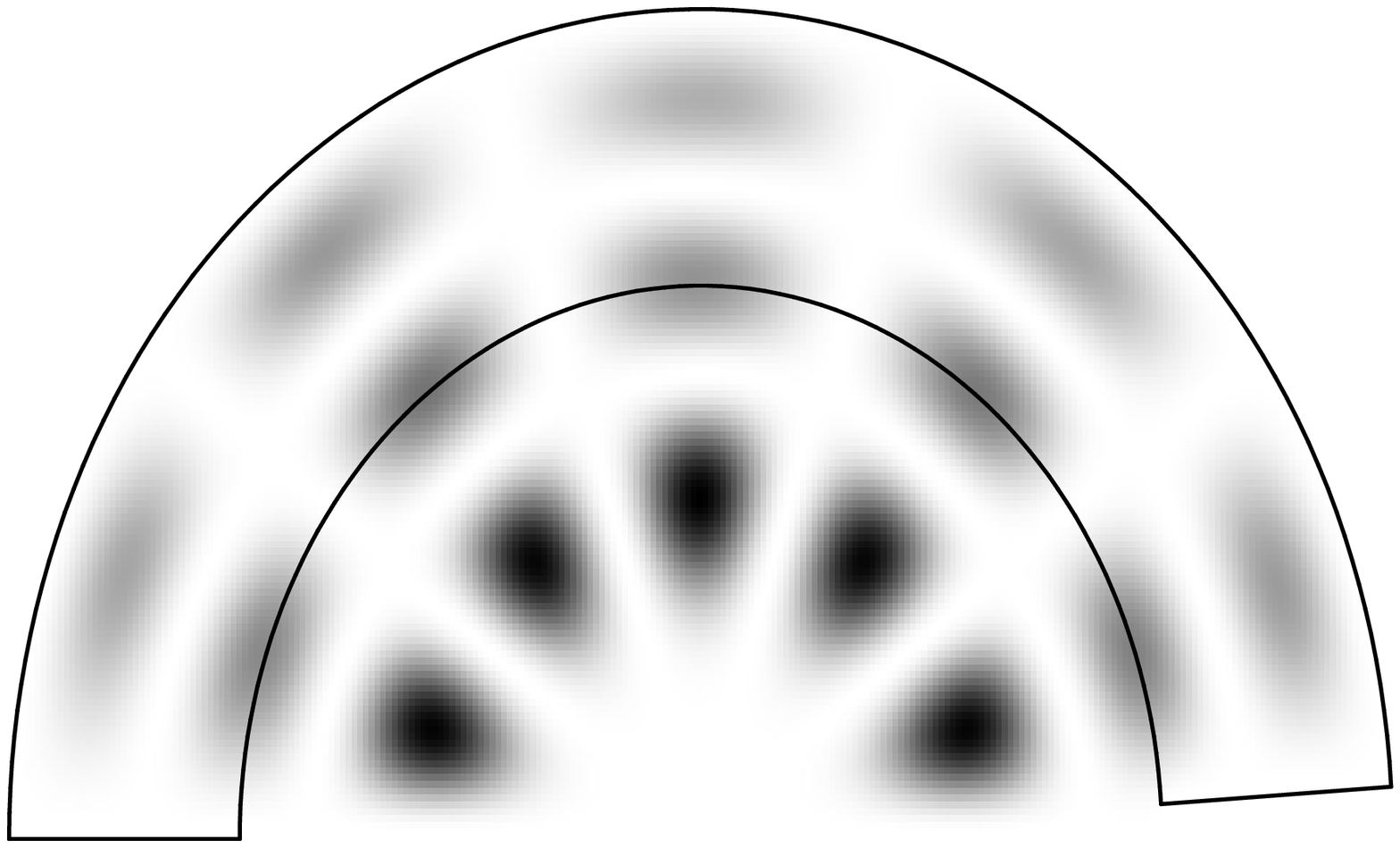}{8cm}

\vspace*{0.25cm}

     \end{center}
     }
     {Normal derivative functions $u_n(s)$ 
      corresponding to the 
      correct eigenvalue with $E=663.88\dots$ (left)
      and the spurious one with $E= 691.77\dots$ (right).
      Here $l_1=r_1 \alpha$, $l_2=r_1 \alpha + r_2-r_1$
      and $l_3=r_1 \alpha + r_2-r_1 + \alpha r_2$.
      One clearly sees the discontinuity in $u_n(s)$ 
      for the spurious solution.
      This is also reflected in the structure of the eigenfunction
      which for the spurious solution has its main contribution
      outside of the billiard.
      Notice that in both cases the eigenfunction
      has been computed according to \eqref{eq:omit-J_0}
      inside and outside of $\Omega$.
      The fact that for the correct eigenfunction
       $\psi_n(\BFq)=0$ (within the numerical accuracy)
       for $\BFq \in \R^2\backslash\Omega$ 
      is another test of the accuracy of the
      eigenvalue computations and eigenfunctions.
           }
     {fig:nwf-kreisringsektor}

More generally, spurious solutions can be understood by  a second
look at the boundary integral equations.
Namely, for the interior Dirichlet problem we have the 
{\it single layer} equation, eq.~\eqref{eq:int-darstellung},
and the {\it double layer} equation, eq.~\eqref{eq:Fredholm}.
On the other hand, the single-layer equation for the outside
scattering problem with Neumann boundary conditions at
$\partial\Omega$ is also given by 
the double layer equation \eqref{eq:Fredholm}.
(see e.g.~\cite{KleRoa74,BurMil71}).
As a consequence, scattering solutions
of the outside scattering problem with Neumann boundary
conditions at $\partial \Omega$ may become relevant
for real $k$. Namely, for resonances 
with small imaginary part they can 
lead to additional solutions for the double layer equation
which are numerically indistinguishable
from the correct solutions.
However, these solutions do not correspond
to solutions of the interior problem and they do not
fulfill the single layer equation.
So a possibility to distinguish spurious solutions
for the interior Dirichlet problem is to check
the validity of the single layer equation as well, which is only
fulfilled simultaneously for correct solutions
of the interior Dirichlet problem.

A common approach (see e.g.\ \cite{BurMil71} and
references therein)
to incorporate this from the beginning
is by combining the single layer and double layer equation
using a linear superposition.
By this the solutions of the outside problem with Neumann
boundary conditions can be removed.
Because of the singular kernel in the single layer equation
special care has to be taken with the implementation.
For the more difficult case of billiards with magnetic field
see \cite{HorSmi2000}.

\FloatBarrier
\subsubsection{Derived quantities in terms of the normal derivative function}

As the normal derivative function contains all information
to determine the eigenfunction, it is interesting
to see if this approach can be used to
compute other quantities of interest.
For example,
if one wants to calculate expectation values $\langle \psi | A |\psi\rangle$
of some operator $A$ in the state $\psi$,
one has to ensure that the eigenfunction $\psi$
is normalized, 
$\langle \psi | \psi \rangle = \int_\Omega | \psi(\bfq)|^2 \;\ud^2q =1$.
In principle this could be done by considering
$\left( \langle \widetilde{\psi}|\widetilde{\psi}\rangle\right)^{-1} 
  \widetilde{\psi}(\bfq)$
of an unnormalized eigenfunction $\widetilde{\psi}$.
However, an accurate computation of 
$\langle \widetilde{\psi}|\widetilde{\psi}\rangle$
using \eqref{eq:wave-function-Y0} is quite time consuming.
Fortunately, there is a simpler way to achieve a normalized $\psi$:
If $\psi$ is a normalized eigenfunction with eigenvalue $E=k^2$
and $u(s)$ is the corresponding normal derivative
then we have the following normalization condition for $u(s)$ 
\cite{BerWil84,Boa94}
\begin{equation} \label{eq:normalization-of-psi}
  \frac{1}{2} \oInt_{\partial \Omega} 
   \bfn(s)\bfq(s) \, | u(s) |^2 \,\ud s = k^2 \;\;.
\end{equation}
If $\widetilde{u}(s)$ is an unnormalized normal derivative,
then one obtains by
\begin{equation}
  u(s) = \frac{\sqrt{2} \, k}
      {\sqrt{\oInt_{\partial\Omega} 
\bfn(s)\bfq(s) \, | \widetilde{u}(s) |^2 \,\ud s}}  \widetilde{u}(s) \;\
\end{equation}
a normalized one.
Starting with a normal derivative normalized in this way, any 
other quantities (e.g.\ expectation values) 
determined in terms of 
$u(s)$ have the correct normalization.

\newcommand{\eindreiDbildkardioide}[3]{

\vspace*{-0.5cm}

   \hspace*{-0.25cm}
   \begin{minipage}{9cm}
     \centerline{\PSImagx{result_#1.ps}{8.3cm}}
   \end{minipage}
   \begin{minipage}{9cm}
     \centerline{\PSImagx{ft_3d_kardioide_#1.ps}{7.8cm}}
   \end{minipage}

\vspace*{-7cm}
\vspace*{-#3}

   \centerline{$n=#1$, odd symmetry}

\vspace*{7cm}


   \hspace*{-0.75cm}
   \begin{minipage}{9cm}
     \centerline{\PSImagx{wfk_kardioide_d_full_#1.ps}{5.0cm}}
   \end{minipage}
   \begin{minipage}{9cm}
     \centerline{\hspace*{0.5cm}\PSImagx{ft_kardioide_#1.ps}{6.5cm}}
   \end{minipage}


   \hspace*{-0.75cm}
   \begin{minipage}{9cm}
     \centerline{{\PSImagx{odd_n__ft_vert_#1.ps}{8.5cm}}}
   \end{minipage}
   \begin{minipage}{9cm}
      \centerline{\PSImagx{husimi_odd_n__husimi2_pss0_#1.ps}{8.5cm} }
   \end{minipage}

}

\newcommand{\einbildkardi}[3]{
   {#3)  $n=#1$, odd symmetry}

\vspace*{0.25cm}

   \centerline{\PSImagx{wfk_kardioide_d_full_#1.ps}{4.4cm}\hspace*{1.5cm}} 
 
   \vspace*{0.25cm}

   \centerline{\PSImagx{ft_kardioide_#1.ps}{6cm}\hspace*{1.5cm}}

   \hspace*{-0.25cm}
   \PSImagx{odd_n__ft_vert_#1.ps}{7.5cm} 

   \hspace*{-0.25cm}
   \PSImagx{husimi_odd_n__husimi2_pss0_#1.ps}{8.0cm} 
}

\newcommand{\zweibildkardi}[4]{
  \hspace*{0.25cm}
  \begin{minipage}{9.25cm}
    \einbildkardi{#1}{#2}{a}
  \end{minipage}  
  \begin{minipage}{9.25cm}
    \einbildkardi{#3}{#4}{b} 
  \end{minipage}  
 
}

\BILDD{t}
     {

       \eindreiDbildkardioide{}{.}{0cm}
     }
     {Three--dimensional plots of $|\psi_{567}(\BFx)|^2$, 
     $|\widehat{\psi}_{567}(\BFp)|^2$, their corresponding grey--scale 
     pictures and the plot of the radially integrated momentum
     distribution $I_{567}(\varphi)$.
     The momentum distribution $|\widehat{\psi}_{567}(\BFp)|^2$
     is concentrated around the energy shell, which is indicated by the 
     inner circle. 
     This state is localized along the shortest unstable periodic orbit,
     leading to an enhancement of
     $|\widehat{\psi}_{567}(\BFp)|^2$ near to $\varphi=\pi/2,3\pi/2$,
     also seen in the plot of  $I_{567}(\varphi)$ near to the momentum
     direction $\varphi=\pi/2$ (marked by a triangle).
     This localization is also clearly visible in the Husimi representation.
 }
     {fig:cardid-ThreeD-A}

\BILD{t}
     {
      \eindreiDbildkardioide{}{.}{1cm}
     }
     {Same as in the previous figure but for $n=116$. 
     In this case there is 
     no prominent localization neither in position nor in momentum space.}
     {fig:cardid-ThreeD-B}

\BILD{t}
     {
     \zweibildkardi{}{}{}{}
     }
     {The  eigenfunction in a) shows localization along the shortest
      unstable orbit which
      is also reflected in the momentum distributions
      and in the Husimi function. 
      The  eigenfunction in b) is an example which appears to be
      quite delocalized both in position and in momentum space. The pictures 
      look like those  expected (according to the quantum ergodicity theorem)
      for a `typical' eigenfunction.}
     {fig:kardi-two-A}

\afterpage{\clearpage}

This is just the first example out of many highlighting the
importance of the normal derivative for numerical computations
of quantities related to eigenfunctions. For example,
there are explicit expressions in terms of $u_n(s)$ to compute the
\begin{itemize}
  \item normalization of $\psi$, eq.~\eqref{eq:normalization-of-psi},
         \cite{BerWil84,Boa94}
  \item eigenfunction $\psi$, 
        eq.~\eqref{eq:wave-function-Y0}
  \item momentum distribution 
\begin{equation}
           \widehat{\psi}_n(\BFp) 
=\frac{1}{2\pi}\Int_{\Omega}\ue^{-\ui \BFp\BFx}\psi_n (\BFx)\, 
    \ud^2\noBFx 
=-\frac{\ui}{4\pi p_n^2}
       \Int_{\partial\Omega}\ue^{-\ui \BFp\BFx(s)}  
  \BFp\BFx(s)u_n(s)\,\ud s \, ,
\end{equation}
        and radially integrated momentum distribution \cite{Zyc92,BaeSch99}
        \begin{equation}
          I(\varphi) :=\Int_0^{\infty}\left|\widehat{\psi}_n(r,\varphi )
                               \right|^2 r\; \ud r \;\; , 
        \end{equation}
        see         \cite{BaeSch99} for details.
  \item Husimi functions (see e.g.\ \cite{TuaVor95,SimVerSar97})
  \item autocorrelation function of eigenstates \cite{BaeSch2002b}.
\end{itemize}

In figs.~\ref{fig:cardid-ThreeD-A}--\ref{fig:kardi-two-A} 
we show for the cardioid billiard
examples of eigenfunctions in position space, the corresponding
momentum distributions, the angular momentum distributions
(for further details and examples see \cite{BaeSch99})
and the corresponding Husimi functions $H_n(s,p)$.
The first example in fig.~\ref{fig:cardid-ThreeD-A}
shows an example of a scarred state,
i.e.\ an eigenstate which shows localization round an unstable
periodic orbit \cite{Hel84}. Below the three--dimensional
plot of the state is the corresponding density plot (black corresponding to 
high intensity) in which the localization is clearly visible.
Also the corresponding three-dimensional plot of
the momentum distribution $\widehat{\psi}_{567}(\BFp)$
reveals enhanced contributions in the directions $\varphi=\pi/2,3\pi/2$.
This is also seen in the plot of $I_{567}(\varphi)$
which shows that the probability to find the particle with momentum
near $\pi/2$ is significantly enhanced compared to the mean of $1/(2\pi)$.
Another representation is
the Husimi--Poincar\'e representation $H_n(s,p)$
where $s$ corresponds to the arclength coordinate
along the billiard boundary and $p$ corresponds to
the projection of the unit velocity vector after the reflection
on the tangent in the point $s$.
In this picture the localization around 
the unstable orbit is maybe most clearly
seen; the places of high intensity are on the line $p=0$
(perpendicular reflection)
and match perfectly with the position of the orbit.

The second example shown in fig.~\ref{fig:cardid-ThreeD-B} 
is an `ergodic' state,
i.e.\ a state which does not show any significant localization
(as much as something like this is possible at low energies)
neither in position nor in momentum space
(apart from the localization on the energy shell).
This is nicely reflected in the various representations.
Two further examples are shown in fig.~\ref{fig:kardi-two-A} 
where a) is a higher lying scar and b) is another
`typical' state (in the sense of the quantum ergodicity theorem).

\FloatBarrier
\section{Concluding remarks --- or what's left ?}

There are many more issues related to scientific computing
in quantum chaos which I did not mention in these notes.
They for example include visualization techniques,
programming of parallel computers (e.g.\ using PVM or MPI),
or using vector computers etc.
Also the more implementation specific aspects,
including the choice of a programming language have not been
discussed.
A good starting point to learn about computing in quantum chaos
are quantum maps
as their numerics is much easier (one can use a black-box
routine to get all eigenvalues at once)
than for billiard systems, where
more complicated methods have to be used.

\vspace*{0.5cm}

\noindent
{\bf Acknowledgements}

\vspace*{0.25cm}

\noindent
I would like to thank the organizers
of the summer school  
{\it The Mathematical aspects of Quantum Chaos I}
in Bologna,
Mirko Degli Esposti
and Sandro Graffi for all their effort, 
especially Mirko for dealing with
everything (including food and wine ;-) 
in the `Italian way'!
Moreover, I am grateful to Grischa Haag
for many discussions on quantum maps, 
and Ralf Aurich for discussions on the boundary
element method. I would like to thank
Professor Frank Steiner and Silke F\"urstberger
for useful comments on the manuscript
and Professors Uzy Smilansky and Andreas Knauf for useful 
remarks on an earlier version of this text.
I would like to thank Fernando Perez for pointing out
some speed improvements of the {\tt Python} implementation.
Most parts of this work have been done during my stay at the
School of Mathematics, University of Bristol,
and the Basic Research Institute in the Mathematical Sciences,
Hewlett-Packard Laboratories Bristol, UK.
In particular I would to thank Professors Jonathan Keating
and Sir Michael Berry for all their support.
Moreover, I am particularly indebted to Professor Steiner
and the University of Ulm for their support throughout the last months.
I acknowledge partial support by the 
Deutsche Forschungsgemeinschaft under contract No. DFG-Ba 1973/1-1.

\section*{Appendix: Computing eigenvalues of quantum maps}

The first thing when thinking of solving a certain problem
numerically is to decide on the programming language.
There are numerous  possibilities, ranging from Assembler, Fortran, Pascal,
C, C++, Java, etc.
to using packages like Octave, Matlab, Maple or Mathematica.
Here I will use the quite recent scripting language {\tt Python} \cite{PYTHON}.
Of course it is beyond the scope of this text to 
give an introduction to this language; 
several excellent introductions can be found
on the Python homepage.
In addition to the basic Python installation you will
also need the {\tt Numeric} package \cite{Numeric},
which is also simple to install. 
The following programs together with further 
information can be obtained from \cite{MyHomepage}.
If you have been wondering about
the name -- yes it originates from  Monty Python's flying circus,
and at several places the documentation refers to more or less
famous Monty Python sketches.

So here is {\tt pert\_cat.py}
(the full version can be obtained via \cite{MyHomepage}):
{\small
\begin{verbatim}
#!/usr/bin/env python

import cmath
from Numeric import zeros,Float,Complex
from math import sin,pi,sqrt
import LinearAlgebra

def quantum_cat(N,kappa):
    """For a given N and kappa this functions returns the corresponding
       unitary matrix U of the quantized perturbed cat map.
    """
    mat=zeros((N,N), Complex)     # complex matrix with NxN elements
    I=1j                          # predefine sqrt(-1)
    # now fill each matrix element
    # (note: this can be done much faster, see the on-line version)
    for k in range(0,N):          
        for l in range(0,N):
            mat[k,l]=cmath.exp(2.0*I*pi/N*(k*k-k*l+l*l)+ \
                           I*kappa*N/2.0/pi*sin(2.0*pi/N*l))/sqrt(N)
    return(mat)

       
def compute_evals_pcat(N,kappa):
    """For a given N and kappa this functions returns the eigenvalues 
       and eigenphase of the unitary matrix U filled via quantum_cat(N,kappa).
    """
    matU=quantum_cat(N,kappa)     # fill matrix U_N
    evals=LinearAlgebra.eigenvalues(matU) # determine eigenvalues of U_N
        
    # determine phase \in [0,2\pi] from the eigenvalues
    phases_N = arctan2(evals.imag,evals.real) + pi
    # useful to determine level-spacing
    phases = concatenate([phases_N,[phases_N[0]+2.0*pi]])
    return(evals,phases)    
           
### Main (used if pert_cat.py is called as script)
if __name__ == '__main__':
    from string import atoi,atof
    import sys

    # Determine eigenvalues and eigenphases
    (evals,phases)=compute_evals_pcat(atoi(sys.argv[1]),atof(sys.argv[2]))

    for k in range(0,N):           # print eigenvalues
        print("% e % e   % e   % e ") %  \
             (evals[k].real,evals[k].imag,phases[k],abs(evals[k])) 
\end{verbatim}
}

The only drawback of the above code is that the loop to fill the matrix
is slower than a corresponding code in C or Fortran
(notice that there are some very nice ways of overcoming
this by inlining of code or on--the--fly compilation which are presently
being developed for example in the context of SciPy \cite{SciPy}).
However, as {\tt diagonalize} uses the LAPACK library
the most time--consuming part (at least for larger $N$)
is done in an efficient way
(not taking into account the possibility of using ATLAS \cite{ATLAS}
for further speed improvements).

As a first test do (for $N=101$ and $\kappa=0.3$)
\begin{verbatim}
  python pert_cat.py 101 0.3
\end{verbatim}
It will output the (complex) eigenvalues as a sequence $x,y$ pairs.
As a test, whether these all lie on the unit circle
the third column is the absolute value of the eigenvectors.
To plot the resulting data you may use 
\begin{verbatim}
  python pert_cat.py 101 0.3 > pcat_101_0.3.dat
\end{verbatim}
which redirects the output of the program to the file 
{\tt pcat\_101\_0.3.dat}.
To plot the resulting file use your favourite plotting program,
e.g.\ for {\tt gnuplot} \cite{gnuplot} just do
\begin{verbatim}
  plot "pcat_101_0.3.dat" using 1:2 with points
\end{verbatim}

\noindent
Now we would like to compute the level spacing distribution.
To do this let us use an interactive Python session
in which we do
{\small
\begin{verbatim}
from Numeric import *                    # Numeric package
from pert_cat import compute_evals_pcat  # the above pert_cat routines 
from AnalyseData import *                # histogram routine (see below)

N=53
kappa=0.3 
(evals,phases)=pert_cat.compute_evals_pcat(N,kappa);
# sort and unfold phases
s_phases=Numeric.sort(phases)*N/(2.0*pi)

# determine Level spacing
# (by computing the difference of the shifted eigenphases)
spacings=s_phases[1:]-s_phases[0:N]

(x_histogram,y_histogram)=histogram(spacings,0.0,10.0,100)
store_histogram(x_histogram,y_histogram,"histogram.dat")
\end{verbatim}
} 

\noindent
Then use your favourite plotting program to plot the level spacing
distribution. For gnuplot you could do
{\small
\begin{verbatim}
goe_approx(x)=pi/2.0*x*exp(-pi/4*x*x)
gue_approx(x)=32/pi/pi*x*x*exp(-4/pi*x*x)
plot "histogram.dat" w l,goe_approx(x),gue_approx(x),exp(-x)
\end{verbatim}
}

\noindent
Here the routines to compute and store the histogram are in 
{\tt AnalyseData.py} whose core reads 
{\small
\begin{verbatim}
 def histogram(data,min,max,nbins):
    from Numeric import *
    # first select only those which lie in the interval [min,max]
    hdat=compress( ((data<max)*(data>min)),data)
    bin_width=(max-min)/nbins
    # define the bins
    bins=min+bin_width*arange(nbins)
    # determine indices
    inds=searchsorted(sort(hdat),bins)
    inds=concatenate([inds,[len(hdat)]])
    # return bins and normalized histogram
    return(bins,(inds[1:]-inds[:-1])/(bin_width*len(hdat)))

def store_histogram(x_distrib,y_distrib,outdat):
    bin_width=x_distrib[1]-x_distrib[0]
    f=open(outdat,"w")                     # open file for writing
    for k in range(0,len(x_distrib)):
        f.write("% e % e \n" %  (x_distrib[k],y_distrib[k]))
        f.write("% e % e \n" %  (x_distrib[k]+bin_width,y_distrib[k]))
    f.close()     
\end{verbatim}
}

\noindent
Again, for further details and full routines see \cite{MyHomepage}.

\end{document}